\documentclass[12pt]{article}
\pdfoutput=1

\usepackage{draft} 
\usepackage[bookmarks=false]{hyperref}
\hypersetup{
    colorlinks=true, 
    linktoc=all,     
    linkcolor=blue,  
    citecolor=blue,
    urlcolor=blue
}

\usepackage{graphicx,color,subfig}
\usepackage{cite}
\usepackage{mciteplus}
\usepackage{skak}
\usepackage{empheq}
\DeclareFontFamily{OT1}{pzc}{}
\DeclareFontShape{OT1}{pzc}{m}{it}{<-> s * [1.10] pzcmi7t}{}
\DeclareMathAlphabet{\mathpzc}{OT1}{pzc}{m}{it}

\usepackage{amsmath}
\usepackage{mathtools}

\DeclarePairedDelimiter\floor{\lfloor}{\rfloor}

\newcommand\half{{1\over 2}}
\newcommand\eps{\epsilon}

\newcommand{\beq}{\begin{equation}}
\newcommand{\eeq}{\end{equation}}
\newcommand{\al}[1]{\begin{align} #1\end{align}}
\newcommand{\als}[1]{\begin{align}\begin{split} #1\end{split}\end{align}}
\newcommand{\crr}[1]{\left\langle #1 \right\rangle}
\newcommand{\Fcal}{\mathcal{F}}
\newcommand{\PP}{\mathcal{P}}

\begin{document}

\unitlength = .8mm

\begin{titlepage}

	\rightline{ CALT-TH 2016-040 }
	
\begin{center}

\hfill \\
\hfill \\
\vskip 1cm

\title{Bootstrapping the Spectral Function: \\ On the Uniqueness of Liouville \\ and the Universality of BTZ}
	
\author{Scott Collier$^1$, Petr Kravchuk$^2$, Ying-Hsuan Lin$^2$,  Xi Yin$^1$}

\address{
$^1$ Jefferson Physical Laboratory, Harvard University, \\
Cambridge, MA 02138 USA
\\
$^2$ Walter Burke Institute for Theoretical Physics, Caltech, \\
Pasadena, CA 91125, USA
}

\email{scollier@physics.harvard.edu, 
pkravchuk@caltech.edu,
yinhslin@gmail.com,
xiyin@fas.harvard.edu}

\end{center}

\abstract{We introduce spectral functions that capture the distribution of OPE coefficients and density of states in two-dimensional conformal field theories, and show that nontrivial upper and lower bounds on the spectral function can be obtained from semidefinite programming. We find substantial numerical evidence indicating that OPEs involving only scalar Virasoro primaries in a $c>1$ CFT are necessarily governed by the structure constants of Liouville theory. Combining this with analytic results in modular bootstrap, we conjecture that Liouville theory is the unique unitary $c>1$ CFT whose primaries have bounded spins. We also use the spectral function method to study modular constraints on CFT spectra, and discuss some implications of our results on CFTs of large $c$ and large gap, in particular, to what extent the BTZ spectral density is universal.
}

\vfill

\end{titlepage}

\eject

\begingroup
\hypersetup{linkcolor=black}
\tableofcontents
\endgroup

%

\section{Introduction} 

Enormous progress in the conformal bootstrap program has been made in recent years based on semidefinite programming \cite{Rattazzi:2008pe, Rychkov:2009ij, Poland:2010wg, Poland:2011ey, ElShowk:2012ht,Kos:2013tga, Beem:2013qxa, Beem:2014zpa, El-Showk:2014dwa,Chester:2014fya,Chester:2014mea,Chester:2014gqa,Bae:2014hia,Chester:2015qca,Iliesiu:2015qra,Kos:2015mba,Beem:2015aoa,Lemos:2015awa,Lin:2015wcg,Collier:2016cls,Lin:2016gcl}. Typically, one aims to bound the scaling dimensions and OPE coefficients of the first few operators in the spectrum based on unitarity and crossing invariance of the 4-point function. Such methods are most powerful in constraining CFTs with simple low lying spectrum, but become less constraining when the spectrum becomes dense. 

In this paper we introduce the {\it spectral function method}, which allows for constraining not just the gap or the first few OPE coefficients but the distribution of OPE coefficients over a wide range of scaling dimensions. While the method can be applied to CFTs in any spacetime dimension, we focus on two-dimensional unitary CFTs, where the spectral functions are defined by truncating the Virasoro conformal block decomposition of a 4-point function in the scaling dimension of the internal primaries, evaluated at the self-dual cross ratio, or by truncating the Virasoro character decomposition of a partition function in the scaling dimension of the primaries, evaluated at the self-dual modulus. More precisely, consider a scalar 4-point function\footnote{The notation $\phi'(\infty)$ stands for $\lim_{z,\bar z\to \infty}z^{2h_\phi}\bar z^{2\bar h_\phi}\phi(z,\bar z)$.}
\ie
g(z,\bar z) \equiv \langle \phi(0) \phi(1) \phi(z,\bar z) \phi'(\infty)\rangle = \sum_{s,\Delta} C_{s,\Delta}^2 {\cal F}_{s,\Delta}(z,\bar z),
\fe
where ${\cal F}_{s,\Delta}(z,\bar z)$ are Virasoro conformal blocks for an internal primary of dimension $\Delta$ and spin $s$. The corresponding spectral function is defined by truncating the Virasoro conformal block decomposition of the four-point function
\ie
f(x) = {1\over g(z=\bar z={1\over 2})} \sum_{s,\Delta\le x} C_{s,\Delta}^2 {\cal F}_{s,\Delta}(z=\bar z = {1\over 2}).
\fe
Of course, for a compact CFT with a discrete spectrum, $f(x)$ will be composed of step functions. If the CFT is non-compact, then typically $f(x)$ will be a monotonically increasing smooth function that takes value between 0 and 1. We will see that the crossing equation
\ie\label{crossingeqn}
\sum_{\Delta,s} C_{s,\Delta}^2 \left[ {\cal F}_{s,\Delta}(z,\bar z) - {\cal F}_{s,\Delta}(1-z,1-\bar z) \right] = 0
\fe
combined with additional assumptions on the spectrum lead to upper and lower bounds on $f(x)$. Numerically, the crossing equation can be utilized by applying to (\ref{crossingeqn}) linear functionals spanned by the basis $\partial_z^n \partial_{\bar z}^m|_{z=\bar z=1/2}$, for odd $n+m\leq N$. The resulting upper and lower bounds on $f(x)$ (which are rigorous bounds although not optimal) will be denoted $f^+_N(x)$ and $f^-_N(x)$.

If we make the assumption that the CFT contains {\it only} scalar Virasoro primaries, we find that $f_N^+(x)$ and $f_N^-(x)$ become closer as $N$ increases, for various values of the central charge $c>1$. We {\bf conjecture} that both converge to the spectral function of Liouville theory, which can be computed by integrating the square of the DOZZ structure constants \cite{Dorn:1994xn,Zamolodchikov:1995aa} times the Virasoro conformal blocks. 
Note that this approach can be extended to the 4-point function involving a pair of different primaries, leading to spectral functions that encode the most general structure constants of the CFT.

Convergence of the upper and lower bounds $f^\pm_N(x)$ to the same value $f_\infty(x)$ is related to the completeness of the derivatives of scalar Virasoro blocks in a suitable space of functions. Conversely, this completeness statement implies the uniqueness of the solution to the crossing equations. We propose a numerical test for the completeness and find compelling evidence suggesting that it holds. 
We moreover obtain numerical approximations to the (conjecturally) unique solution to the crossing equations, which reproduce the DOZZ spectral function with high accuracy. In contrast to the semidefinite methods, this linear approach does not rely on the assumption that the OPE coefficients are real. The linear and semidefinite results above therefore lead us to conjecture that \emph{the DOZZ structure constants are the unique solution to the crossing equations for (not necessarily unitary) CFTs with only scalar primaries (of non-negative scaling dimensions) and }$c>1$.

Interestingly, we find that the bounds on the spectral function $f_N^\pm(x)$ exist for external operator dimensions $\Delta_\phi\geq {c-1\over 16}$ (${3\over 4}$ of the Liouville threshold), and converge to a step function when $\Delta_\phi$ is equal to ${c-1\over 16}$. When $\Delta_\phi<{c-1\over 16}$, the crossing equation cannot be satisfied with only scalar internal primaries, ruling out the possibility of such operators.\footnote{This was also observed in unpublished work of Balt van Rees.} We will see that all of these are in agreement with the analytic continuation of Liouville 4-point functions.

A caveat in the above uniqueness claim is that we have assumed a non-degenerate scalar spectrum. If degeneracies are allowed, then the operator algebra of Liouville CFT tensored with a topological quantum field theory (TQFT) (or equivalently, a finite dimensional commutative non-unital Frobenius algebra) would also solve the crossing equation. In fact, such a TQFT can always be ``diagonalized" by a basis change, and amounts to superselection sectors. We will give partial arguments suggesting that under our assumptions, ``Liouville$\,\otimes\,$TQFT" is the only possibility.

If we further invoke modular invariance, it will turn out that demanding that a unitary CFT contains only primaries of spins in a finite range ($s\leq s_{max}$ for some finite $s_{max}$) implies that the CFT must have a non-compact spectrum with only scalar primaries, and that the spectral density $\rho(\Delta)$ must be that of Liouville theory, namely
\ie
\rho(\Delta) \propto {1\over \sqrt{\Delta - {c-1\over 12}}}.
\fe
This leads us to conjecture that {\it Liouville theory is the unique unitary CFT with $c>1$ whose primaries have bounded spins.} 

The spectral function method also can be applied to  modular bootstrap.  In this context, we write the torus partition function as
\ie
Z(\tau,\bar\tau) \equiv {\rm Tr} \, q^{L_0-{c\over 24}} \bar q^{\bar L_0 - {c\over 24}} = \sum_{\Delta,s} d_{\Delta,s} \chi_{\Delta,s}(\tau,\bar\tau),
\fe
where $\chi_{\Delta,s}$ is the Virasoro character associated with a primary of dimension $\Delta$ and spin $s$ and $d_{\Delta,s}$ is the degeneracy. The modular spectral function is defined by truncating the Virasoro character decomposition of the partition function
\ie
f_{\rm mod}(x) = {1\over Z(\tau=-\bar\tau=i)} \sum_{s,\Delta\le x} d_{\Delta,s} \chi_{\Delta,s}(\tau=-\bar\tau=i).
\fe
Once again, upper and lower bounds $f_{{\rm mod},N}^\pm(x)$ can be obtained by acting on the modular crossing equation
\ie
\sum_{\Delta,s} d_{\Delta,s} \left[\chi_{\Delta,s}(\tau,\bar\tau) - \chi_{\Delta,s}(-1/\tau,-1/\bar\tau) \right]=0
\fe
with linear functionals spanned by the basis $(\tau\partial_\tau)^n(\bar\tau\partial_{\bar\tau})^m|_{\tau=-\bar\tau=i}$, for odd $n+m\leq N$. In \cite{Collier:2016cls} (improving upon \cite{Hellerman:2009bu, Friedan:2013cba}), an upper bound $\Delta_{\rm mod}(c)$ on the gap in the scaling dimenions was computed numerically as a function of the central charge $c$. When this bound is saturated, the entire spectrum is fixed by modular invariance, and is determined by the zeroes of the optimal linear functional acting on the Virasoro characters. We will see in examples of small $c$ (between 2 and 8) that under the assumption of maximal dimension gap, $f_{mod,N}^+(x)$ and $f_{mod,N}^-(x)$ converge with increasing $N$ to step functions, corresponding to the spectral functions of known theories.

For larger values of $c$, even when the dimension gap is maximized, the convergence of the bounds $f_{\rm mod}^\pm(x)$ to a sum of step functions is difficult to see numerically, because a good approximation of the optimal linear functional requires larger values of $N$, and because the step function feature becomes invisible due to an exponentially large spectral density. Nonetheless, for $50\leq c\leq 300$, we find empirically that the horizontal average $\overline{f_{{\rm mod},N}}(x)$ of the upper and lower bounds converges rather quickly with $N$, and the result is in good agreement with the total contribution from thermal $AdS_3$ and BTZ black hole \cite{Banados:1992wn} to the gravity partition function, which results in the modular spectral function
\ie\label{modbtz}
f_{\rm mod}^{\rm BTZ}(x) = {3\over 4} + {1\over 4} {\rm Erf}\left(\sqrt{6\pi\over c}(x - {c\over 6}) \right) + \left({1\over c}~{\rm corrections}\right).
\fe
Note that this asymptotic spectral function at large $c$ is nontrivial when the dimension $x$ lies in a window of width $\sim \sqrt{c}$ around $c/6$. The agreement with the numerical bounds confirms the validity of the effective field theory of pure gravity in $AdS_3$ in the canonical ensemble, for temperatures of order 1 in AdS units.

Curiously, BTZ black holes corresponding to operators of scaling dimension $\Delta$ in the range ${c\over 12}<\Delta<{c\over 6}$ never dominate the canonical ensemble, and yet have macroscopic (AdS scale) horizon, provided that $\Delta-{c\over 12}$ scales with $c$. While the naive expectation from effective field theory is that the Bekenstein-Hawking entropy formula should be a valid counting of the microstates of such BTZ black holes, it is unclear to us whether this is a universal property of CFTs with sufficiently large gap.\footnote{Such a universality would in particular require the dimension gap bound $\Delta_{\rm mod}(c)$ to have asymptotic slope ${1\over 12}$, namely ${d\Delta_{\rm mod}(c)\over dc}\to {1\over 12}$, $c\to \infty$, which is not ruled out by the result of \cite{Collier:2016cls} but remains unproven (with no numerical evidence either).}
In principle, the modular spectral function bounds at large $c$ should either confirm or disprove such statements. To probe the density of states in the regime $\Delta = y c$ for ${1\over 12}<y<{1\over 6}$ and large $c$ would require exponential precision in determining the modular spectral function, which is beyond our current numerical capability.

This paper is organized as follows. In section 2 we introduce the spectral function for the scalar 4-point function in a 2D CFT, and explain how to obtain upper and lower bounds $f_N^\pm(x)$ from semidefinite programming. We then specialize to the case where only scalar primaries are present, and demonstrate the convergence of the bounds toward the Liouville spectral function. In section 3, we examine the completeness of scalar Virasoro conformal blocks which would be implied by the aforementioned convergence, and we give numerical evidence that the completeness indeed holds. We then present analytic arguments based on modular invariance that a unitary CFT with $c>1$ and Virasoro primaries of bounded spin must be a non-compact CFT with the same spectral density as that of Liouville. This together with the result of section 2 strongly supports the conjecture that Liouville theory is the only CFT with bounded spins. In section 4, we analyze the numerical bounds on the modular spectral function in a number of examples. We conclude with a discussion on the universality of the BTZ spectral density in large-$c$ CFTs with large gaps.

\section{Spectral function bounds from semidefinite programming}

\subsection{A sphere four-point spectral function}\label{sec:4ptSpectral}

We begin by considering the conformal block decomposition of the sphere four-point function of a pair of scalar Virasoro primary operators $\phi_1,\phi_2$ of dimensions $\Delta_1,\Delta_2$,
\ie\label{mixedc}
g_{12}(z,\bar z) =& \langle \phi_1(z,\bar z)\phi_2(0)\phi_2(1)\phi_1'(\infty)\rangle\\
=& \sum_{s=0}^\infty\sum_{\Delta\in\mathcal{I}_{12;s}}C_{12;s,\Delta}^2\mathcal{F}_{12;s,\Delta}(z,\bar z).
\fe
Here $\mathcal{I}_{12;s}$ is the set of scaling dimensions of spin-$s$ primary operators in the $\phi_1 \phi_2$ OPE and $C_{12;s,\Delta} = C_{\phi_1\phi_2\mathcal{O}}$ is the OPE coefficient corresponding to the fusion of $\phi_1$ and $\phi_2$ into the primary $\mathcal{O}$ with dimension $\Delta$ and spin $s$.\footnote{When the operator spectrum is degenerate, $C_{12;\Delta,s}^2$ would be replaced by the sum of squares of OPE coefficients of all primaries of dimension $\Delta$ and spin $s$.} The OPE coefficients are real in a unitary CFT. The conformal block ${\cal F}_{12;\Delta,s}$ takes the form
\ie
\mathcal{F}_{12;s,\Delta}(z,\bar z) =& F^{\rm Vir}_c\left({\Delta_1\over 2},{\Delta_2\over 2},{\Delta_2\over 2},{\Delta_1\over 2};{\Delta+s\over 2};z\right) \bar F^{\rm Vir}_c\left({\Delta_1\over 2},{\Delta_2\over 2},{\Delta_2\over 2},{\Delta_1\over 2};{\Delta-s\over 2};\bar z\right) \\
&+F^{\rm Vir}_c\left({\Delta_1\over 2},{\Delta_2\over 2},{\Delta_2\over 2},{\Delta_1\over 2};{\Delta-s\over 2};z\right) \bar F^{\rm Vir}_c\left({\Delta_1\over 2},{\Delta_2\over 2},{\Delta_2\over 2},{\Delta_1\over 2};{\Delta+s\over 2};\bar z\right),
\fe
where $F^{\rm Vir}_c(h_1,h_2,h_3,h_4;h;z)$ is the holomorphic Virasoro conformal block with external primaries of weight $h_i$ and an internal primary of weight $h$, in a CFT with central charge $c$. Note that in writing the four-point function this way we have assumed a parity-invariant spectrum.\footnote{In what follows we specialize to the case where the spectrum only has scalar primary operators, so this distinction is trivial.}  An efficient method for computing Virasoro conformal blocks is Zamolodchikov's recurrence relation \cite{Zamolodchikov:1985ie,Zamolodchikov:1995aa}, which we review in Appendix~\ref{App:Zam}.  It computes the blocks as expansions in the ``nome'' $q(z)$, defined as
\ie
q(z) \equiv \exp(i\pi\tau(z)),
\quad
\tau(z) \equiv {i F(1-z) \over F(z)}, \quad F(z) = {}_2F_1({1/2}, {1/2}, 1 | z).
\fe
Note that as $z$ ranges over the complex plane, $q(z)$ takes value in an eye-shaped region on the unit disc, and the expansion of a conformal block in $q$ converges on the entire unit disc. In the numerical approach, we apply Zamolodchikov's recurrence relation up to a finite depth $d_q$, which generates the correct $q$-series coefficients up to order $q^{d_q}$. We then truncate the conformal block to this order as an approximation of the exact block. 

It follows from the associativity of OPE that the four-point function is crossing symmetric, which amounts to the crossing equation
\ie
\sum_{s=0}^\infty \sum_{\Delta\in\mathcal{I}_{12;s}}C^2_{12;s,\Delta}\left[\mathcal{F}_{12;s,\Delta}(z,\bar z) - \mathcal{F}_{12;s,\Delta}(1-z,1-\bar z)\right]=0.
\fe
This relation puts highly nontrivial constraints on the spectrum and OPE coefficients of the CFT, some of which were analyzed in \cite{Kim:2015oca, Chang:2015qfa, Lin:2015wcg, Chang:2016ftb, Lin:2016gcl}. In previous works, one typically either focuses on a limit of the crossing equation in the cross ratio and extracts asymptotic properties of the spectrum, or numerically bounds the scaling dimension and OPE coefficients of the first few operators from the positivity assumption on $C_{12;\Delta,s}^2$.

We now introduce a ``spectral function'' that captures the distribution of OPE coefficients over a range of scaling dimensions of primaries in the $\phi_1\phi_2$ OPE, defined through the conformal block decomposition of the four-point function evaluated at the crossing-symmetric point  $z=\bar z = \half$, truncated on the dimension of internal primary operators:
\ie
f(\Delta_*)\equiv {1\over g_{12}(1/2,1/2)}\sum_{s=0}^{\floor{\Delta_*}}\sum_{\Delta\in\mathcal{I}_{12;s},\Delta\le\Delta_*}C^2_{12;s,\Delta}\mathcal{F}_{12;s,\Delta}(1/2,1/2).
\fe
Note that due to the unitarity bound, $f(\Delta_*)$ receives no contribution from primary operators with spin $s>\Delta_*$. By definition, obviously, $f(\Delta_*)$ is a non-decreasing function that takes value between 0 and 1.

One can place bounds on the spectral function using semidefinite programming as follows. We would like to either maximize or minimize the spectral function subject to the crossing equation expanded around $z=\bar z={1\over 2}$
\ie
0 =& \sum_{s=0}^\infty\sum_{\Delta\in\mathcal{I}_{12;s}}\left.C^2_{12;s,\Delta}\partial_z^m\partial_{\bar z}^n\mathcal{F}_{12;s,\Delta}(z,\bar z)\right|_{z=\bar z=\half},~m+n\text{ odd}.
\fe
Note that $z={1\over 2}$ corresponds to the nome $q = e^{-\pi}$, thus the $q$-expansion of conformal blocks converges rather quickly at this point.
Consider a set of coefficients $y_{0,0}$ and $y_{m,n}$ ($m+n$ odd) such that
\ie\label{eq:FourPtUpperInequality}
(y_{0,0}-\Theta(\Delta_*-\Delta))\mathcal{F}_{12;s,\Delta}(1/2,1/2) + \sum_{m+n\text{ odd}}\left.y_{m,n}\partial_z^m\partial_{\bar z}^n\mathcal{F}_{12;s,\Delta}(z,\bar z)\right|_{z=\bar z= \half}\ge 0. 
\fe
Here $\Theta(\Delta_*-\Delta)$ is the step function that takes value 1 for $\Delta\leq\Delta_*$ and 0 otherwise. $(\Delta, s)$ runs through all possibly allowed values of dimension and spin in the OPE. We could place additional assumptions on the spectrum by restricting the range of $(\Delta,s)$ in (\ref{eq:FourPtUpperInequality}). For instance, if we are to impose a dimension gap $\Delta_{\rm gap}$ or twist gap $t_{\rm gap}$, then we have respectively $\Delta\geq \text{max}(s,\Delta_{\rm gap})$ or $\Delta\geq s +t_{\rm gap}$ for the spin-$s$ (non-vacuum) primaries.\footnote{In the case of a compact CFT, one must take care to additionally impose~\eqref{eq:FourPtUpperInequality} and~\eqref{eq:FourPtLowerInequality} on the vacuum block.}

We shall seek the \emph{minimal} $y_{0,0}$ such that (\ref{eq:FourPtUpperInequality}) holds, which we denote by $y_{0,0}^{\rm min}$. 
It follows that
\ie
f(\Delta_*) =& {1\over g_{12}(1/2,1/2)}\sum_{s=0}^\infty \sum_{\Delta\in \mathcal{I}_{12;s}}C^2_{12;s,\Delta}\mathcal{F}_{12;s,\Delta}(1/2,1/2)\Theta(\Delta_*-\Delta)
\\
\le & {1\over g_{12}(1/2,1/2)}\sum_{s=0}^\infty\sum_{\Delta\in \mathcal{I}_{12;s}} C^2_{12;s,\Delta}
\\
&\times \left[y_{0,0}^{\rm min}\mathcal{F}_{12;s,\Delta}(1/2,1/2)+\sum_{m+n\text{ odd}}\left.y_{m,n}\partial_z^m\partial_{\bar z}^n\mathcal{F}_{12;s,\Delta}(z,\bar z)\right|_{z=\bar z = \half}\right]\\
=& y_{0,0}^{\rm min},
\fe
where we have invoked unitarity by making use of the non-negativity of the squared structure constants, and applied the crossing equation. In other words, $y_{0,0}^{\rm min}$ is an upper bound on the value of the spectral function at $\Delta_*$.

Likewise, if we minimize $w_{0,0}$ subject to
\ie\label{eq:FourPtLowerInequality}
(w_{0,0}+\Theta(\Delta_*-\Delta))\mathcal{F}_{12;s,\Delta}(1/2,1/2) + \sum_{m+n \text{ odd}}\left. w_{m,n}\partial_z^m\partial_{\bar z}^n\mathcal{F}_{12;s,\Delta}(z,\bar z)\right|_{z=\bar z=\half}\ge0,
\fe
then 
\ie
f(\Delta_*) \ge
& -w_{0,0}^{\rm min},
\fe
i.e., $-w_{0,0}^{\min}$ is a lower bound on the value of the spectral function at $\Delta_*$.

To obtain these bounds numerically we need to restrict to a finite subset of linear functionals acting on the crossing equation. We will do so by restricting the sums in~\eqref{eq:FourPtUpperInequality} and~\eqref{eq:FourPtLowerInequality} to odd  $m+n\le N$; we refer to $N$ as the ``derivative order.'' The upper and lower bounds on the spectral function derived from the above minimization procedure using linear functionals up to derivative order $N$ will be denoted $f^+_N(\Delta_*)$ and $f^-_N(\Delta_*)$, respectively. While these bounds at every $N$ are rigorous by themselves, the optimal bounds are obtained by extrapolating to the $N\to\infty$ limit.

The numerical implementation of the above procedure is performed using the SDPB package \cite{Simmons-Duffin:2015qma}, with two practical modifications. Firstly, we will need to truncate the spectrum: while the application of SDPB does not require cutting off the dimension spectrum from above, a sufficiently large but finite truncation on the spin is necessary. In principle, the spin truncation means that we would not be taking into account all inequalities obeyed by the coefficients $y_{0,0}$ and $y_{m,n}$, resulting in stronger-than-correct bounds on the spectral function.\footnote{Indeed, the discussion in section~\ref{sec:ModularAnalytic} shows precisely why it is dangerous to truncate spectra of primaries on their spins.} Nonetheless, working at a fixed derivative order $N$, we generally find that the spectral function bounds stabilize to within numerical precision once the maximal spin $s_{\rm max}$ is taken to be sufficiently large (empirically, $s_{\rm max}$ at order $N$ is sufficient). For the application to theories with only scalar primaries in the next few subsections, of course, we do not need to worry about the spin truncation being sufficiently large. In this case, however, we must be especially careful in taking the truncation on the $q$-series of the conformal blocks to be sufficiently large, as the corrections to the approximate blocks would introduce nonzero spin primary contributions.

Secondly, since SDPB deals with the question of whether there exists a linear combination of polynomials $p_i(x)$ that is non-negative for all $x\ge 0$, the above minimization problem must be recast in the form of inequalities on polynomial functions of $\Delta$ on a semi-infinite line.
For instance, suppose we impose a lower bound $\Delta_s^*$ on the dimension of spin-$s$ primaries as part of the a priori assumptions on the spectrum, then (\ref{eq:FourPtUpperInequality}) is equivalently written as
\ie\label{eq:2Inequalities}
y_{0,0}\mathcal{F}_{12;s,\Delta}(1/2,1/2)+\sum_{m+n\text{ odd}}y_{m,n}\partial_z\partial_{\bar z}^n\left.\mathcal{F}_{12;s,\Delta}(z,\bar z)\right|_{z=\bar z=\half}\ge & 0,~\Delta\ge \Delta_*, \\
(y_{0,0}-1)\mathcal{F}_{12;s,\Delta}(1/2,1/2)+\sum_{m+n\text{ odd}}y_{m,n}\partial_z\partial_{\bar z}^n\left.\mathcal{F}_{12;s,\Delta}(z,\bar z)\right|_{z=\bar z=\half}\ge & 0,~\Delta_s^* \leq \Delta<\Delta_*.
\fe
By default, $\Delta_s^*$ can be set to the unitarity bound.
While the first inequality in (\ref{eq:2Inequalities}) can be implemented in SDPB by a simple shift in the variable $\Delta$, the second inequality which holds for $\Delta$ in an interval is more subtle. It is handled\footnote{This trick is due to David Simmons-Duffin.} by converting the inequality to one on the semi-infinite line by a change of variable $\Delta = {(\tilde{\Delta}\Delta_*+\Delta_s^*)/(\tilde\Delta+1)}$; now $\Delta_s^*\leq\Delta<\Delta_*$ amounts to $\tilde\Delta\geq 0$.

\subsection{Bounding the spectral function in a CFT with only scalar primaries}
\label{ulbounds}

We now specialize to the case of 
 CFT with only scalar primary operators. 
We do not specify the normalization of the primaries; as far as the spectral function is concerned, the external primaries are effectively normalized through the 4-point function (thus capturing relative OPE coefficients). This allows us to deal simultaneously with compact and non-compact CFTs. (By a non-compact CFT, we mean one with continuous spectrum and no $SL(2,\bR)\times SL(2,\bR)$-invariant vacuum.) As alluded to in the introduction, there is only one known unitary CFT with $c>1$ of this type, namely Liouville theory, and we will compare our bounds to the Liouville spectral function which can be obtained by numerically integrating the known OPE coefficients (given by DOZZ formula \cite{Dorn:1994xn,Zamolodchikov:1995aa}, as reviewed in Appendix~\ref{dozzreview}) with the Virasoro conformal blocks.

We can write the four-point function involving a pair of primaries $\phi_1$, $\phi_2$ as
\ie
g_{12}(z,\bar z) =&\int_0^\infty d\Delta~C^2_{12;0,\Delta}\mathcal{F}_{12;0,\Delta}(z,\bar z),
\fe
and the spectral function as
\ie
f(\Delta_*) = {1\over g_{12}(1/2,1/2)}\int_0^{\Delta_*}d \Delta~C^2_{12;0,\Delta}\mathcal{F}_{12;0,\Delta}(1/2,1/2).
\fe
This accommodates both continuous and discrete spectra (in the latter case the integral will receive contributions from delta-functions). To place bounds on $f(\Delta_*)$, we simply solve the minimization problem (\ref{eq:FourPtUpperInequality}), (\ref{eq:FourPtLowerInequality}) for $s=0$ only. This is implemented with SDPB with a given set of $c$, $\Delta_1$ and $\Delta_2$, while scanning over a range of $\Delta_*$, at increasing derivative orders $N$.

\begin{figure}[h!]
\centering
\subfloat[]{\includegraphics[width = .49\textwidth]{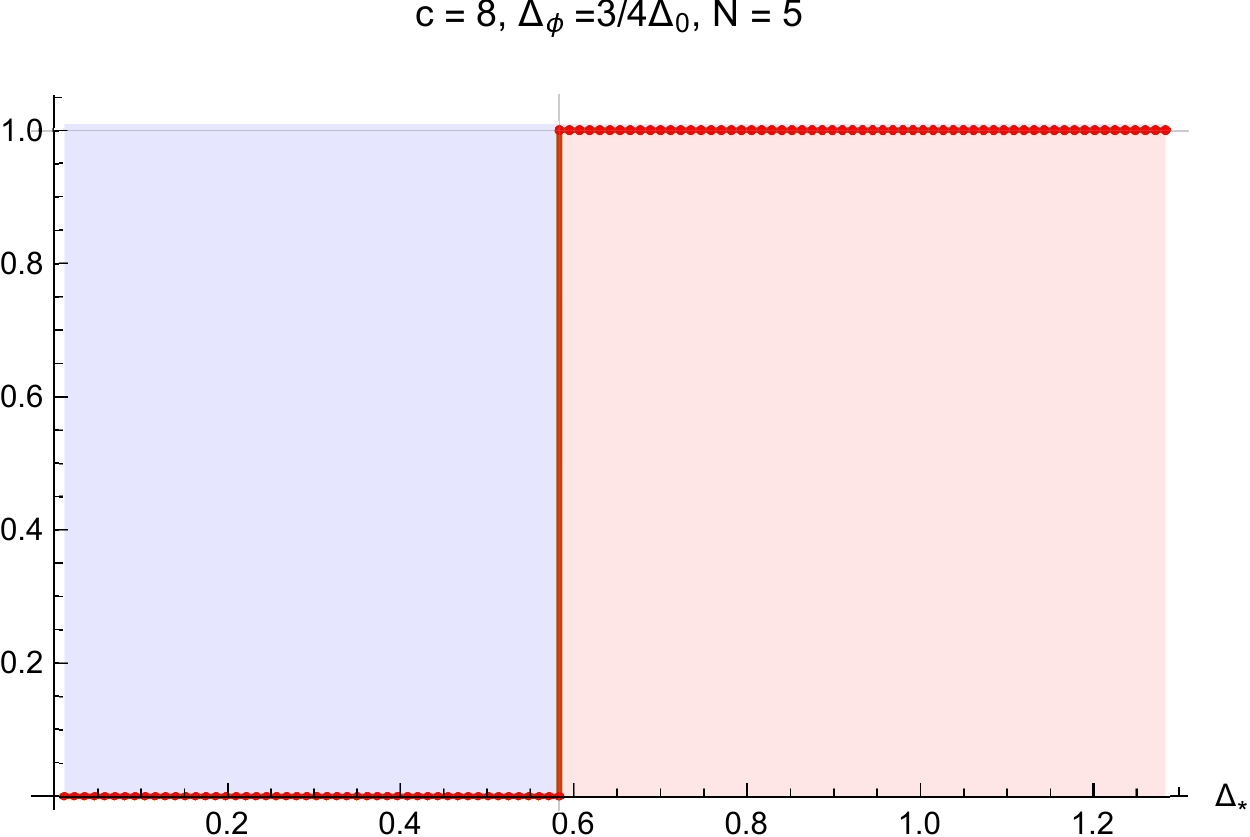}}~
\subfloat[]{\includegraphics[width=.49\textwidth]{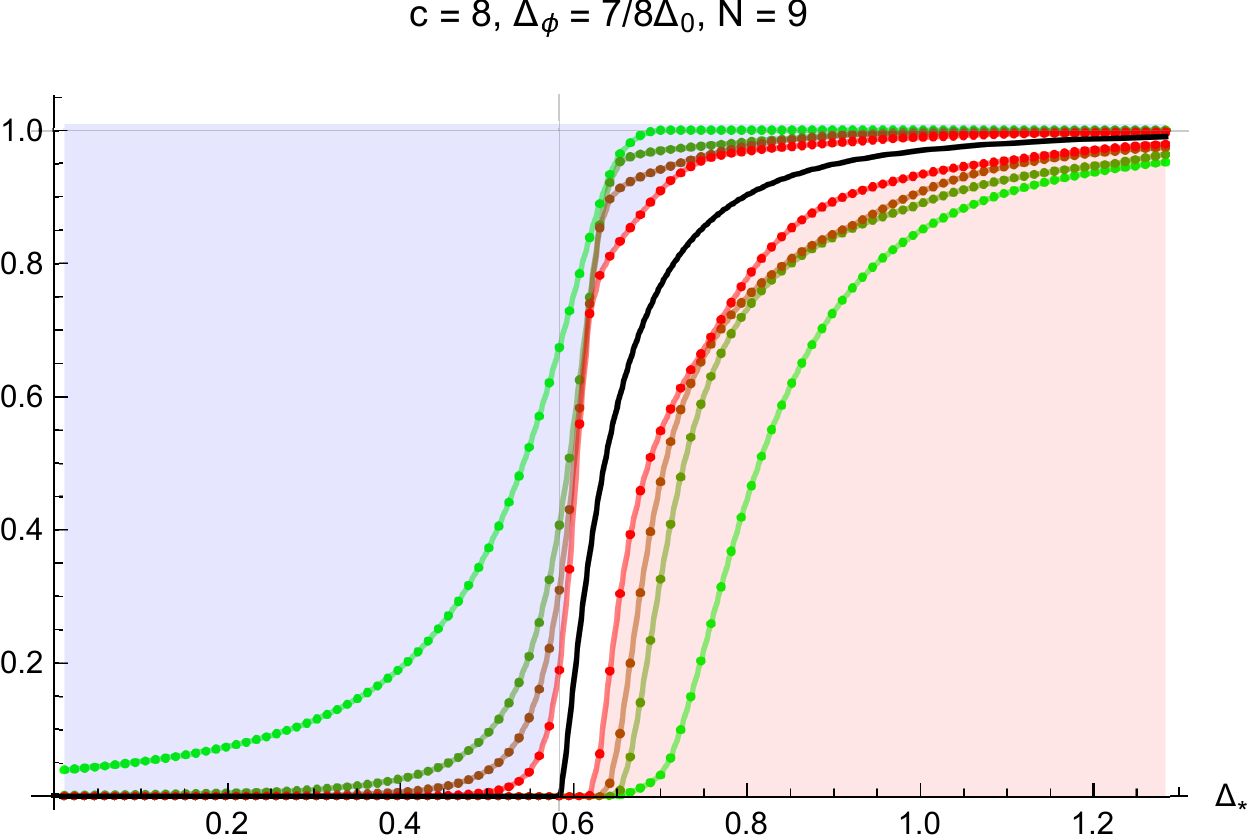}}
\\
\subfloat[]{\includegraphics[width=.49\textwidth]{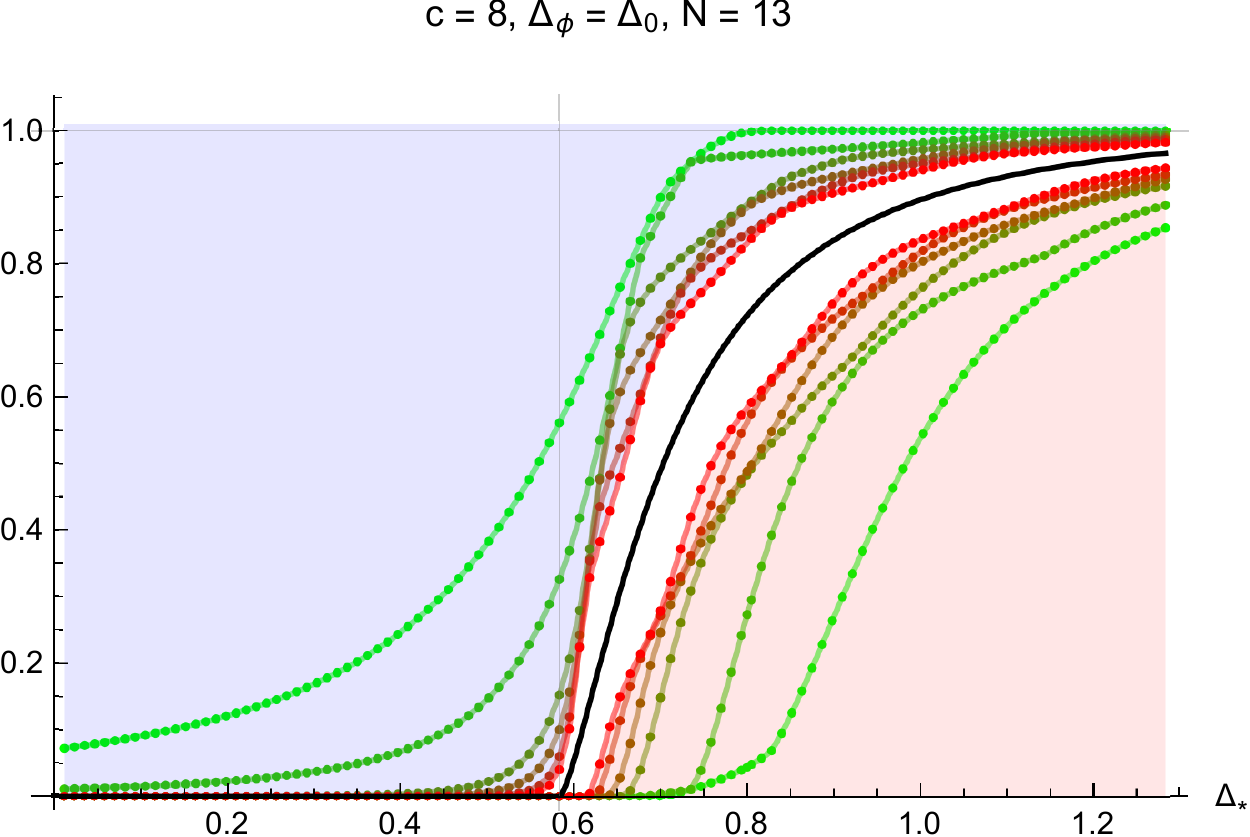}}~
\subfloat[]{\includegraphics[width=.49\textwidth]{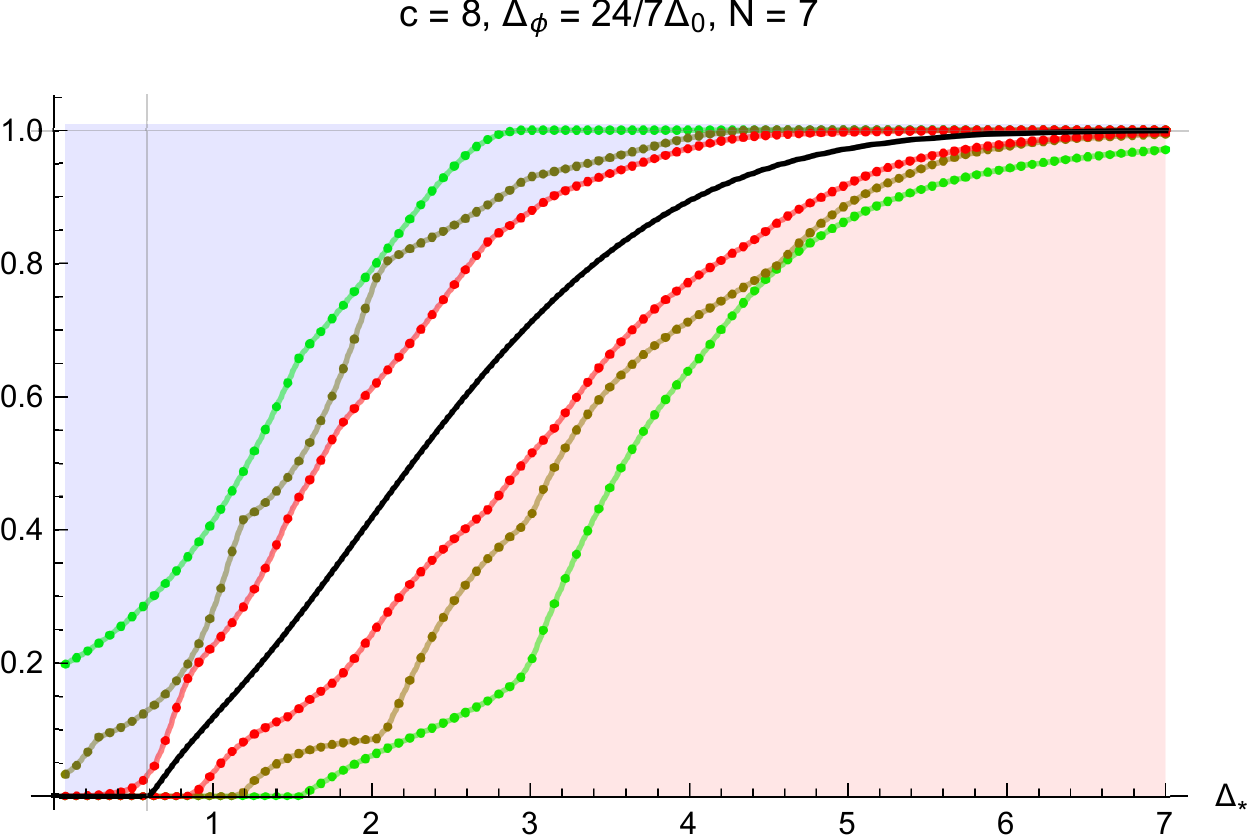}}
\caption{Upper and lower bounds on the spectral function from linear functionals of increasing derivative order (from green to red), assuming only scalar primaries for $c=8$ with $\Delta_\phi /  \Delta_0 = {3 \over 4}, {7 \over 8}, 1, {24 \over 7}$.
In all cases, the shaded regions are excluded and the black curve denotes the corresponding spectral function of (analytically continued) Liouville theory.}\label{fig:LiouvilleBounds1}
\end{figure}

First, we consider the case where all external operators in the four-point function have the same scaling dimension (above or below the Liouville threshold, $\Delta_0 \equiv 2\xi$). Our results for $c=8$ are summarized in Figure~\ref{fig:LiouvilleBounds1}.
We observe that as the derivative order $N$ increases, the upper and lower bounds approach one another, narrowing the allowed range of the spectral function. Both bounds appear to be converging upon the spectral function of Liouville theory (whose background charge $Q$ is related to $c$ by $c=1+6Q^2$), which sits in the middle of the allowed window.

There exist solutions to the scalar-only crossing equations when the external operator dimension drops below the Liouville threshold, so long as $\Delta_{\phi}\ge {3\over 4}\Delta_0$. For $\Delta_{\phi}< {3\over 4}\Delta_0$, solutions to the crossing equations with only scalar primaries in the OPE are excluded by our numerical analysis. When $\Delta_{\phi} = {3\over 4}\Delta_0$, we find that the upper and lower bounds on the spectral function converge quickly to a step function, i.e., $f^+_N(\Delta_*)\approx f_-^N(\Delta_*)\approx\Theta(\Delta_*-2\xi)$, already at small derivative order $N$. This case and an example where $\Delta_\phi$ lies in between ${3\over 4}\Delta_0$ and the Liouville threshold are included in Figure~\ref{fig:LiouvilleBounds1}.

In fact, for $\Delta_\phi\in({c-1\over 16},{c-1\over 12})$, our bounds on the spectral function are entirely consistent with the {\it analytic continuation} of the Liouville spectral function to external operator dimensions below the Liouville threshold. Indeed, such analytically continued Liouville correlators arise in the study of certain normalizable BPS correlators in super-Liouville theory \cite{Lin:2015wcg} as a result of a relation due to Ribault and Teschner between $SL(2)$ WZW model correlators and Liouville correlators \cite{Ribault:2005wp}. A priori, the crossing invariant Liouville 4-point function involves external primaries of scaling dimension $\Delta_i = 2\A_i (Q-\A_i)$, and an integration over internal primaries of scaling dimension $\Delta = 2\A (Q-\A)$, where both $\A_i$ and $\A$ lie on the half line ${Q\over 2} + i\mathbb{R}_{\geq0}$. We can analytically continue $\A_i$ to the real axis, away from ${Q\over 2}$, provided that no pole in the structure constant $C(\alpha_1,\alpha_2,\alpha)$ as a function of $\A$ crosses the integration contour ${Q\over 2} + i\mathbb{R}$. This is possible for ${Q\over 2} < \alpha_1+\alpha_2<{Q}$, but fails for $\A_1+\A_2\leq {Q\over 2}$ when a pole in $\A$ crosses the contour and the 4-point function picks up a residue contribution that violates unitarity. Indeed, $\A_1 = \A_2 = {Q\over 4}$ corresponds to $\Delta_\phi = {3\over 4}\Delta_0$, and we find the step function behaviour demonstrated in Figure~\ref{fig:LiouvilleBounds1} whenever $\alpha_1 + \alpha_2 = {Q\over 2}$.\footnote{This step function behavior is consistent with the fact that the 4-point conformal block with $\A_1+\A_2={Q\over 2}$ and internal primary with $\A={Q\over 2}$ is crossing invariant by itself. This conformal block is the same as the holomorphic part of the 4-point function $\langle e^{2\A_1\phi(z)}e^{2\A_2\phi(0)}e^{2\A_2\phi(1)}e^{2\A_1\phi(\infty)}\rangle$ in the linear dilaton CFT with background charge $Q$. Note that in the linear dilaton theory the closure of the OPE demands a non-unitary spectrum.}

\begin{figure}[h!]
\centering
\subfloat{\includegraphics[width=.49\textwidth]{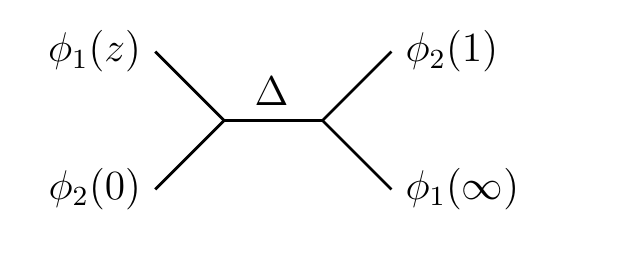}}~\addtocounter{subfigure}{-1}\subfloat[]{\includegraphics[width=.49\textwidth]{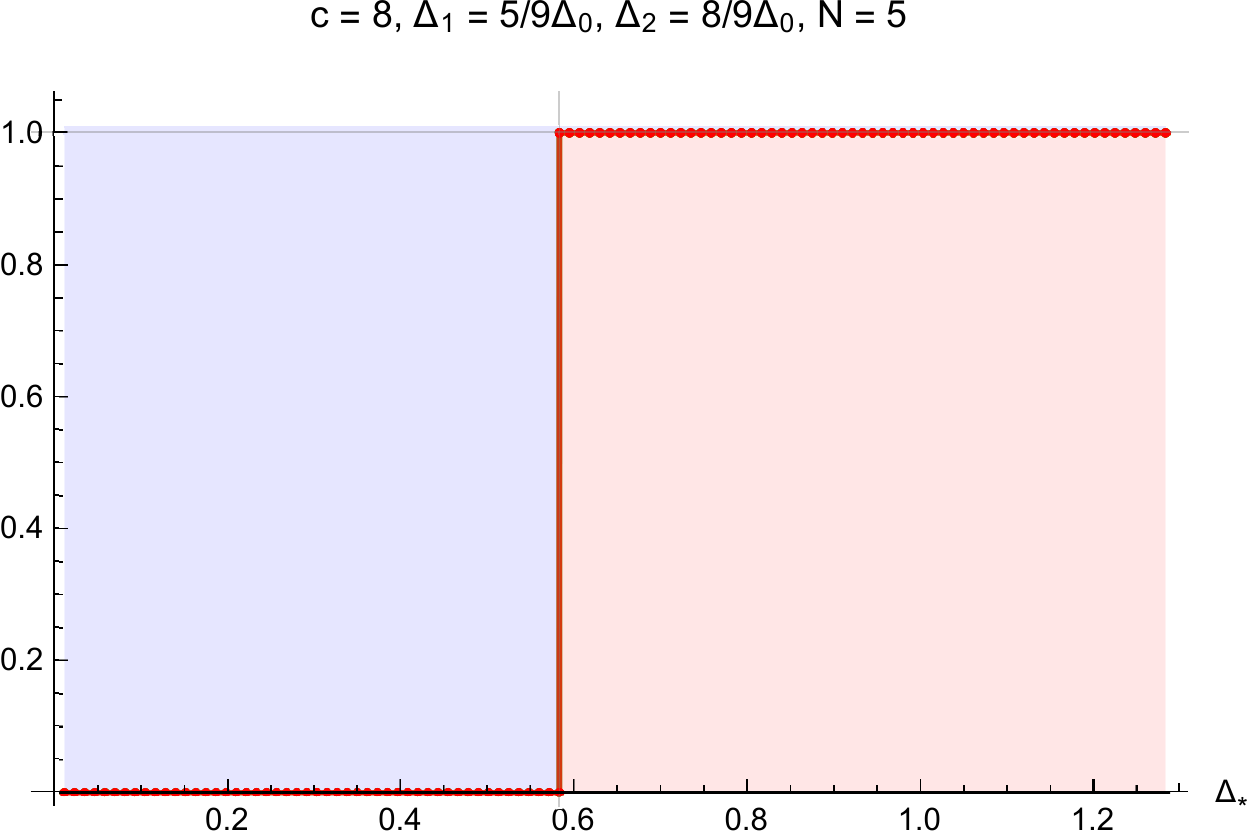}}
\\
\subfloat[]{\includegraphics[width=.49\textwidth]{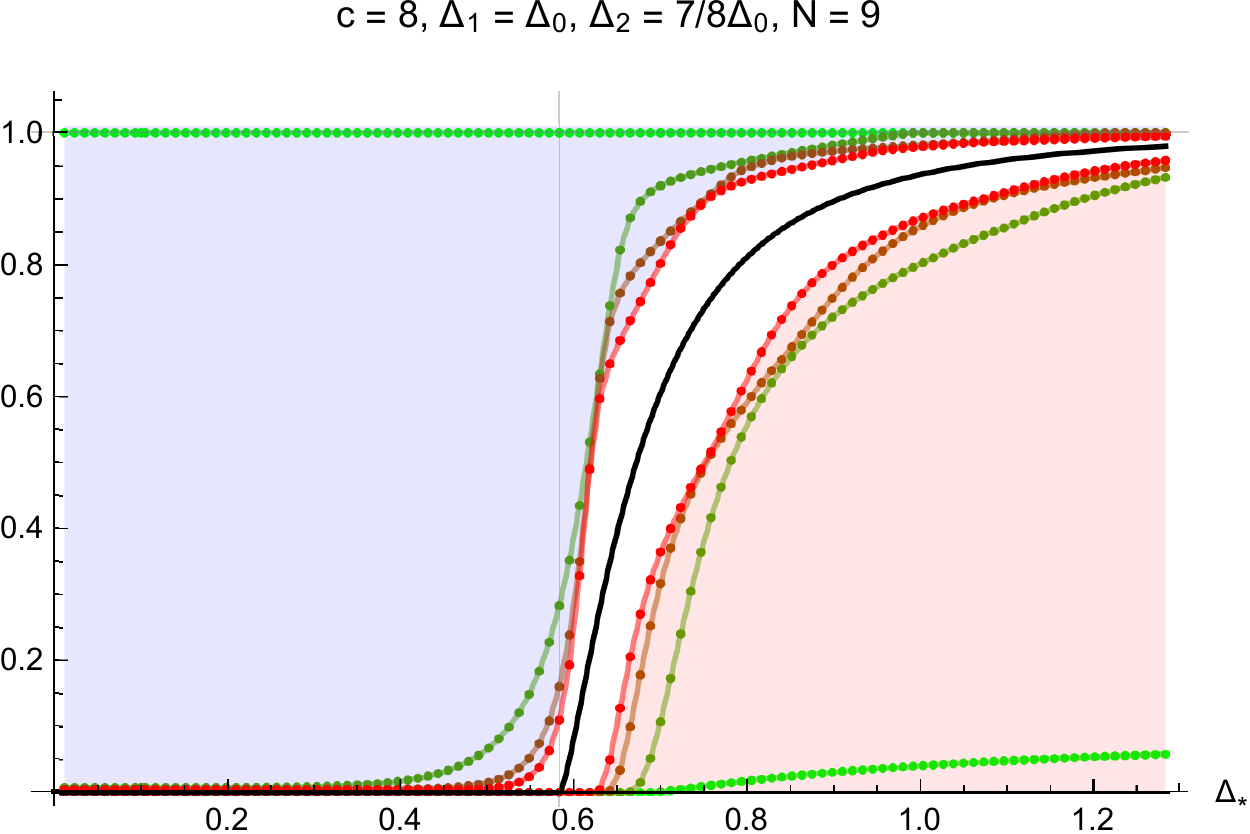}}~
\subfloat[]{\includegraphics[width=.49\textwidth]{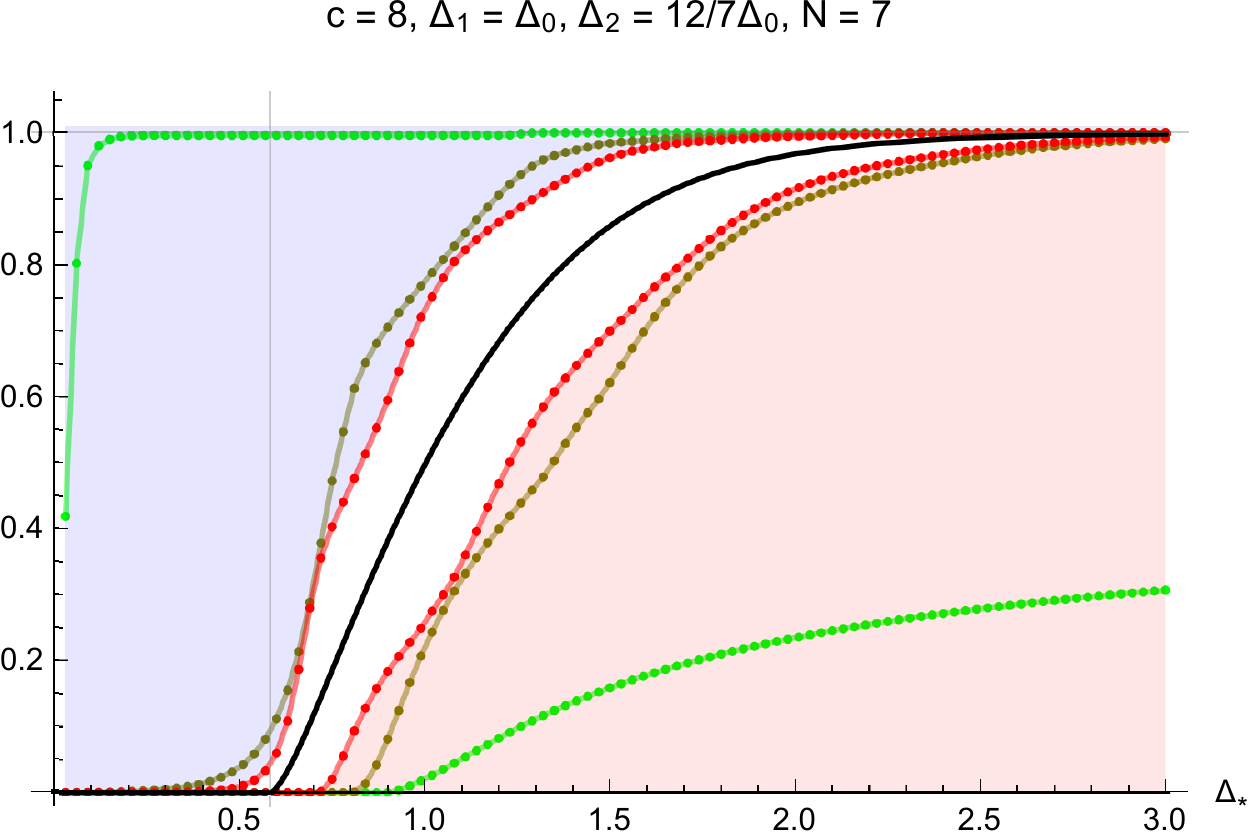}}
\caption{Upper and lower bounds on the mixed correlator spectral function for $c=8$ and $({\Delta_1\over\Delta_0},{\Delta_2\over\Delta_0})=({5\over 9},{8\over 9}),(1,{7\over 8}),(1,{12\over 7})$. The black curve denotes the (analytically continued) DOZZ spectral function. In (c), a small gap of $\Delta_{\rm gap}=0.01$ has been imposed to explicitly exclude the vacuum channel which would correspond to a singular conformal block for the mixed correlator.}\label{fig:LiouvilleBoundsMixed}
\end{figure}

Next, we study the bounds on the spectral function for the 4-point function involving a pair of primaries $\phi_1$ and $\phi_2$ of different scaling dimensions, of the form (\ref{mixedc}).
Note that for a non-compact CFT with only scalar primaries, such spectral functions capture the complete set of structure constants for three primaries of arbitrary weights. In Figure~\ref{fig:LiouvilleBoundsMixed} we plot the upper and lower bounds on the mixed correlator spectral function for $c=8$ with external primaries of various dimensions $(\Delta_1,\Delta_2)$. Once again, the bounds narrow down the allowed window towards the spectral function of Liouville theory.

Apart from the case of $\alpha_1+\alpha_2 = {Q\over 2}$, our numerical upper and lower bounds have not quite converged convincingly to the (analytic continuation of) the Liouville spectral function, due to the computational complexity of computing bounds at high derivative order $N$. Our results nonetheless suggest such a convergence in the $N\to \infty$ limit, supporting our conjecture that {\it the DOZZ structure constants $C(\alpha_1,\alpha_2,\alpha_3)$ are the unique solution to the crossing equations for unitary CFTs with $c>1$ and only scalar primaries}.

Note that the convergence of the bounds on the $\langle\phi\phi\phi\phi\rangle$ spectral function would determine the $\phi\phi$ OPE up to normalization; if this holds for all $\Delta_\phi$, it would then determine, assuming a non-degenerate spectrum, the conformal block decomposition of $\langle \phi_1 \phi_1 \phi_2\phi_2\rangle$ as well. This then determines the most general $\phi_1\phi_2$ OPE, up to normalization. Compatibility with all crossing equations fixes the normalizations of OPE coefficients to be DOZZ up to an overall scale factor which cannot be fixed for a non-compact CFT.\footnote{\label{foot:overallscale} This is because we can always tensor with a non-unital Frobenius algebra $\mathcal{G}_\alpha$ with a single generator $e$, $(e,e)=1$, $e^2=\alpha e$ for any $\alpha\in \mathbb{R}$.} Thus, in order to establish our conjecture for the uniqueness of the DOZZ solution for the scalar-only crossing equations in the non-degenerate case, it suffices to consider the OPE of pairs of identical primaries, and then the result for mixed correlators would follow.

One can notice that the bounds appear to change slowly with $N$ in certain regions of the plots. We also observed in the numerical studies of spectral functions in modular bootstrap that the convergence of upper and lower bounds is relatively slow for continuous spectra as compared to discrete spectra (see section~\ref{sec:modularscalaronly}) in the cases where we know that the solution to the modular crossing equation is unique. It appears to be quite difficult numerically to push these bounds to higher derivative orders $N$, due to the need to substantially increase the truncation order $d_q$ on the $q$-expansion of the Virasoro conformal blocks. This is discussed in Appendix~\ref{app:semidefinite}. In the next subsection, we consider an alternative method of directly solving the linear system that determines the spectral function assuming that the optimal upper and lower bounds coincide. This method in fact {\it does not rely on the assumption of reality of the OPE coefficients} and appears to converge much faster to the DOZZ spectral function.

\section{The linear method and the uniqueness of Liouville theory}

\subsection{Solution of the linear constraints on the spectral function}
\label{linearprob}

For CFTs with only scalar primaries, if the upper and lower bounds on the spectral function indeed converge (thereby pinning down the Liouville spectral function as the only solution), namely $y_{0,0}^{\rm min}$ in (\ref{eq:FourPtUpperInequality}) agrees with $-w_{0,0}^{\rm min}$ in (\ref{eq:FourPtLowerInequality}), we would have a solution to the linear equation
\ie
\label{eq:thetainbasis}
\Theta(\Delta_*-\Delta)\mathcal{F}_{12;0,\Delta}(1/2,1/2) = y_{0,0}\mathcal{F}_{12;0,\Delta}(1/2,1/2) + \sum_{m+n\text{ odd}}\left.y_{m,n}\partial_z^m\partial_{\bar z}^n\mathcal{F}_{12;0,\Delta}(z,\bar z)\right|_{z=\bar z= \half}.
\fe
That is to say, on a certain vector space of functions in $\Delta$, the function $\Theta(\Delta_*-\Delta)\mathcal{F}_{12;0,\Delta}(1/2,1/2)$ can be decomposed on the basis spanned by $\mathcal{F}_{12;0,\Delta}(1/2,1/2)$ and $\partial_z^m\partial_{\bar z}^n\mathcal{F}_{12;0,\Delta}(z,\bar z)|_{z=\bar z= \half}$. Since the step functions are themselves complete, our conjecture of the DOZZ structure constants as the unique solution is related to the completeness of this basis on a suitably defined Hilbert\footnote{It is not obvious that the Hilbert space structure is the fundamentally correct one; for example, it might be that the correct notion is denseness in some Banach space.}
space of functions in $\Delta$.\footnote{Note that the linear independence of $\mathcal{F}_{12;0,\Delta}(1/2,1/2)$ from $\partial_z^m\partial_{\bar z}^n\mathcal{F}_{12;0,\Delta}(z,\bar z)|_{z=\bar z= \half}$ as functions of $\Delta$ is guaranteed by the existence of DOZZ structure constants as a solution to the crossing equation.} While we do not have a proof of this statement, we can analyze the linear problem directly in an attempt to solve for the coefficient $y_{0,0}$ (for a truncated system). The stability of the solution and its convergence to the Liouville spectral function will provide strong evidence for the conjecture.

Another way to arrive at~\eqref{eq:thetainbasis} is the following. In a non-compact CFT with only scalar primaries the crossing symmetry equations, together with a normalization condition $g_{12}(1/2,1/2)=1$,~\eqref{eq:thetainbasis} can be written as
\als{\label{eq:linearsystemfunction}
	\int_0^\infty d\Delta C^2_{12;0,\Delta} \Fcal_{12;0,\Delta}(1/2,1/2)&=1,\\
	\int_0^\infty d\Delta C^2_{12;0,\Delta} \partial^n_z\partial^m_{\bar z}\Fcal_{12;0,\Delta}(1/2,1/2) &=0,\quad n+m\text{ odd}.
}
We may equivalently express these equations as
\als{\label{eq:linearsystemvector}
	\crr{v,p_{0,0}}&=1,\\
	\crr{v,p_{n,m}}&=0,\quad n+m\text{ odd},
}
where the vectors $v$, $p_{n,m}$ represent the functions
\al{
v(\Delta)&=C^2_{12;0,\Delta}/f_v(\Delta),\\
p_{n,m}(\Delta)&=\partial^n_z\partial^m_{\bar z}\Fcal_{12;0,\Delta}(1/2,1/2)/f_p(\Delta).
}
for some suitable choices of $f_v(\Delta)$ and $f_p(\Delta)$ (see Appendix~\ref{app:linear} for details),
while the inner product is defined by
\beq
	\crr{x,y}=\int_0^\infty x^*(\Delta)y(\Delta)d\mu(\Delta),
\eeq
with the measure $d\mu(\Delta)=f_v(\Delta)f_p(\Delta)d\Delta$.

We now hope for completeness of the set $p_{n,m}$ (from hereon we only consider $n=m=0$ or $n+m$ odd), assuming that all functions in question have finite norm. We truncate by $n+m\leq N$ and consider the approximation of $v$ by its orthogonal projection $v_N=P_N v$ onto $\PP_N=\overline{\mathrm{span}\{p_{n,m}\}_{n+m\leq N}}$.  
Note that despite the notation $v_N$, because of the equations~\eqref{eq:linearsystemvector}, $v_N$ is independent of a particular solution $v$. It can be computed by evaluating the Gram matrix of $p$ vectors and taking its inverse,
\beq
	v_N=\sum_{n,m}^{n+m\leq N}\sum_{n',m'}^{n'+m'\leq N} \crr{v,p_{n,m}}(G_N^{-1})^{n,m}_{n',m'}p_{n',m'}=\sum_{n',m'}^{n'+m'\leq N} (G_N^{-1})^{0,0}_{n',m'}p_{n',m'},
\eeq
where 
\beq
(G_N)^{n,m}_{n',m'}=\crr{p_{n,m},p_{n',m'}},\quad n+m\leq N,\,n'+m'\leq N.
\eeq
The spectral function can be computed as the inner product
\beq	
	f(\Delta_*)=\crr{v,\theta_{\Delta_*}}=\int_0^{\Delta^*} d\Delta C^2_{12;0,\Delta} \Fcal_{12;0,\Delta}(1/2,1/2),
\eeq
where
\beq
	\theta_{\Delta_*}(\Delta)=\Theta(\Delta_*-\Delta)p_{0,0}(\Delta).
\eeq
We have an estimate,
\beq
	\crr{v,\theta_{\Delta_*}}=\crr{v,P_N\theta_{\Delta_*}}+\crr{v,(1-P_N)\theta_{\Delta_*}}=\crr{v_N,\theta_{\Delta_*}}+R_N(\Delta_*),
\eeq
where
\beq
	|R_N(\Delta_*)|^2=|\crr{v,(1-P_N)\theta_{\Delta_*}}|^2\leq |v|^2|(1-P_N)\theta_{\Delta_*}|^2=E_N |v|^2|\theta_{\Delta_*}|^2.
\eeq
Note that $E_N=|(1-P_N)\theta_{\Delta_*}|^2/|\theta_{\Delta_*}|^2$ is also independent of a particular solution $v$ and is computable from~\eqref{eq:linearsystemvector}. 

If~\eqref{eq:thetainbasis} holds in the norm induced from $\crr{\cdot,\cdot}$, then $E_N\to 0$. Conversely, if we show for all $\Delta_*$ that $\lim_{N\to\infty}E_N=0$, it will imply that any normalizable solution to~\eqref{eq:linearsystemvector} and thus to~\eqref{eq:linearsystemfunction} is equal to the limit $\lim_{N\to\infty} v_N$, which is unique if exists. Our strategy would be therefore to evaluate $v_N$ and $E_N$ numerically and estimate their limits.

\begin{figure}[h!]
\centering
\subfloat[]{\includegraphics[width=.49\textwidth]{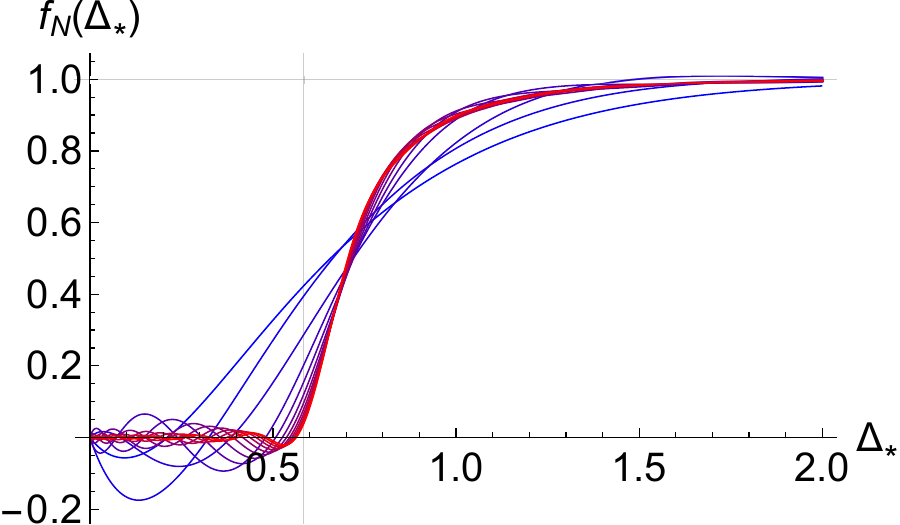}}~
\subfloat[]{\includegraphics[width=.49\textwidth]{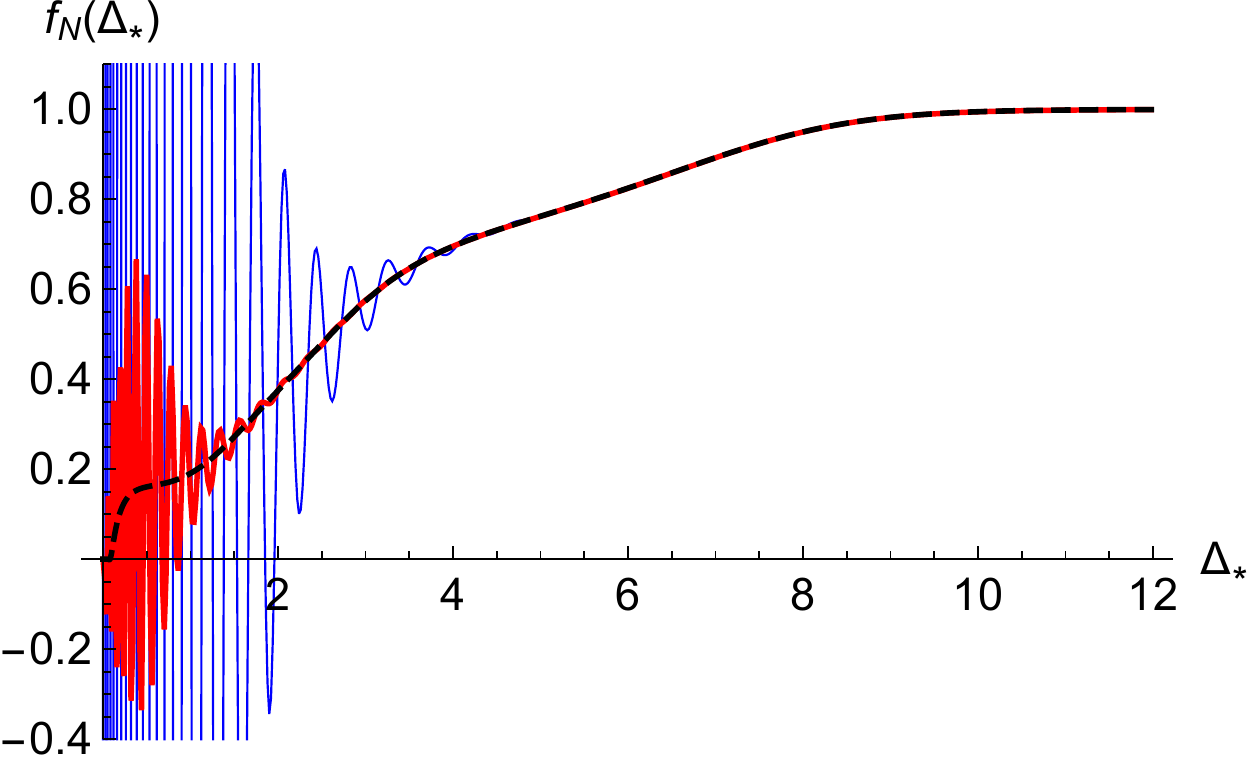}}
\caption{(a) Plot of $f_N(\Delta_*)$ for $c=8$, $\Delta_\phi = {7\over 12}$, as $N$ ranges from $N=1$ (blue) to $N=25$ (red) with step of 2. (b) Comparison of $f_N(\Delta_*)$ ($N=27$ in solid blue and $N=33$ in solid red) with the exact DOZZ spectral function (dashed, black) for $c=2$, $\Delta_\phi = {55\over 12}$.}\label{fig:linear1}
\end{figure}

We first numerically evaluate $f_N(\Delta_*)=\crr{v_N,\theta_{\Delta_*}}$ and find that it converges to the Liouville spectral density in the limit $N\to\infty$. For example, in Figure~\ref{fig:linear1}(a) the approximation $f_N$ is plotted at successive odd values of $N$ up to $N=25$ for $c=8$ and $\Delta_\phi=7/12$. We can see that the curves exhibit the expected convergence. Another example where the external operator dimension is far above the Liouville threshold is shown in Figure~\ref{fig:linear1}(b), where we studied $c=2$, $\Delta_\phi=55/12$, up to $N=33$ and $d_q=200$. While $f_N(\Delta_*)$ oscillates wildly at smaller $N$ (the case $N=27$ is shown for comparison), the oscillation settles down substantially as $N$ is increased.

\begin{figure}[h!]
\centering
\subfloat[$c=8$, $\Delta_\phi = \Delta_0$, $N=25$]{\includegraphics[width=.49\textwidth]{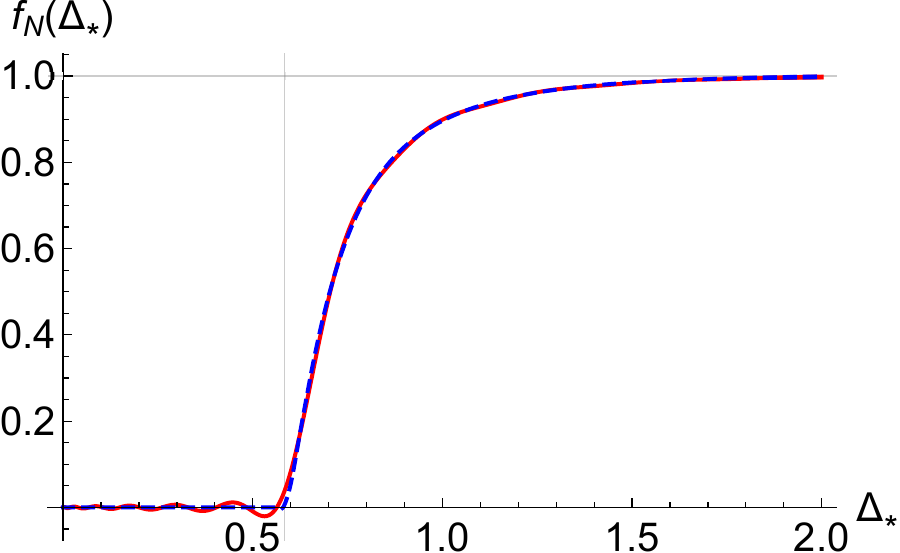}}~
\subfloat[$c=30$, $\Delta_\phi = \Delta_0$, $N=19$]{\includegraphics[width=.49\textwidth]{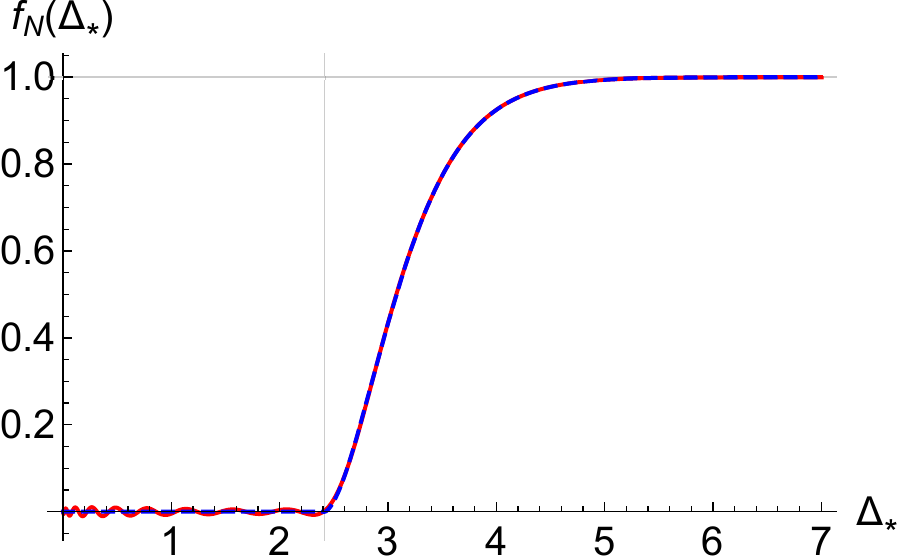}}
\\
\subfloat[$c=8$, $\Delta_\phi = {24\over 7}\Delta_0$, $N=19$]{\includegraphics[width=.49\textwidth]{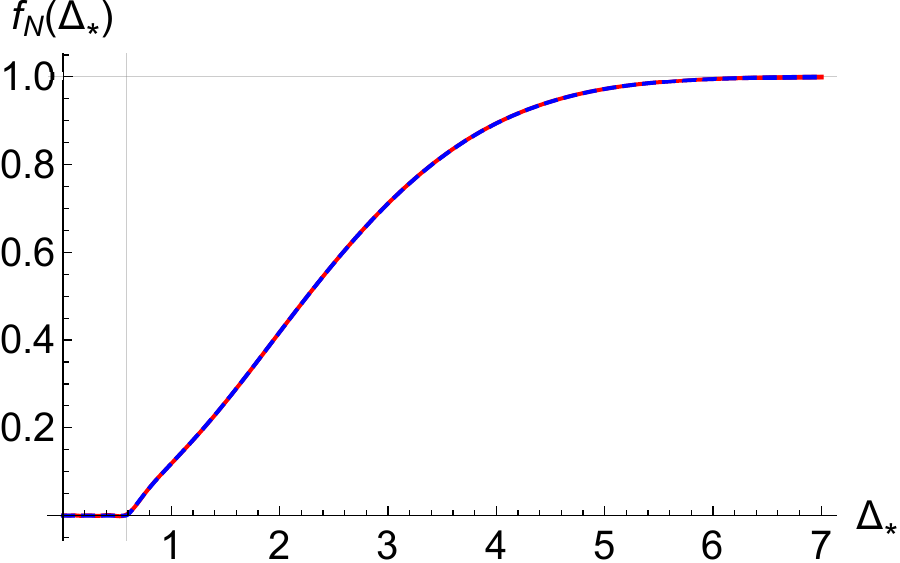}}~
\subfloat[$c=8$, $\Delta_1=\Delta_0$, $\Delta_2 = {12\over 7}\Delta_0$, $N=13$]{
\includegraphics[width=.49\textwidth]{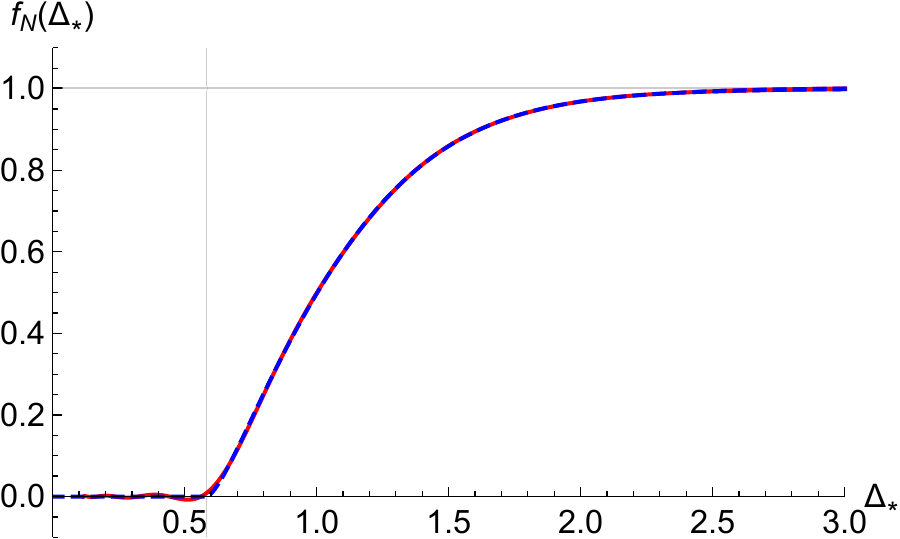}}
\caption{Comparison of $f_{N}(\Delta_*)$ (solid, red) with the exact DOZZ spectral function (dashed, blue) for external operator dimension $\Delta_\phi$, and in the mixed correlator case, external operator dimensions $\Delta_1$ and $\Delta_2$ ($\Delta_0 \equiv {c-1\over 12}$ is the Liouville threshold as before). }\label{fig:linear2}
\end{figure}

In Figure~\ref{fig:linear2} we compare $f_{N}(\Delta_*)$ with the DOZZ spectral function for $c=8$ and $c=30$, with $\Delta_\phi$ at or above the Liouville threshold, as well as an example of a mixed correlator spectral function\footnote{The mixed correlator spectral function was obtained under a technical assumption of a lower dimension bound of $\frac{\Delta_0}{5}$ (note that this is below the Liouville threshold $\Delta_0$), see appendix~\ref{app:linear} for details.} with two different values of external operator dimensions. In all cases we find good agreement.

\begin{figure}[h!]
\centering
\subfloat{\includegraphics[width=.49\textwidth]{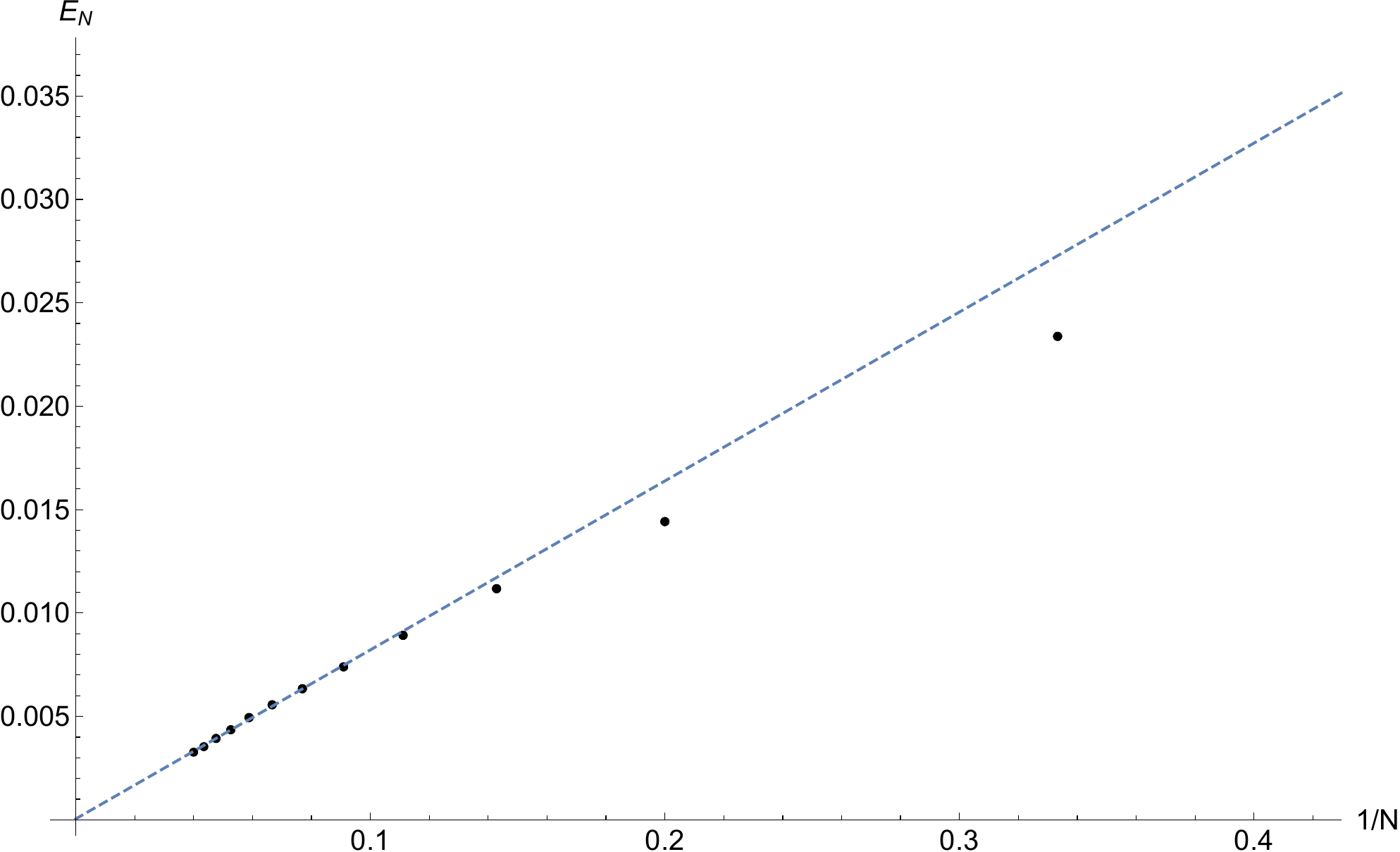}}
\caption{Plot of $E_N$ for as a function of $1/N$, $c=8$, $\Delta_\phi={7\over12}$, $\Delta_*=0.8$, $N\leq 25$. The dashed curve is a linear fit for $N\geq 11$.}\label{fig:linear3}
\end{figure}

To further support the conjecture, we numerically compute the error estimate $E_N$ as a function of $N$. For example, in Figure~\ref{fig:linear3} we show $E_N$ as a function of $1/N$ for $\Delta_\phi=\frac{c-1}{12}$, $\Delta_*=\frac{c}{10}$, and $c=8$. In the figure we also show a linear fit using $N\geq 11$. Empirically, we find that the result is consistent with $E_N\sim N^{-1}$. We study $E_N$ in more detail in appendix~\ref{app:linear_completeness}.

The discussion above depends on the assumption that $v$ has finite norm. This assumption itself depends on the choice of measure. We describe our choice of measure and details of our implementation in appendix~\ref{app:linear}. Here we simply note that with our choice, $v$ has finite norm if $C^4_{12;0,\Delta}$ is locally integrable on $[0,\infty)$, and the OPE expansion is convergent in the region $|z|<1$. Discrete spectra have infinite norm since $C^4_{12;0,\Delta}$ involves squares of delta-functions, but such spectra are excluded by modular invariance.

\subsection{Constraints from modular invariance}\label{sec:ModularAnalytic}

Strong constraints on the primary spectrum, especially in the scalar-only case, follow from modular invariance alone.
In fact, there is a simple argument that shows any 2D CFT with $c>1$ and primary operators with {\it bounded spin} must have a spectrum identical to that of Liouville theory: that is, the spectrum is non-compact, has scalar primaries only, and has a spectral density that is uniformly distributed in Liouville momentum $P = \sqrt{2(\Delta - {c-1\over 12})}$.

Suppose the primaries have spins no greater than $s_{\rm max}$. We can write the reduced torus partition function in the following way:
\ie
\tilde{Z}(\tau,\bar\tau) =& \tau_2^{\half}|\eta(\tau)|^2 Z(\tau,\bar\tau)\\
=&\tau_2^{\half}\left(|q^{-\xi}(1-q)|^2+\sum_{h+\tilde{h}>0}d(h,\tilde h) q^{h-\xi}\bar q^{\tilde{h}-\xi}\right)\\
=& \sum_{s,|s|\le s_{\rm max}}e^{2\pi i s\tau_1}f_s(\tau_2),
\fe
where $q=e^{2\pi i \tau}$, $\xi = {c-1\over 24}$, $d(h,\tilde h)$ is the degeneracy of primary operators in the spectrum with conformal weights $(h,\tilde h)$ and $f_s(x) = \sum_{\Delta\ge |s|}d({\Delta+s\over 2},{\Delta-s\over 2}) x^{\half}e^{-2\pi x(\Delta-2\xi)}$. For now we assume that the CFT is compact, and the vacuum character is degenerate and so $s_{\rm max}\ge 1$. The non-compact CFTs may be viewed as limiting cases, where the spectral density diverges and we divide the partition function by an infinite normalization factor which removes the vacuum contribution.

Now consider the following change of variables
\ie
x = \tau_2,~y= {\tau_2\over |\tau|^2},
\fe
chosen so that the modular $S$ transformation exchanges $x$ and $y$. We can then write the modular crossing equation in terms of these variables as
\ie
\sum_s e^{2\pi i s\sqrt{{x\over y}-x^2}}f_s(x) = \sum_s e^{2\pi i s\sqrt{{y\over x}-y^2}}f_s(y)
\fe
Of course, the functions above have branch cuts at $x= y^{-1}$, but since the sum over spins is finite by assumption, the analytic continuation around the branch is straightforward. Furthermore, $f_s(y)$ is an analytic function for ${\rm Re}(y)>0$. To proceed, we fix $x=re^{-i\alpha}$ with $r>0$ and $0<\alpha<{\pi \over 2}$ and $y = \eps$ with $\eps\to 0^+$, so that the modular crossing equation becomes
\ie
\sum_s e^{2\pi s\sqrt{r\over \eps}(\sin{\alpha\over 2}+i\cos{\alpha\over 2})}f_s(x) \approx& \sum_s f_s(\eps)\\
=& \sum_s f_s(\eps^{-1}),
\fe
where in the first line we dropped the phase factors $e^{2\pi i s\sqrt{{\eps\over r}e^{i\alpha}-\eps^2}}$ (which are close to 1 due to the boundedness of $s$) in front of $f_s(\eps)$; this is a valid approximation since $f_s(\epsilon)$ is positive for all $s$. In the second line we again invoked modular invariance (this particular equality is realized as the modular crossing equation with $\tau_1 = 0,\tau_2 = \eps$).
In the case that the CFT is compact, the right-hand side is dominated by the contribution of the vacuum, in particular
\ie
\sum_s f_s(\eps^{-1})\approx \eps^{-\half}e^{{4\pi\xi\over \eps}}.
\fe
By comparing to the $\eps\to 0$ limit of the left-hand side, which is dominated by the term with maximal spin 
\ie\label{eq:LHSMaximalSpin}
\sum_s e^{2\pi s\sqrt{r\over \eps}(\sin{\alpha\over 2}+i\cos{\alpha\over 2})}f_s(re^{-i\alpha}) \approx e^{2\pi s_{\rm max}\sqrt{r\over \eps}(\sin{\alpha\over 2}+i\cos{\alpha\over 2})}f_{s_{\rm max}}(re^{-i\alpha}),
\fe
we arrive at a contradiction and deduce that unitary 2D CFTs with primary operators of bounded spin must have non-compact spectra: namely, there is no $SL(2,\bR)\times SL(2,\bR)$-invariant vacuum and the dimension of the lowest-lying primary operator obeys $\Delta_{\rm min}>0$.

In fact, this same logic allows us to conclude that the dimension of the lowest-lying operator must obey $\Delta_{\rm min}\ge {c-1\over 12}$. In the $\eps\to 0$ limit, we have
\ie
e^{2\pi s_{\rm max}\sqrt{r\over \eps}(\sin{\alpha\over 2}+i\cos{\alpha\over 2})}f_{s_{\rm max}}(re^{-i\alpha})\approx & \sum_s f_s(\eps^{-1})\\
=& \tilde{Z}(\tau_1=0,\tau_2 = \eps^{-1})\\
=& \eps^{-\half}\int_0^\infty d\Delta \rho(\Delta)e^{-{2\pi\over \eps}(\Delta-2\xi)},
\fe
where $\rho(\Delta)$ is the density of primary operators in the spectrum with dimension $\Delta$ (of any spin). The two sides of the equation are clearly incompatible if the minimum scaling dimension for which $\rho(\Delta)$ is nonzero is smaller than $2\xi$. Furthermore, by non-negativity of the spectral density, the right-hand side can grow no faster than $\eps^{-\half}$ as $\eps\to 0^+$. On the other hand, the absolute value of (\ref{eq:LHSMaximalSpin}) grows like $e^{2\pi s_{\rm max}\sqrt{r\over \eps}\sin{\alpha\over 2}}$ in this limit. Modular invariance thus demands that $s_{\rm max} = 0$: that is, a unitary 2D CFT with primary operators of bounded spin must in fact have only scalar primary operators in addition to having a non-compact spectrum. Moreover in this case the modular crossing equation becomes
\ie
f_0(x) = f_0(y),
\fe
which demands that $f_0(x)$ is a constant. Thus the required spectral density is nothing other than that of Liouville theory, namely $\rho(\Delta) = \rho_{\rm Liouville}(\Delta) \propto (\Delta-2\xi)^{-\half}\Theta(\Delta-2\xi)$,\footnote{To normalize the reduced partition function of Liouville theory to 1, the constant of proportionality is $\sqrt{2}$.} completing the argument. In particular, the dimension of the lowest-lying operator must be exactly $\Delta_{\rm min} = 2\xi$.

By this result, our conjecture that the DOZZ structure constants are the unique solution to the crossing equations for a unitary 2D CFT with central charge $c>1$ and only scalar primaries, as supported by substantial numerical evidence in sections~\ref{ulbounds} and~\ref{linearprob} leads us to conjecture that Liouville theory is the unique unitary $c>1$ CFT with Virasoro primaries of bounded spin. \footnote{Note that we have not made use of the torus 1-point function. A priori, the modular invariance of the torus 1-point function puts nontrivial constraints on the structure constants with a pair of primaries identified. For the purpose of establishing our conjecture regarding the uniqueness of Liouville, once the OPE coefficients are pinned down to those of DOZZ by the crossing equation, the torus 1-point functions are already modular invariant \cite{Hadasz:2009sw}.}

\subsection{Degenerate spectrum and TQFT}\label{caveat}

In our analysis of the crossing equation so far, we have implicitly assumed that the scalar primaries are labeled by a continuous parameter, namely the scaling dimension $\Delta_\phi$, without further degeneracy. If this assumption is relaxed, one can construct more examples of (non-compact) $c>1$ CFTs with only scalar primaries, by taking the tensor product of Liouville CFT with a topological quantum field theory (TQFT); the latter has a finite dimensional Hilbert space on the circle and its structure constants are governed by those of a commutative Frobenius algebra \cite{Abrams:1996ty}.\footnote{To be precise, we do not need to require the TQFT to have a vacuum state (or the algebra to be unital).} We conjecture that this is the only possibility.

Let us assume that the scalar primaries are labeled by their scaling dimension $\Delta$ and an extra index $i$, and denote the structure constants by 
\ie
C_{ijk}(\Delta_1,\Delta_2,\Delta_3) = A_{ijk}(\Delta_1,\Delta_2,\Delta_3) C^{\rm DOZZ}(\Delta_1,\Delta_2,\Delta_3),
\fe
where we have explicitly factored out the DOZZ structure constants. Our numerical results in the previous sections on the spectral function of mixed correlators of the form $\langle\phi_1\phi_2\phi_2\phi_1\rangle$ indicate that for a CFT with degenerate scalar-only primary spectrum,
\ie
\sum_k (A_{ijk}(\Delta_1,\Delta_2,\Delta))^2 = B_{ij}(\Delta_1,\Delta_2)
\fe
is independent of $\Delta$. In fact, we can strengthen this result slightly. Let us consider a mixed correlator $\langle \phi_i \phi_j \phi_k \phi_\ell\rangle$ where $\phi_i, \phi_\ell$ have scaling dimension $\Delta_1$, $\phi_j,\phi_k$ have scaling dimension $\Delta_2$, and the crossing equation
\ie\label{crosseq}
&\sum_m \sum_\Delta C_{ijm}(\Delta_1, \Delta_2, \Delta) C_{k\ell m}(\Delta_1,\Delta_2,\Delta) {\cal F}_{12;0,\Delta}(z,\bar z) 
\\
&= \sum_m \sum_\Delta  C_{i\ell m}(\Delta_1, \Delta_2, \Delta) C_{kj m}(\Delta_1,\Delta_2,\Delta) {\cal F}_{12;0,\Delta}(1-z,1-\bar z) .
\fe
By taking the part of (\ref{crosseq}) that is odd under $z\to 1-z, \bar z\to 1-\bar z$, our earlier claim of the uniqueness of scalar-only solution to the crossing equation implies that
$$\sum_m A_{ijm}(\Delta_1,\Delta_2,\Delta) A_{k\ell m}(\Delta_1,\Delta_2,\Delta) + (j\leftrightarrow \ell)$$
is independent of $\Delta$. On the other hand, for the even part of (\ref{crosseq}) under $z\to 1-z, \bar z\to 1-\bar z$, the numerical analysis described in appendix~\ref{app:linear_completeness} is consistent with the conjecture 
 that $\{\left.\partial_z^n \partial_{\bar z}^m F_{12;0,\Delta}\right|_{z=\bar z= {1\over 2}}, n,m\in \mathbb{Z}_{\geq 0}, n+m~{\rm even}\}$ form a complete basis on the space functions of $\Delta$ on the positive real axis defined by the same norm as in section~\ref{linearprob}, which implies that
$
\sum_m A_{ijm}(\Delta_1,\Delta_2,\Delta) A_{k\ell m}(\Delta_1,\Delta_2,\Delta) = \sum_m A_{i\ell m}(\Delta_1,\Delta_2,\Delta) A_{kj m}(\Delta_1,\Delta_2,\Delta) 
$
for every $\Delta>0$, and thus 
\ie\label{aab}
\sum_m A_{ijm}(\Delta_1,\Delta_2,\Delta) A_{k\ell m}(\Delta_1,\Delta_2,\Delta) = B_{ijk\ell}(\Delta_1,\Delta_2)
\fe
is independent of $\Delta$.

It is likely that by analyzing a system of crossing equations for multiple scalar correlators involving $\phi_i, \phi_j, \phi_k, \phi_\ell$ of generally different scaling dimensions, one could establish that the spectral function for $\langle \phi_i \phi_j \phi_k \phi_\ell \rangle$ (with only scalar Virasoro primaries in the OPEs) is proportional to that of Liouville CFT, which would be equivalent to the statement that
\ie\label{aaid}
\sum_m A_{ijm}(\Delta_1,\Delta_2,\Delta) A_{k\ell m}(\Delta_3,\Delta_4,\Delta) 
=& \sum_m A_{i\ell m}(\Delta_1,\Delta_4,\Delta) A_{kj m}(\Delta_3,\Delta_2,\Delta) 
\\
=& B_{ijk\ell}(\Delta_1,\Delta_2,\Delta_3,\Delta_4)
\fe
is independent of $\Delta$, extending (\ref{aab}). We leave the numerical bootstrap of the spectral function with four generic external weights to future work. We now argue that if (\ref{aaid}) holds, then our conjecture follows.

To each pair-of-pants decomposition of a genus $g$ Riemann surface, represented by a trivalent graph, we may associate a sum of product of $A_{ijk}$'s, with indices contracted and scaling dimensions identified along each edge of the graph, which we denote by $\widehat Z_g$. (\ref{aaid}) implies the crossing relation between graphs with fixed weights on the edges, and by applying crossing one can always turn the trivalent graph into one that does not contain tadpole subgraphs.\footnote{That is to say, modular constraints on the general torus 1-point function are not needed for the argument presented here.} (\ref{aaid}) further implies that $\widehat Z_g$ is independent of the scaling dimension on every edge that connects a pair of distinct vertices, and thus the genus $g$ partition function of the CFT is equal to $\widehat Z_g$ times the Liouville partition function. It then follows from modular invariance that $\widehat Z_g$ is independent of the pair-of-pants decomposition, and depends on the genus $g$ only.

To proceed, pick a finite set of scaling dimensions $\Delta_a$, $a=1,\ldots,M$ and let $N_\Delta$ be the number of degenerate primaries of dimension $\Delta$, which we will assume to be finite. Set $N=\max N_{\Delta_a}$ and extend the ranges of the discrete labels to run up to $N$ for all $\Delta_a$ by setting the previously undefined structure constants to zero. Then the totality of $A_{ijk}(\Delta_a,\Delta_b,\Delta_c)$ gives an element $\mathcal{A}$ in $\mathcal{C}=S^3(\oplus_{a} \mathbb{R}^{N})$. The space $\mathcal{C}$ is equipped with an action of $\prod_{a}O(N)$, corresponding to changes of basis for the discrete labels. $\widehat{Z}_g$ regarded as polynomials generate the algebra of $\prod_{a}O(N)$ invariants on $\mathcal{C}$. It follows that $\mathcal{A}$ is equivalent to any other $\mathcal{A}'$ with the same values of $\widehat{Z}_g$ by a $\prod_{a}O(N)$ reparametrization. In particular, since $\mathcal{A}$ is such that values of $\widehat{Z}_g$ on it are independent of the internal labels, $\mathcal{A}$ is equivalent to $\mathcal{A}_0$ in which all $A_{ijk}(\Delta_a,\Delta_b,\Delta_c)$ are replaced by $a_{ijk}=A_{ijk}(\Delta_1,\Delta_1,\Delta_1)$. It then follows that we can choose $N_{\Delta_a}=N$.

By taking various finite sets of scaling dimensions sharing the dimension $\Delta_1$, we find that $N_\Delta=N$ is independent of $\Delta$ and thus the density of states is given by $N$ copies of Liouville density. Furthermore, for any such finite set we have
\beq
	A_{ijk}(\Delta_a,\Delta_b,\Delta_c)=a_{ijk}
\eeq
up to a reparametrization of finite labels. Note that such reparametrizations depend on the choice of our finite set of $\Delta_a$'s, being defined only up to automorphisms of $a_{ijk}$. As we show below, these automorphisms are scarce. Compatibility between different $\Delta_a$ then completely fixes them after we fix the reparametrization for $\Delta_1$.
Thus we can pass to the full continuous set of scaling dimensions, and conclude that the CFT in question is a tensor product of Liouville with a TQFT defined by the structure constants $a_{ijk}$ (or the partition functions $\widehat{Z}_g$).

In fact, we can always find a basis in which the structure constants $a_{ijk}$ are diagonalized. To see this, note that the crossing equation for $a_{ijk}$ implies that the matrices $M_i$ with entries $(M_i)_{jk}=a_{ijk}$ are mutually commuting $N\times N$ symmetric matrices, and thus can be simultaneously diagonalized by some $O(N)$ matrix $R$, namely $\Lambda'_{ijk}=\sum_{mn}R_{jm}R_{kn}a_{imn}$ are diagonal in $jk$. Then $\Lambda_{ijk}=\sum_m R_{im}\Lambda'_{mjk}$ is still diagonal in $jk$ and  completely symmetric, and thus $\Lambda_{ijk}=\delta_{ijk}\lambda_k$. Multiplying by a diagonal matrix with $\pm 1$ entries if necessary, we can set $\lambda_m>0$. (If some $\lambda_m=0$ they do not contribute to the correlators and we can obviously add or remove such $\lambda$'s at will.) It is straightforward to check that the automorphisms of $\Lambda_{ijk}$ are just the permutations preserving the $\lambda$'s. The partition functions are $\widehat{Z}_g=\sum_n\lambda_n^{g-1}$.

This diagonalization implies that the algebra defined by $a_{ijk}$ is given by $\oplus_{n} \mathcal{G}_{\lambda_n}$. Here $\mathcal{G}_\lambda=\mathbb{R}e$ with $(e,e)=1$ and $e^2=\lambda e$. Forming the tensor product ${\rm Liouville}\otimes\mathcal{G}_\lambda$ corresponds to rescaling all OPE coefficients by $\lambda$. The overall scale of OPE coefficients cannot be fixed in the absence of the vacuum, and thus we can regard all these theories as isomorphic to Liouville. Therefore, the TQFT structure amounts to superselection sectors.

\section{The modular spectral function}
\label{sec:modularspectralfunction}

\subsection{The minimization problem}

We now consider the decomposition of the reduced torus partition function of a compact, unitary CFT (assumed to be parity-invariant)\footnote{As in \cite{Collier:2016cls}, the bounds we derive here assuming a parity-invariant spectrum can be applied to parity non-invariant theories as well by considering the parity-positive projection of the partition function.} with no conserved currents into non-degenerate Virasoro characters
\ie
\hat{Z}(\tau,\bar\tau) =& |\tau|^{\half}|\eta(\tau)|^2 Z(\tau,\bar\tau)\\
=& \hat{\chi}_0(\tau)\hat{\bar{\chi}}_0(\bar\tau) + \sum_{s\ge 0}\sum_{\Delta\in \mathcal{I}_s}d_{\Delta,s}\left(\hat{\chi}_{\Delta+s\over 2}(\tau)\hat{\bar{\chi}}_{\Delta-s\over 2}(\bar\tau) + \hat{\chi}_{\Delta-s\over 2}(\tau)\hat{\bar{\chi}}_{\Delta+s\over 2}(\bar\tau)\right),
\fe
where $\mathcal{I}_s$ is the discrete spectrum of dimensions of primary operators, $d_{\Delta,s} = d({\Delta+s\over 2},{\Delta-s\over 2})=d({\Delta-s\over 2},{\Delta+s\over 2})$, and the reduced characters are given by
\ie
\hat{\chi}_0(\tau) =& (-i\tau)^{1\over 4}q^{-\xi}(1-q)\\
\hat{\chi}_h(\tau) =& (-i\tau)^{1\over 4}q^{h-\xi}.
\fe
Analogously to the four-point spectral function introduced in section~\ref{sec:4ptSpectral}, we define a ``modular spectral function'' by truncating the Virasoro character decomposition of the reduced partition function up to a cutoff dimension $\Delta_*$, evaluated at the self-dual modulus $\tau=-\bar\tau=i$,
\ie
f_{\rm mod}(\Delta_*) =& {1\over \hat{Z}(i,-i)}\left[\hat{\chi}_0(i)\hat{\bar{\chi}}_0(-i) + \sum_{s= 0}^{\floor{\Delta_*}}\sum_{\Delta\in\mathcal{I}_s,\Delta\le \Delta_*}d_{\Delta,s}\left(\hat{\chi}_{\Delta+s\over 2}(i)\hat{\bar{\chi}}_{\Delta-s\over 2}(-i)+\hat{\chi}_{\Delta-s\over 2}(i)\hat{\bar{\chi}}_{\Delta+s\over 2}(-i)\right)\right].
\fe
 
As with the four-point spectral function, it is straightforward to place bounds on $f_{\rm mod}(\Delta_*)$ due to modular invariance using semidefinite programming. Defining $\hat{Z}_{\Delta,s}(\tau,\bar\tau) = \hat{\chi}_{\Delta+s\over 2}(\tau)\hat{\bar{\chi}}_{\Delta-s\over 2}(\bar\tau)+\hat{\chi}_{\Delta-s\over 2}(\tau)\hat{\bar{\chi}}_{\Delta+s\over 2}(\bar\tau)$ and $\hat{Z}_{0,0}(\tau,\bar\tau) = \hat{\chi}_0(\tau)\hat{\bar{\chi}}_0(\bar\tau)$, the modular crossing equation demands that
\ie
0 =& \partial_z^m\partial_{\bar z}^n\left.\left[\hat{Z}_{0,0}(\tau,\bar\tau) + \sum_{s=0}^{\infty}\sum_{\Delta_s\in\mathcal{I}_s}d_{\Delta,s}\hat{Z}_{\Delta,s}(\tau,\bar\tau)\right]\right|_{z=\bar z = 0},~m+n\text{ odd}
\fe
where we have redefined $\tau = ie^z$, $\bar\tau=-ie^{-\bar z}$. We then seek to minimize $y_{0,0}$ subject to the inequalities
\ie\label{eq:ModularUpperInequality}
(y_{0,0}-1)\hat{Z}_{0,0}(i,-i)+\sum_{m+n\text{ odd}}\left.y_{m,n}\partial_z^m\partial_{\bar z}^n\hat{Z}_{0,0}(\tau,\bar\tau)\right|_{z=\bar   z = 0}\ge & 0\\
(y_{0,0}-\Theta(\Delta_*-\Delta))\hat{Z}_{\Delta,s}(i,-i) + \sum_{m+n\text{ odd}}\left.y_{m,n}\partial_z^m\partial_{\bar z}^n\hat{Z}_{\Delta,s}(\tau,\bar\tau)\right|_{z=\bar z=0}\ge& 0,~ \Delta\ge \Delta_s^*,~s\ge 0,
\fe
for arbitrary coefficients $y_{m,n}$. In the first line we have singled out the inequality involving the vacuum primary. In the second line, we made the extra assumption of a gap $\Delta_s^*$ in the spin-$s$ sector of the spectrum, as will be useful in later applications.
As before, the minimal such $y_{0,0}$ gives an upper bound on the modular spectral function, since
\ie\label{eq:ModularUpperBound}
f_{\rm mod}(\Delta_*) \le &{1\over \hat{Z}(i,-i)}\Bigg[y^{\rm min}_{0,0}\left(\hat{Z}_{0,0}(i,-i)+\sum_{s,\Delta}d_{\Delta,s}\hat{Z}_{\Delta,s}(i,-i)\right)\\
&\left.+\sum_{m+n\text{ odd}}y_{m,n}\partial_z^m\partial_{\bar z}^n\left(\hat{Z}_{0,0}(\tau,\bar\tau)+\sum_{s,\Delta}d_{\Delta,s}\hat{Z}_{\Delta,s}(\tau,\bar\tau)\right)\Bigg]\right|_{z=\bar z=0}\\
=& y^{\rm min}_{0,0}.
\fe
Similarly, the minimal $w_{0,0}$ subject to the constraints
\ie\label{eq:ModularLowerInequality}
(w_{0,0}+1)\hat{Z}_{0,0}(i,-i)+\sum_{m+n\text{ odd}}\left.w_{m,n}\partial_z^m\partial_{\bar z}^n\hat{Z}_{0,0}(\tau,\bar\tau)\right|_{z=\bar   z = 0}\ge & 0\\
(w_{0,0}+\Theta(\Delta_*-\Delta))\hat{Z}_{\Delta,s}(i,-i) + \sum_{m+n\text{ odd}}\left.w_{m,n}\partial_z^m\partial_{\bar z}^n\hat{Z}_{\Delta,s}(\tau,\bar\tau)\right|_{z=\bar z=0}\ge& 0,~ \Delta\ge \Delta_s^*,~s\ge 0,
\fe
provides a nontrivial lower bound on the modular spectral function
\ie
f_{\rm mod}(\Delta_*) \ge -w_{0,0}^{\rm min}.
\fe
Working up to a finite derivative order $m+n\le N$, we denote the corresponding upper and lower bounds obtained in this way $f^+_{{\rm mod},N}(\Delta_*)$ and $f^-_{{\rm mod},N}(\Delta_*)$ respectively.

\subsection{Some consistency checks}

\subsubsection{Extremal spectra with maximal gap}
In \cite{Collier:2016cls}, an upper bound $\Delta_{\rm mod}(c)$ on the gap in the scaling dimension of primary operators due to modular invariance of the torus partition function was computed numerically as a function of the central charge. Given a dimension gap $\Delta_{\rm gap}$($\leq \Delta_{\rm mod}(c)$), an upper bound on the degeneracy of primaries at dimension $\Delta_{\rm gap}$ can be obtained provided $\Delta_{\rm gap}>{c-1\over 12}$. When this upper bound on the degeneracy at the gap is saturated, the entire modular invariant spectrum is determined by the locations of the zeros of the optimal linear functional (optimized with respect to the degeneracy bound) acting on the Virasoro characters. Such (candidate) CFT spectra were dubbed `extremal.' Furthermore, it is expected that for each given $c>1$, there is a unique modular invariant spectrum (imposing positivity but not the integral condition on the degeneracy of primaries) whose dimension gap saturates the upper bound $\Delta_{\rm mod}(c)$ \cite{ElShowk:2012hu}.

 In \cite{Collier:2016cls}, a number of examples of CFTs with spectra that saturated the bound on the dimension gap were identified at small values of the central charge. Here, we study the bounds on the modular spectral function at these values of the central charge assuming the maximal dimension gap. We will find that the resulting bounds indeed pin down the extremal modular spectral functions. To compute the bounds on the modular spectral functions in these cases, we impose (\ref{eq:ModularUpperInequality},\ref{eq:ModularLowerInequality}) with $\Delta_s^* = \text{max}(s,\Delta_{\rm mod}(c))$.
 
\begin{figure}[h!]
 \centering
 \subfloat[]{\includegraphics[width=.49\textwidth]{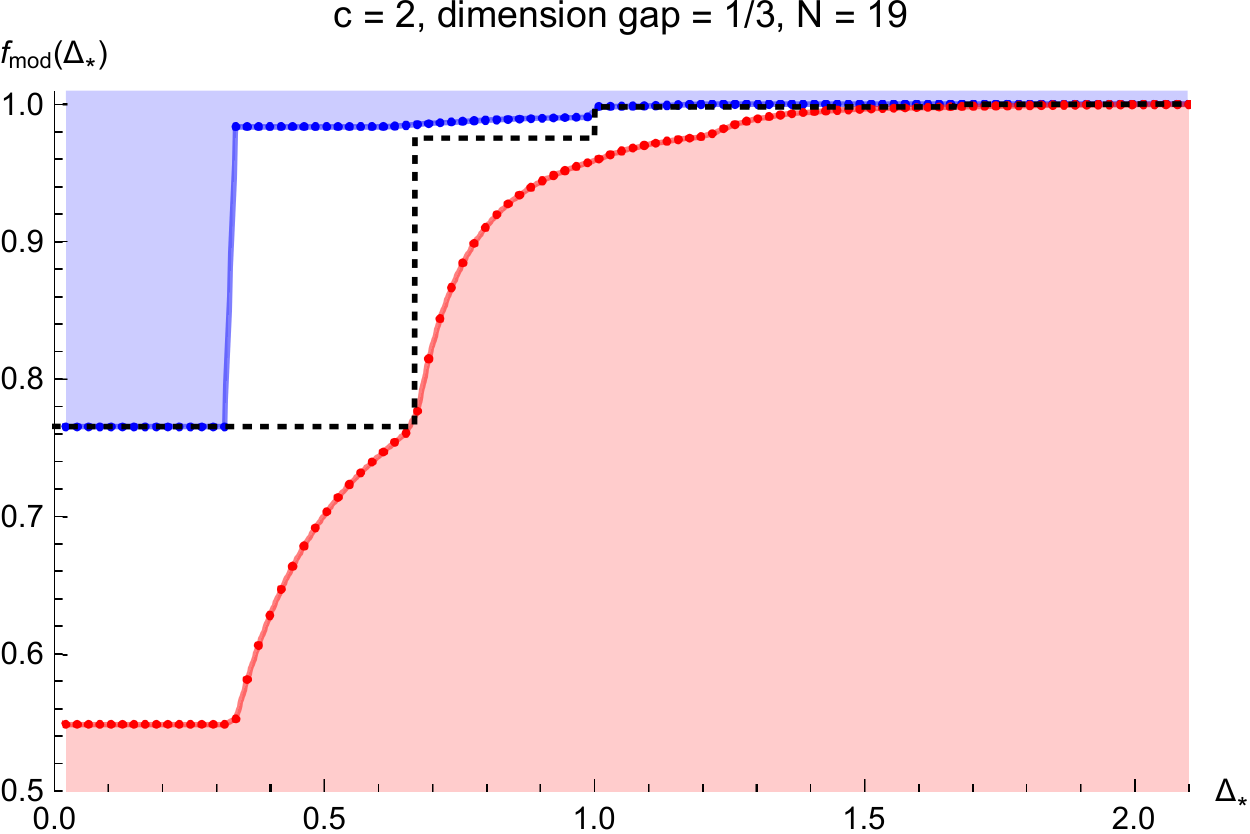}}~
 \subfloat[]{\includegraphics[width=.49\textwidth]{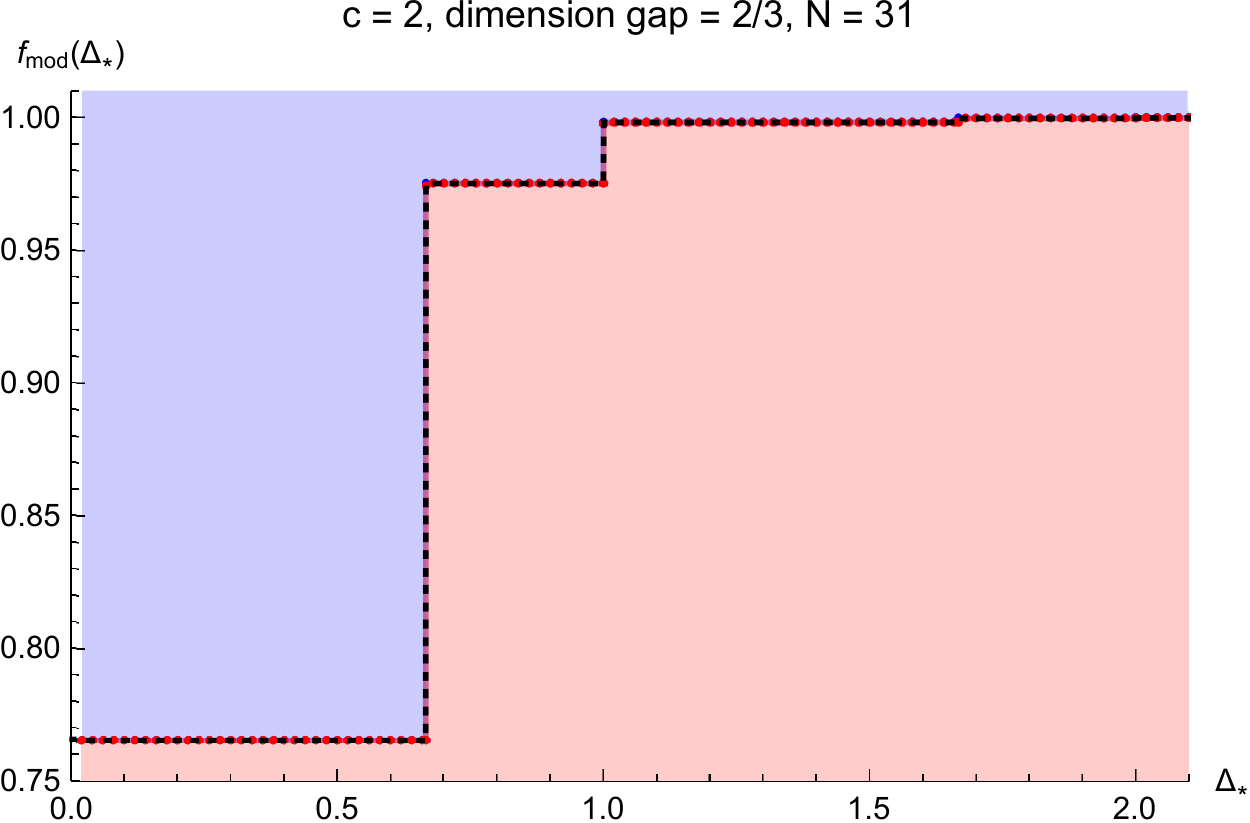}}\\
 \subfloat[]{\includegraphics[width=.49\textwidth]{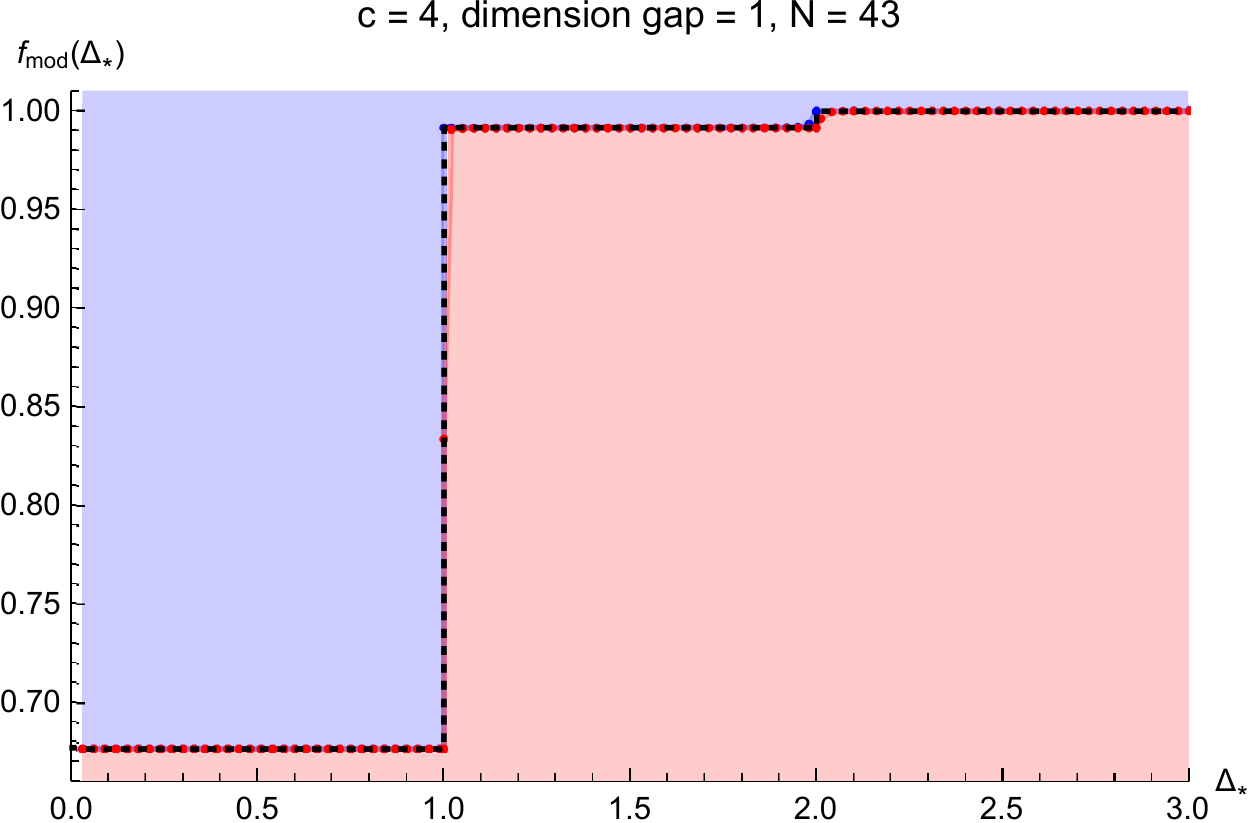}}~
 \subfloat[]{\includegraphics[width=.49\textwidth]{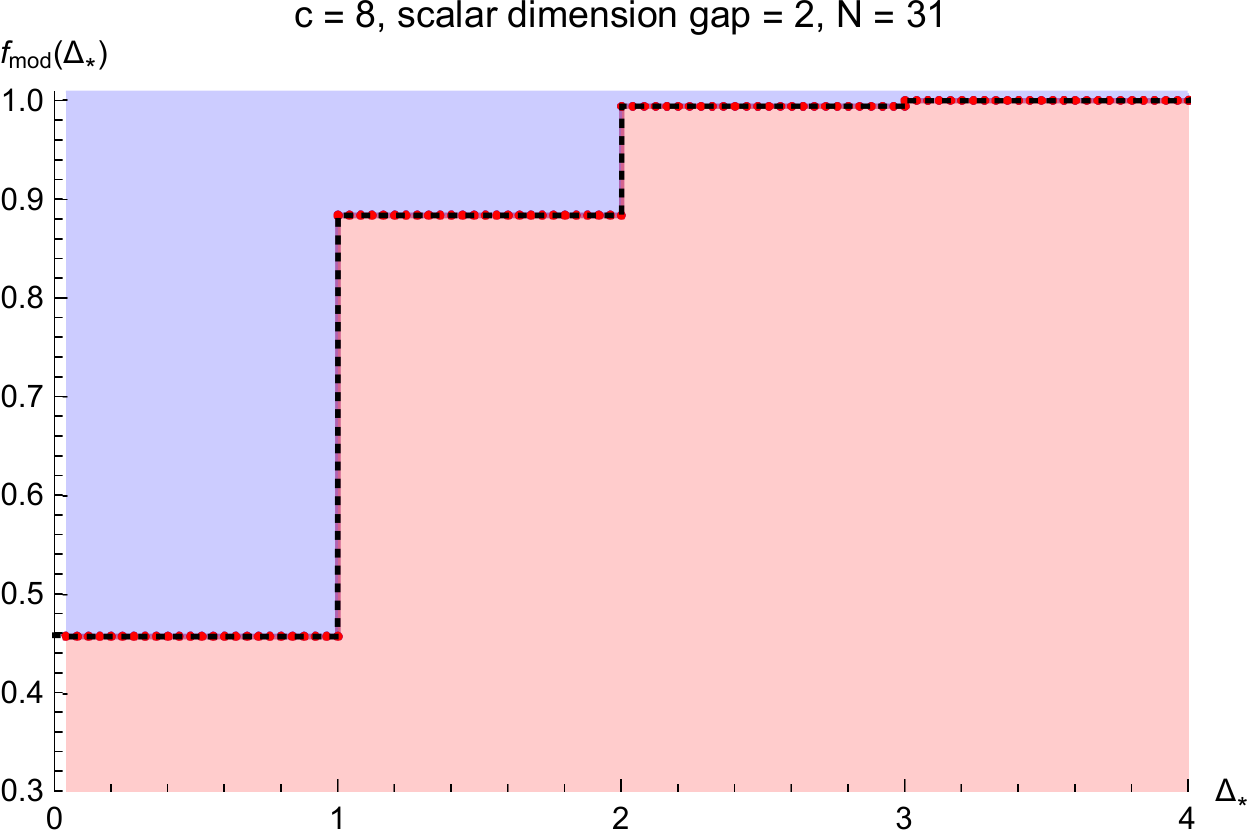}}
 \caption{The upper (blue) and lower (red) bounds on the modular spectral function. {\bf Top:} The bounds for $c=2$ with an assumed dimension gap one-half of (left) and equal to (right) the maximal gap allowed by modular invariance.
 {\bf Bottom:} The bounds on the modular spectral function for $c=4$ with the maximal dimension gap (left) and for $c=8$ with the maximal gap in the spectrum of scalar primaries (right). In all cases, the dotted lines denote the modular spectral function for the corresponding extremal spectrum.}\label{fig:ModularExtremal}
 \end{figure}

 For $c=2$, the dimension gap bound of $\Delta_{\rm mod}(2) = {2\over 3}$ is realized by the spectrum of the $SU(3)$ WZW model at level one. This theory admits a description in terms of free bosons with $T^2$ target space at the $\bZ_3$-invariant point in its complex structure and K\"ahler moduli spaces, with partition function
 \ie
Z_{\rm ext}(2,{2\over 3}) =&\sum_{n_i,w^j\in\bZ}{q^{{\alpha'\over 4}k_L^2}{\bar q}^{{\alpha'\over 4}k_R^2}\over |\eta(\tau)|^4},
 \fe
 where
 \ie
 k_{L,R}^2 =& {G^{mn}\over \alpha'}(n_m + B_{mk}w^k \pm G_{mk}w^k)(n_n + B_{nl}w^l \pm G_{nl}w^l)
 \fe
 for $G = \begin{pmatrix} 1 & \half \\ \half & 1\end{pmatrix},~B = \begin{pmatrix} 0 & \half \\ -\half & 0\end{pmatrix}$. The bounds on the modular spectral function collapse precisely to this extremal modular spectral function when the maximal gap is imposed, as shown in Figure~\ref{fig:ModularExtremal}.
 
 For $c=4$, the dimension gap bound of $\Delta_{\rm mod}(4) = 1$ is realized by the spectrum of the $SO(8)$ WZW model at level 1, which also admits a description in terms of 8 free fermions with diagonal GSO projection. This theory occupied the kink on the curve $\Delta_{\rm mod}(c)$. The partition function of this theory is given by
 \ie
Z_{\rm ext}(4, 1) = {1\over 2}\left(\left|{\Theta_2(\tau)\over\eta(\tau)}\right|^8+\left|{\Theta_3(\tau)\over\eta(\tau)}\right|^8+\left|{\Theta_4(\tau)\over\eta(\tau)}\right|^8\right).
 \fe
 Once again, in Figure~\ref{fig:ModularExtremal} we see that the bounds on the modular spectral function collapse to that of the extremal spectrum.
 
 For $c=8$, there is a nontrivial bound on the dimension gap in the spectrum of scalar primaries, $\Delta_{\rm mod}^{s=0}(8) = 2$. This bound on the scalar gap is saturated by the spectrum of the $E_8$ WZW model at level one. This theory, which occupied the first kink on the bounding curve $\Delta_{\rm mod}^{s=0}(c)$, admits an equivalent description in terms of 8 compact bosons at the holomorphically factorized point in the moduli space; the holomorphic factor is described by the Narain compactification on $\Gamma_8$, the root lattice of $E_8$. The partition function is 
 \ie
 Z_{{\rm ext},s=0}(8,2) = |j(\tau)|^{2\over 3},
 \fe
 where $j(\tau)$ is the elliptic $j$-invariant.
 Figure~\ref{fig:ModularExtremal} shows that the bounds on the modular spectral function (derived using $\Delta_s^*= \Delta_{\rm mod}^{s=0}\delta_{s,0} + s$) collapse to that of the extremal spectrum.

\subsubsection{Only scalar primaries}
\label{sec:modularscalaronly}

\begin{figure}[h!]
\centering
\includegraphics[width=.65\textwidth]{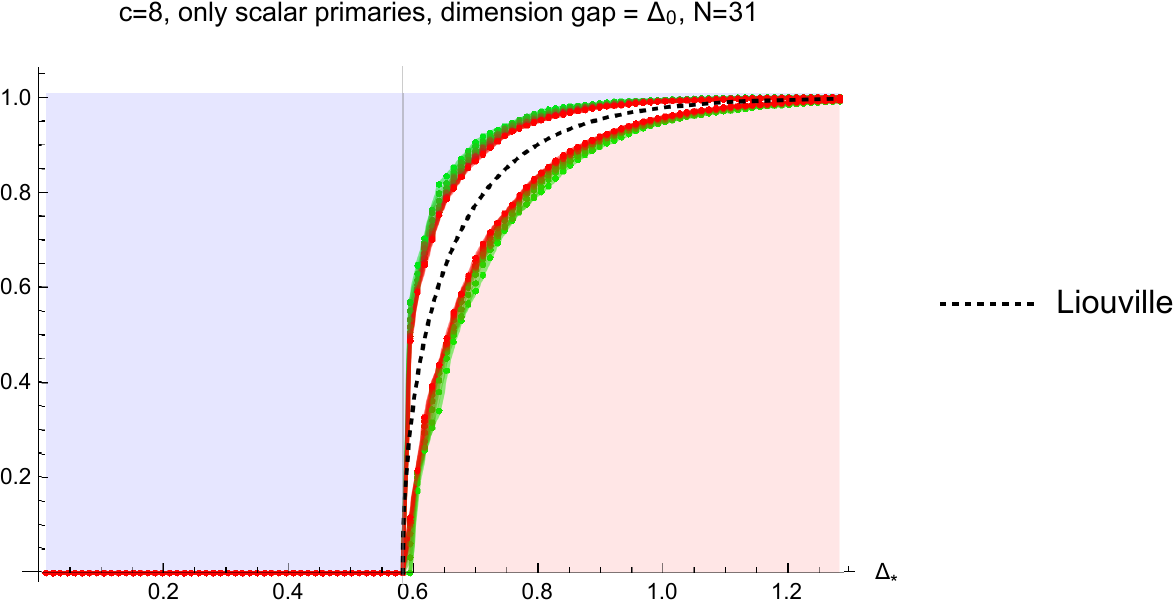}
\caption{Bounds on the modular spectral function assuming only scalar primaries and a dimension gap of $\Delta_{\rm gap} = {c-1\over 12}$ for $c=8$.}\label{fig:ModularScalarsOnly}
\end{figure}

As an additional example to illustrate the convergence of the bounds on the modular spectral function, we revisit the case of a CFT with only scalar primary operators. In section~\ref{sec:ModularAnalytic}, we showed that a unitary $c>1$ CFT with primaries of bounded spins must have a non-compact spectrum with only scalar primaries and a density of states equal to that of Liouville theory. The modular spectral function of Liouville theory is given by
\ie\label{eq:LiouvilleModularSpectral}
f_{\rm mod}^{\rm Liouville}(\Delta_*) 
=& \text{Erf}(\sqrt{2\pi(\Delta_*-2\xi)}).
\fe
Assuming a scalar-only spectrum, and a dimension gap $2\xi$ (that is, we impose (\ref{eq:ModularUpperInequality},\ref{eq:ModularLowerInequality}) for $s=0$ only with $\Delta_0^* = 2\xi$), the numerical results of upper and lower bounds on the modular spectral function are shown in Figure~\ref{fig:ModularScalarsOnly}.
Note that while the bounds do appear convergent toward the Liouville modular spectral function (as they must), the rate of convergence is rather slow compared to the previous examples of discrete extremal spectra at small $c$. On the other hand, such a slow convergence with $N$ is qualitatively similar to our bounds on the 4-point spectral function in the scalar-only case, as analyzed in section~\ref{ulbounds}, where we also expect a continuous spectrum, and also to the non-compact example in the next subsection. 

We thus expect that the slow convergence is associated with continuity of the spectrum. This is natural from the point of view of extremal functionals -- the extremal spectrum should converge to the continuous one as $N\to\infty$, but at any finite $N$ the extremal spectrum has finitely many operators, which therefore should condense. On the other hand, in the discrete case we typically need only a small number of operators to accurately determine the partition function or the correlation function in the neighborhood of the crossing/modular symmetric point.

\subsubsection{No scalar primaries}
In \cite{Collier:2016cls}, following the observation that the bound on the gap of the dimension of scalar primaries diverged as $c\to 25^-$, it was shown that for $c\ge 25$ there exist (non-compact) modular-invariant spectra with no scalar primary operators. This is due to the fact that the modular invariant function
\ie\label{eq:NoScalarPartitionFunction}
Z^{\text{no-scalar}}(\tau,\bar\tau) =& {J(\tau)+\bar J(\bar\tau)\over \tau_2^{\half}|\eta(\tau)|^2},
\fe
where $J(\tau) = j(\tau)-744$, may be interpreted as the partition function of a unitary non-compact CFT with no scalar primary operators, twist gap $c-25\over 12$ and dimension gap $c-13\over 12$ for $c\ge 25$. This spectrum turns out to saturate the bound on the dimension gap in the case that there are no scalar primaries in the spectrum. Writing $J(\tau) = \sum_{s=-1}^\infty j_s q^s$, the modular spectral function takes the form
\ie\label{eq:NoScalarsModularSpectral}
f_{\rm mod}^{\text{no-scalar}}(\Delta_*)={1\over 984}\sum_{s=-1}^{\floor{\Delta_*-2\xi}}j_s \text{Erf}(\sqrt{2\pi(\Delta_*-2\xi-s)}).
\fe

\begin{figure}[h!]
\centering
\includegraphics[width=.5\textwidth]{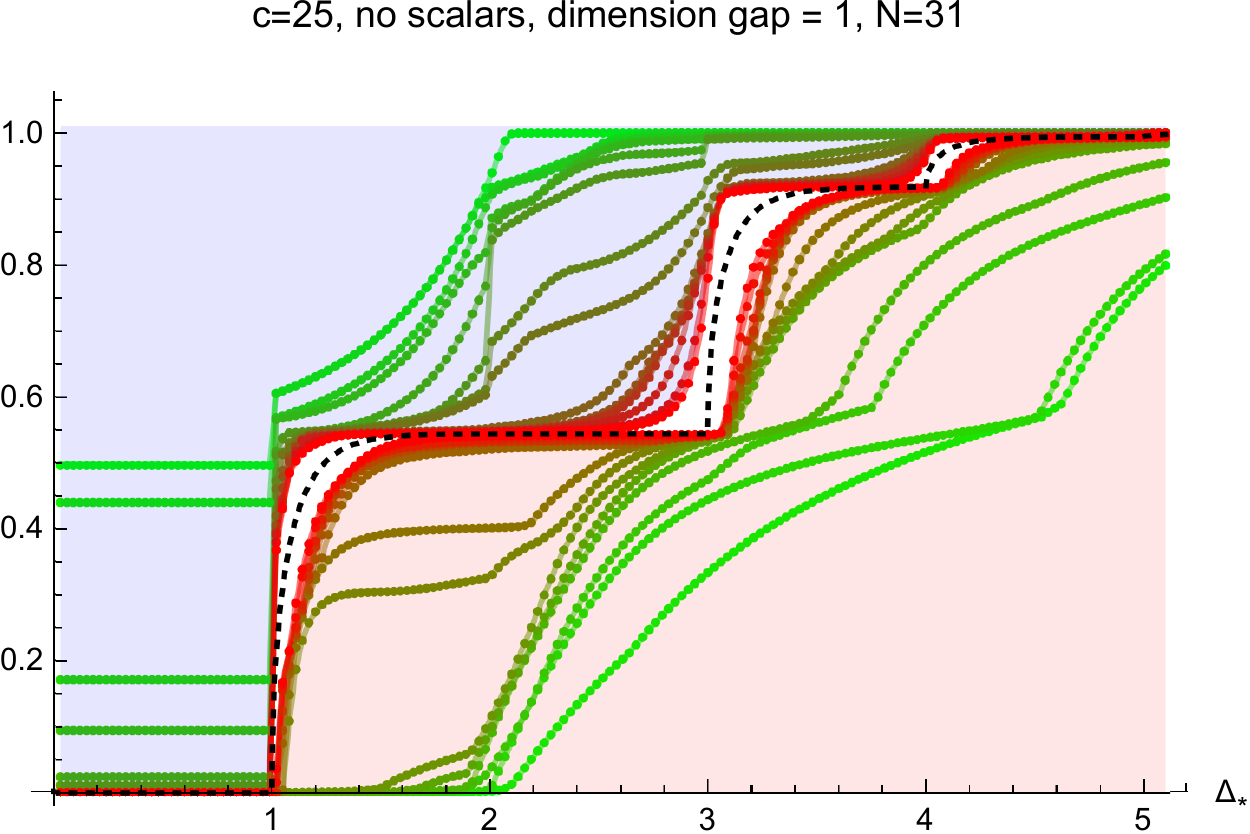}
\caption{Bounds on the modular spectral function with no scalar primaries in the spectrum and a dimension gap of $\Delta_{\rm gap} = {c-13\over 12}$ for $c=25$. The dotted black curve denotes the extremal modular spectral function (\ref{eq:NoScalarsModularSpectral}).}\label{fig:ModularNoScalars}
\end{figure}

To compute the bounds on the modular spectral function in this case, we impose (\ref{eq:ModularUpperInequality},\ref{eq:ModularLowerInequality}) for $s>0$ with $\Delta_s^* = \text{max}(s,2\xi-1)$. As shown in Figure~\ref{fig:ModularNoScalars}, the bounds on the modular spectral function do indeed appear to be converging to (\ref{eq:NoScalarsModularSpectral}) as the derivative order of the linear functional is increased, suggesting that the no-scalar spectrum is unique for $c=25$. Note that for $c=25$ the dimension gap $\frac{c-13}{12}$ coincides with the unitarity bound (since we assume no scalars). We expect that for $c>25$ the uniqueness holds only under the assumption of the dimension gap $\frac{c-13}{12}$.

\subsection{CFTs at large $c$ with large gap}\label{sec:LargeC}
In \cite{Collier:2016cls}, the upper bound on the dimension gap $\Delta_{\rm mod}(c)$ due to modular invariance of the torus partition function was computed numerically for central charge up to $c\sim\mathcal{O}(10^2)$. As $c$ is increased, the convergence of the upper bound with increasing derivative order $N$ slows, and accurate determinations of the optimal bound on the gap require a careful extrapolation to the limit $N\to\infty$. Nonetheless, a conjecture on the monotonicity of the slope of the optimal bounding curve ${d \Delta_{\rm mod}(c)\over d c}$ leads one to conclude that the asymptotic slope is less than $1\over 9$. Potentially, the asymptotic slope could be as small as ${1\over 12}$, a possibility that is natural from the holographic perspective (see the discussion in the next section) but with no direct evidence from the analysis of the modular crossing equation.

\begin{figure}[h!]
\centering
\subfloat[]{\includegraphics[width=.49\textwidth]{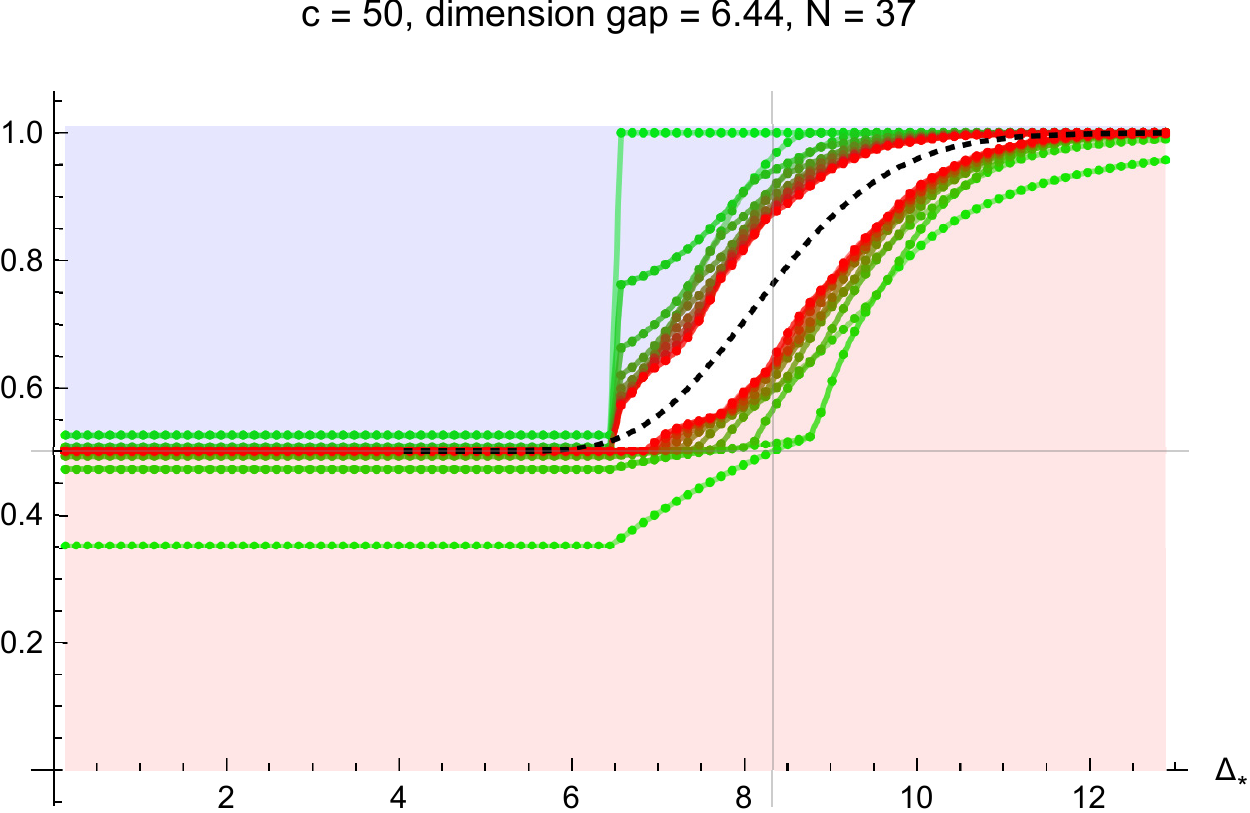}}~
\subfloat[]{\includegraphics[width=.49\textwidth]{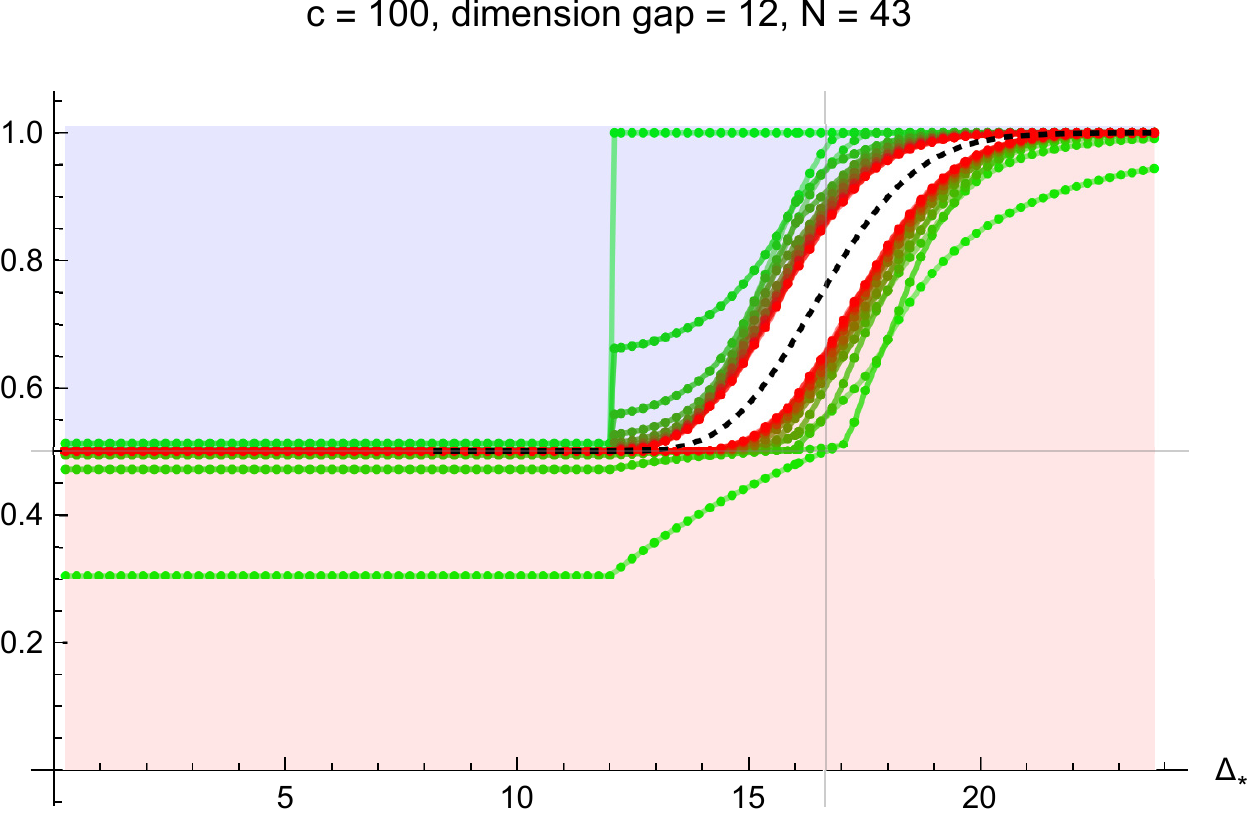}}\\
\subfloat[]{\includegraphics[width=.5\textwidth]{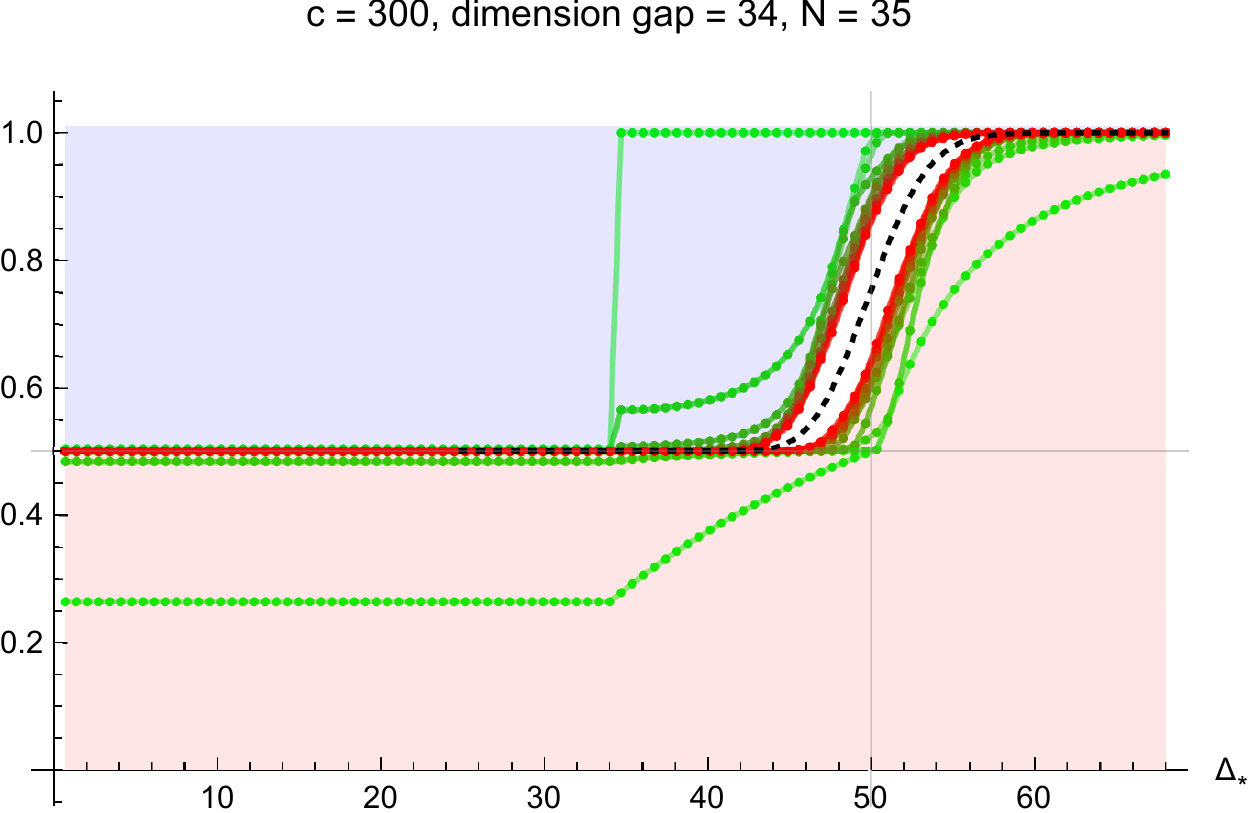}}
\caption{Upper and lower bounds on the modular spectral function in the case that the dimension gap is close to the maximal value allowed by modular invariance for $c=50,100,300$. The dotted black curve shows the modular spectral function of perturbative pure gravity due to the thermal $AdS_3$ and Euclidean BTZ saddles in the gravitational path integral.}\label{fig:ModularLargeC}
\end{figure}

Thus at large $c$ it has been difficult to determine $\Delta_{\rm mod}(c)$ accurately, and furthermore the exponential growth of operator degeneracies (combined with the need to go to very large $N$ to get a good approximation of the optimal linear functional at large $c$) makes it practically impossible to resolve the discreteness of the spectrum by bounding the modular spectral function even when the bound $\Delta_{\rm mod}(c)$ is saturated. Nonetheless, we can study the bounds on the modular spectral function assuming a gap $\Delta_{\rm gap}$ close to $\Delta_{\rm mod}(c)$, at values of $c$ where the value of $\Delta_{\rm mod}(c)$ can be reliably computed by numerical extrapolation of $\Delta_{\rm mod}^{(N)}(c)$ to $N=\infty$. Figure~\ref{fig:ModularLargeC} shows plots of the bounds on the modular spectral function for $c=50,100,300$ with assumed dimension gap $\Delta_{\rm gap}$ close to the bound $\Delta_{\rm mod}(c)$.

The plots reveal several interesting features that we believe are universal at large $c$ assuming a sufficiently large gap $\Delta_{\rm gap}$($>{c-1\over 12}$). Firstly, for $\Delta_*\ll {c\over 6}$, the upper and lower bounds on the modular spectral function converge to $f_{\rm mod}(\Delta_*)={1\over 2}$: that is, modular invariance demands that the vacuum character accounts for exactly half of the partition function at the self-dual temperature. Furthermore, the bounds appear to be convergent upon a smooth function that interpolates between $\half$ and $1$ in a window of size $\sim \sqrt{c}$ about $\Delta_* = {c\over 6}$. 

It is generally expected that 2D CFTs at large $c$ with large gap should be holographically dual to a semiclassical theory of pure gravity in $AdS_3$. To the best of our knowledge, this statement has not been precisely formulated: how large does the gap need to be? If we merely demand that the dimension gap\footnote{In a non-compact CFT where the vacuum is absent, by gap we mean the dimension of the lightest primary.
} grow linearly in $c$, corresponding to a Planckian mass gap in the bulk theory, but with a coefficient less than ${1\over12}$, then the entropy need not follow Bekenstein-Hawking in the entire range $\Delta \geq {c\over6}$.  This is the range of masses for which BTZ black holes dominate the canonical ensemble at its Hawking temperature.  
On the other hand, one might expect that CFTs with gap close to $\Delta_{\rm mod}(c)$ (if they exist) are holographic duals to suitable non-perturbative completions of pure gravity in $AdS_3$, in the sense that observables such as the spectral density are correctly captured by the perturbative expansion around known saddle points of the gravitational path integral in the bulk up to $\exp(-c)$ corrections.

This suggests that we compare the bounds on the modular spectral function to that of pure gravity, which, up to a priori unknown non-perturbative corrections, is computed by the contributions from thermal $AdS_3$ and the Euclidean BTZ black hole saddle points, which are known to be perturbatively 1-loop exact. We derive this modular spectral function in Appendix~\ref{app:BTZSpectral}, see in particular (\ref{eq:BTZSpectral},\ref{eq:BTZSpectralLargeC}).
The bounds shown in Figure~\ref{fig:ModularLargeC} indeed appear to be converging upon the pure gravity result (\ref{modbtz}) for dimensions above the assumed gap.

\begin{figure}[h!]
\centering
\subfloat[]{\includegraphics[width=.49\textwidth]{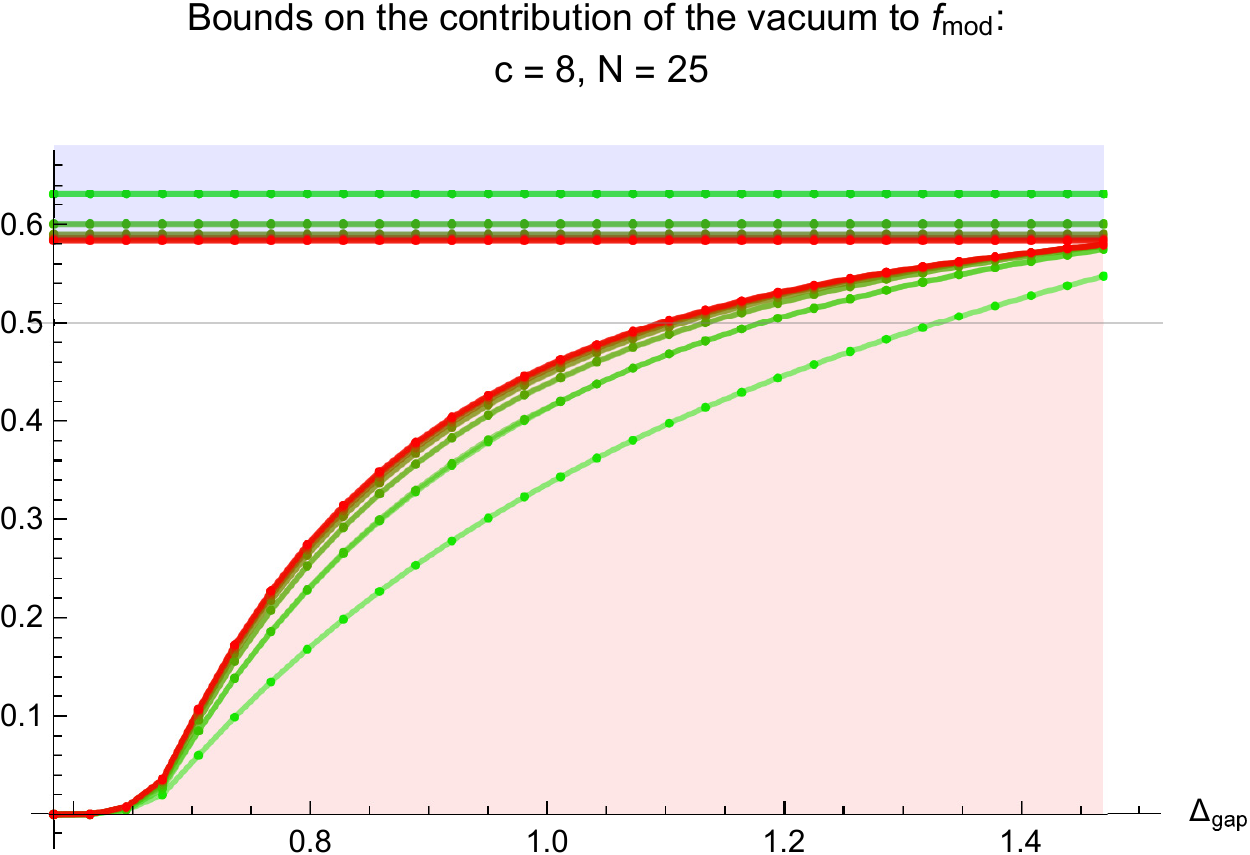}}~
\subfloat[]{\includegraphics[width=.49\textwidth]{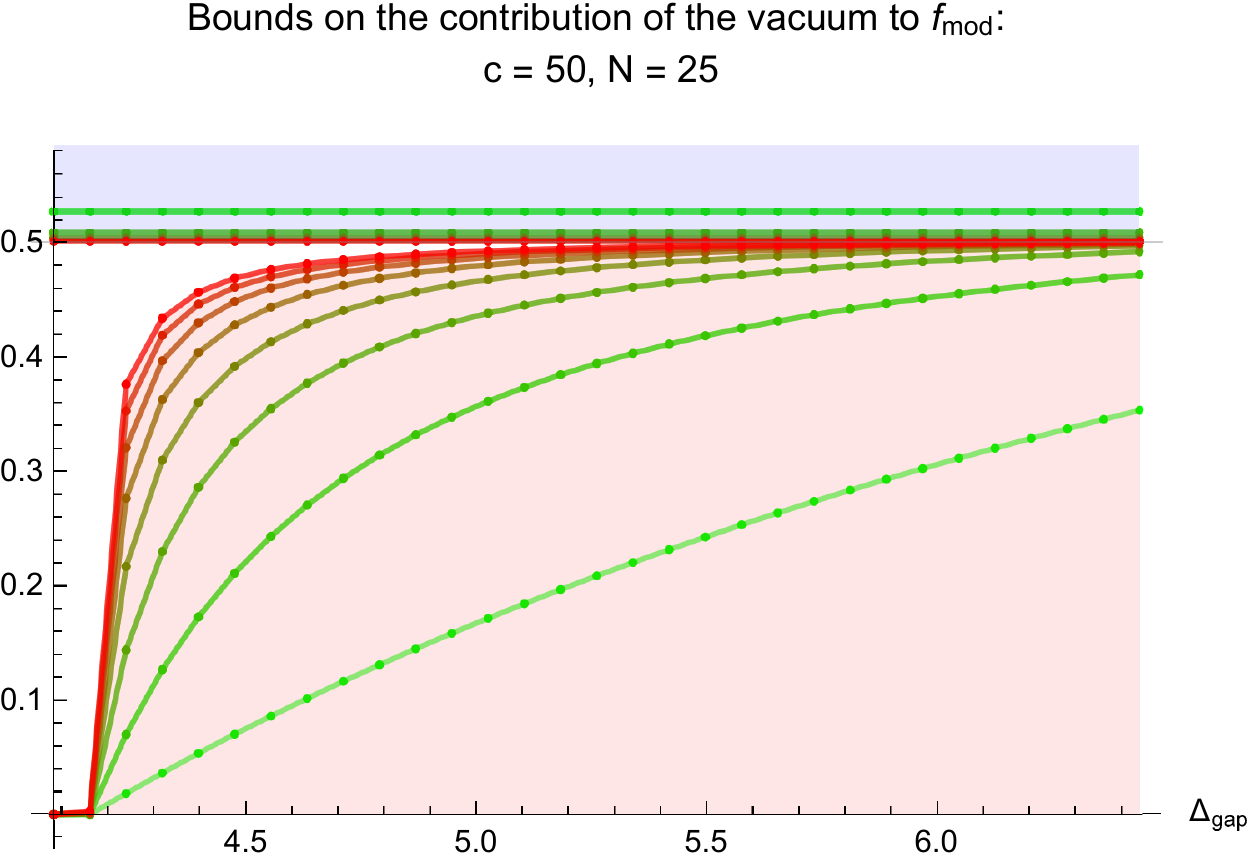}}\\
\subfloat[]{\includegraphics[width=.49\textwidth]{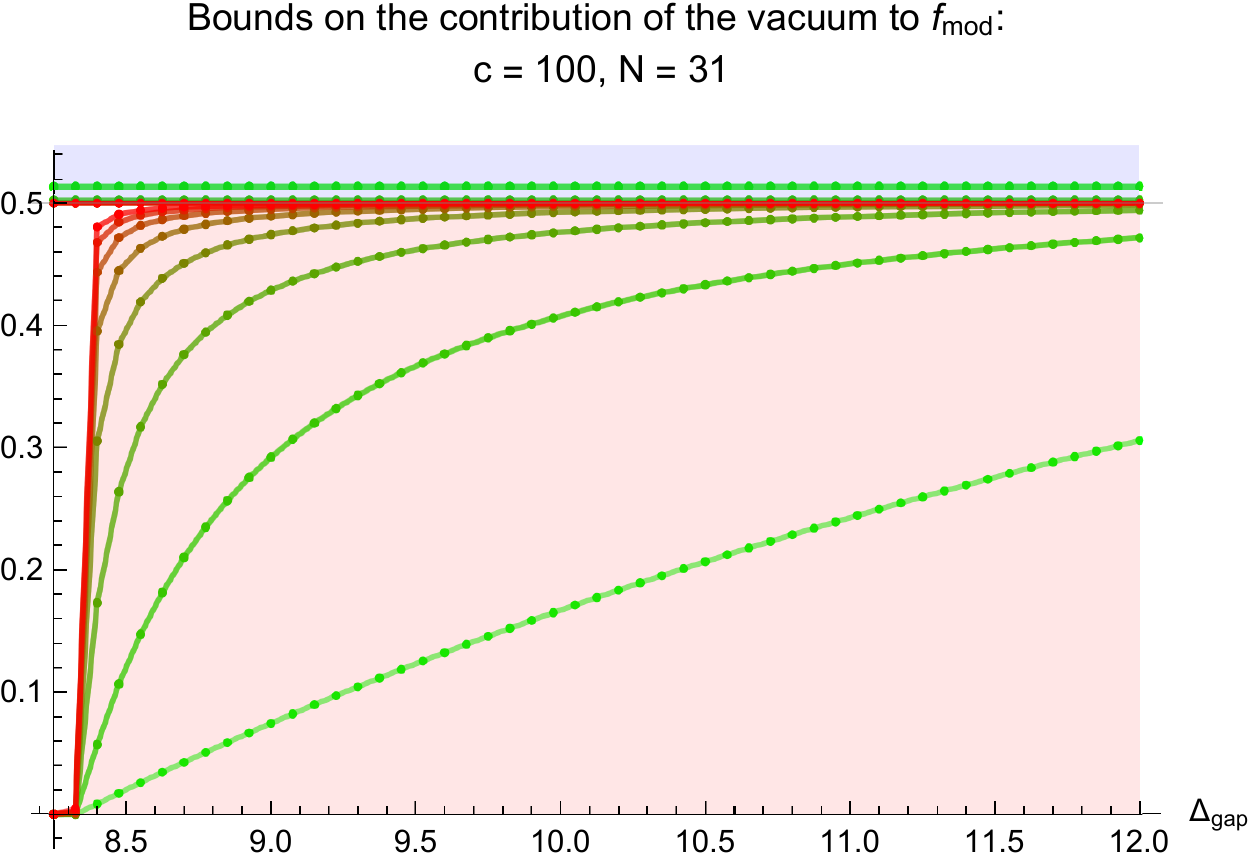}}
\caption{Bounds on the contribution of the vacuum character to the modular spectral function as a function of the imposed gap in the dimensions of primary operators for $c=8,50,100$.}\label{fig:ModularVacuum}
\end{figure}

Note that for $\Delta_*\ll {c\over 6}$, that the bounds on the modular spectral function with large gap converge to $\half$ can be explained by the fact that in the semiclassical limit, the gravitational path integral evaluated at the self-dual temperature is dominated by the contributions of two saddles mentioned above, which are exchanged by the modular $S$ transformation, and thus the vacuum contribution accounts for ${1\over 2}$. Interestingly, at large $c$, the upper and lower bounds on the vacuum contribution already converge to ${1\over 2}$ when the dimension gap is slightly above ${c-1\over 12}$, not necessarily close to $\Delta_{\rm mod}(c)$. This is illustrated in Figure~\ref{fig:ModularVacuum}, where we plot the bounds on the contribution of the vacuum to the modular spectral function as a function of the dimension gap for $c=8,50,100$. Note that the vacuum contribution to the spectral function determines the partition function $Z(\tau,\bar\tau)$ itself at the self-dual temperature ($\tau=-\bar\tau=i$). From the bulk perspective, that the vacuum accounts for ${1\over 2}$ of the modular spectral function amounts to the statement that the thermal $AdS_3$ and BTZ saddle points are the two dominant saddle points in the gravitational path integral, while all other saddle points are exponentially suppressed.\footnote{This property does not hold when the dimension gap is less than or equal to $c-1\over 12$, even if the former is of order $c$, at large $c$. A possible bulk interpretation is that there are singular saddle point contributions (such as the Euclidean continuation of the massless BTZ black hole) to the gravity path integral where the pure gravity perturbation theory breaks down.}

In \cite{Hartman:2014oaa}, it was shown that in theories where the light spectrum is appropriately sparse (a condition of which the maximal gap is an extreme case), the microcanonical entropy is given universally by the Cardy formula to leading order for all operators with dimension $\Delta\ge {c\over 6}$. Note that the sparseness criterion is satisfied by our assumption on the gap, but our statement regarding the modular spectral function and thereby the spectral density extends to the regime of $\Delta$ slightly below ${c\over 6}$ (see further discussion in the next section).

\begin{figure}[h!]
\centering
\includegraphics[width=.75\textwidth]{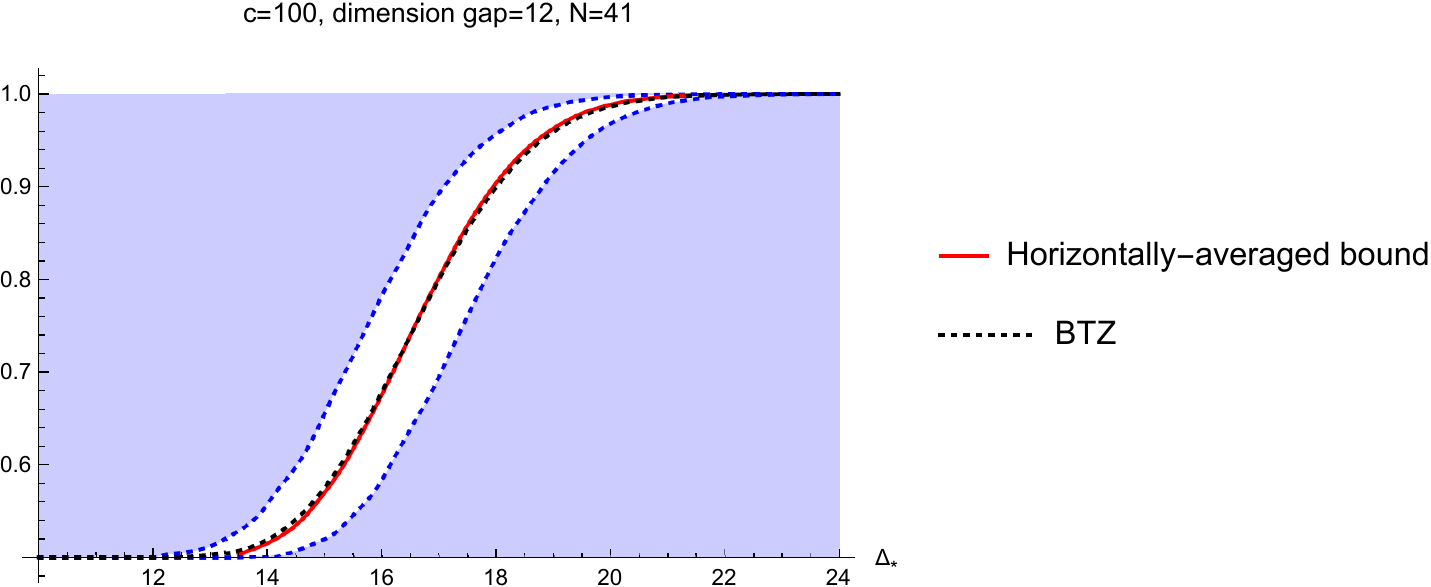}
\caption{The horizontal average of the upper and lower bounds on the modular spectral function for $c=100$ with dimension gap close to the upper bound imposed by modular invariance.}\label{fig:ModularHorizontalAverage}
\end{figure}

We conjecture that in the large $c$ limit, assuming a gap sufficiently close to $\Delta_{\rm mod}(c)$,\footnote{As remarked earlier, it is likely that a gap not too far above $c-1\over 12$ will suffice, not necessarily close to $\Delta_{\rm mod}(c)$.} the bounds $f_{\rm mod}^\pm(c)$ converge onto the modular spectral function of pure gravity described above, up to order $\exp(-c)$ corrections. Indeed, we note that for $c\sim {\cal O}(10^2)$, the horizontal average of the bounds $\overline{f_{{\rm mod},N}}(\Delta_*)$ is already well approximated by the pure gravity modular spectral function at moderate values of $N$, as shown in Figure~\ref{fig:ModularHorizontalAverage}.

\section{On the universality of the BTZ spectral density}

The BTZ black hole in $AdS_3$ has a striking feature that is unlike black holes in other spacetime dimensions (in asymptotically either AdS or flat spacetime): a Planckian mass BTZ black hole has a macroscopic horizon radius (rather than, say, Planckian radius), provided that the mass is an order 1 fraction above the BTZ threshold in Planck units. The microstates of such a black hole would be dual to an operator in the CFT of dimension $\Delta>(1+\epsilon){c\over 12}$, where $\epsilon$ is an order 1 fraction that does not scale with $c$. When there is a sufficiently large mass gap in the spectrum, standard effective field theory reasoning in the bulk would suggest that the the entropy of the BTZ black hole, which captures the degeneracy of microstates, should be computed from the Bekenstein-Hawking formula based on the Einstein-Hilbert action, as any local higher-derivative corrections to the Einstein-Hilbert action of pure gravity in three dimensions can be absorbed by field redefinition. This would predict a degeneracy or spectral density
\ie\label{specbtz}
\rho(\Delta) \sim \exp\left[2\pi \sqrt{{c\over 3} (\Delta - {c\over 12})}\,\right]
\fe
to leading order in the large $c$ limit, for $\Delta/c>{1\over 12}$.

For $\Delta>{c\over 6}$, i.e., above {\it twice} the BTZ threshold, this universal behavior of the spectral density was demonstrated in \cite{Hartman:2014oaa} to be a consequence of the sparseness of the spectrum and modular invariance. From the gravity perspective, this is also the regime in which the Euclidean BTZ black hole solution is the dominant saddle point of the Euclidean pure gravity path integral, i.e., the BTZ black hole dominates the canonical ensemble at its Hawking temperature, and therefore the spectral density must be (\ref{specbtz}) in order to produce the correct free energy above the self-dual temperature.

The regime ${c\over 12}<\Delta<{c\over 6}$ is much more interesting. Here the BTZ black hole does {\it not} dominate the canonical ensemble. Its contribution to the gravitational free energy is non-perturbatively suppressed compared to the thermal $AdS_3$ contribution. A priori, since we do not know the most general non-perturbative contributions to the pure gravity path integral, we cannot draw any reliable conclusion on the spectral density in this regime. This also puts doubt on the validity of the Bekenstein-Hawking formula, despite the macroscopic size of the horizon. If the Bekenstein-Hawking formula is violated in this regime, it then indicates some sort of breakdown of the effective field theory reasoning based on locality.

What conclusion can we draw from our numerical bounds on the modular spectral function for $c\sim {\cal O}(10^2)$? We saw that in a window of size $\sqrt{c}$ around ${c\over 6}$, the modular spectral function is constrained to well approximate the $AdS_3+{\rm BTZ}$ answer, in agreement with the expectation from the known perturbative contributions to the Euclidean gravity path integral. Unfortunately, our numerical results do not have sufficient resolution to allow for distilling a contribution of order $\exp(-c)$, thus preventing us from concluding whether the Bekenstein-Hawking entropy of BTZ correctly accounts for the spectral density in the regime $\Delta = yc$, for ${1\over 12}<y<{1\over 6}$. In fact, if the latter is true, then the asymptotic slope of the modular bound on the dimension gap, $\lim_{c\to \infty} d\Delta_{\rm mod}(c)/dc$, must be equal to ${1\over 12}$, but this has not been shown. Thus, the fate of the small-yet-large BTZ black holes below twice the BTZ threshold remains a mystery.

\section*{Acknowledgements} 

We are grateful to Daniel Harlow, Zohar Komargodski, David Simmons-Duffin, and Balt van Rees for discussions. PK, YL, and XY would like to thank the Center for Quantum Mathematics and Physics at University of California, Davis, PK and XY thank University of Kentucky, PK thanks Harvard University, XY thanks Korea Institute for Advanced Study, Stony Brook University, Indian String Meeting 2016, and Weizmann Institute, for their hospitality during the course of this work. This work is supported by a Simons Investigator Award from the Simons Foundation, by DOE grant DE-FG02-91ER40654, and by DOE grant DE-SC0011632 (PK, YL). SC is supported in part by the Natural Sciences and Engineering Research Council of Canada via a PGS D fellowship. YL is supported in part by the Sherman Fairchild Foundation.
The numerical computations in this work are performed using the SDPB package \cite{Simmons-Duffin:2015qma} on the Odyssey cluster supported by the FAS Division of Science, Research Computing Group at Harvard University.

\appendix

\section{Zamolodchikov's recurrence relation}
\label{App:Zam}

The Virasoro block for a four-point function $\langle {\cal O}_1(z) {\cal O}_2(0) {\cal O}_3(1) {\cal O}_4(\infty) \rangle$ with central charge $c$, external weights $h_i$, and internal weight $h$ can be represented as
\ie
F_c^{Vir}(h_i; h; z) =& {[16 q(z)]^{h - {c-1 \over 24}} z^{{c-1 \over 24} - h_1 - h_2}} (1-z)^{{c-1 \over 24} - h_1 - h_3}
\\
 &\times [\theta_3(q(z))]^{{c-1 \over 8} - 4(h_1 + h_2 + h_3 + h_4)} H( \lambda_i^2, h | q(z)),
\fe
where the nome $q(z)$ is defined as
\ie
q(z) \equiv \exp(i\pi\tau(z)),
\quad
\tau(z) \equiv {i F(1-z) \over F(z)}, \quad F(z) \equiv {}_2F_1({1/2}, {1/2}, 1 | z).
\fe
If we define
\ie
c = 1 + 6Q^2, \quad Q = b + {1 \over b}, \quad h_{m,n} = {Q^2\over 4} - \lambda_{m,n}^2, \quad \lambda_{m,n} = {1\over 2} ({m\over b} + nb),
\fe
then $H( \lambda_i^2, h | q(z))$ satisfies Zamolodchikov's recurrence relation
\ie\label{recH}
H(\lambda_i^2, h | q(z)) = 1 + \sum_{m,n\geq 1} {[q(z)]^{mn} R_{m,n}(\{\lambda_i\}) \over h - h_{m,n} }
H(\lambda_i^2, h_{m,n} + mn| q(z) ),
\fe
where $h_{m,n}$ are the weights of degenerate Virasoro representations, and $R_{m,n}(\{\lambda_i \})$ are
\ie
R_{m,n}(\{\lambda_i \}) = 2 {\prod_{r,s} (\lambda_1+\lambda_2 - \lambda_{r,s}) (\lambda_1-\lambda_2 - \lambda_{r,s}) (\lambda_3+\lambda_4 - \lambda_{r,s}) (\lambda_3-\lambda_4 - \lambda_{r,s}) \over \prod_{k,\ell}' \lambda_{k,\ell} }.
\fe
The product of $(r,s)$ is over
\ie
r &= -m+1,-m+3,\dotsc,m-1,
\\
s &= -n+1,-n+3,\dotsc,n-1,
\fe
and the product of $(k,\ell)$ is over
\ie
k &= -m+1,-m+2,\dotsc,m,
\\
\ell &= -n+1,-n+2,\dotsc,n,
\fe
excluding $(k,\ell)=(0,0)$ and $(k, \ell) = (m,n)$.

\section{Liouville CFT and DOZZ structure constants}
\label{dozzreview}

The Liouville CFT is parameterized by the central charge $c = 1+6Q^2$, where $Q = b+b^{-1}$, and a cosmological constant $- \mu < 0$.  It is governed by the action
\ie
S_{\rm Liouville} = {1\over 4\pi}\int d^2z\sqrt{g}\left(g^{mn}\partial_m\phi\partial_n\phi + QR\phi + 4\pi \mu e^{2b \phi}\right).
\fe 
To study Liouville theory on the sphere, one typically works with a flat reference metric $g_{mn}$ supplemented with the boundary condition
\ie
\phi(z,\bar z)= -2Q\log|z| + \mathcal{O}(1),~|z|\to\infty.
\fe
 The field $\phi(z,\bar z)$ is not a primary operator under holomorphic coordinate transformations $z\to w(z)$. In this case one must take care to regulate the action and introduce boundary terms to ensure that the action is finite and invariant under conformal transformations. 
 
The Hilbert space consists of a continuous spectrum of scalar primary operators $\mathcal{V}_\alpha$ with $\A \in {Q\over2} + i\bR_{\geq0}$ and conformal dimension $\Delta = 2\A(Q-\A)$. Operators with $\A$ outside this range, such as the identity operator, do not correspond to normalizable states and thus do not belong to the Hilbert space.  Making use of a somewhat nonstandard convention (the reason for which will become clear soon), we normalize the primaries so that in the asymptotic regime where the Liouville potential vanishes (the $\phi \to -\infty$ limit) they take the form\footnote{In the literature on the Liouville CFT, usually considered are operators with the asymptotic form ${\cal V}_\A \sim e^{2\A\phi}$ and which do not have standard two-point functions.
}
\ie
\mathcal{V}_\alpha \sim S(\alpha)^{-\half}e^{2\alpha\phi}+S(\alpha)^{\half}e^{2(Q-\alpha)\phi},
\fe
where $S(\alpha)$ is the reflection amplitude
\ie
 S(\A) =& - (\pi\mu \gamma(b^2))^{(Q - 2\A) / b} { \Gamma(1-{(Q - 2\A) / b}) \Gamma(1-{(Q - 2\A) b } ) \over \Gamma(1+{(Q - 2\A) / b}) \Gamma(1+{(Q - 2\A) b} ) }.
\fe  
The torus partition function is the same as that of a single non-compact free scalar.  The sphere two-point function of primary operators is
\ie
& \langle \mathcal{V}_{\alpha_1}(z, \bar z)\mathcal{V}_{\alpha_2}(0) \rangle = {\delta(\alpha_1-\alpha_2)\over |z|^{\Delta_1 + \Delta_2}}.
\fe
Note that with our choice of conventions the two-point function is canonically normalized. 
 The sphere three-point function is given by the DOZZ structure constants \cite{Dorn:1994xn,Zamolodchikov:1995aa}
\ie
\label{dozz}
& \langle \mathcal{V}_{\alpha_1}(z_1, \bar z_1)\mathcal{V}_{\alpha_2}(z_2, \bar z_2) \mathcal{V}_{\alpha_3}(z_3, \bar z_3) \rangle = \left(\prod_{j=1}^3S(\alpha_j)^{-\half}\right){C(\A_1, \A_2, \A_3) \over |z_{12}|^{\Delta_1+\Delta_2-\Delta_3} |z_{23}|^{\Delta_2+\Delta_3-\Delta_1} |z_{31}|^{\Delta_3+\Delta_1-\Delta_2} },
\\
& C(\A_1, \A_2, \A_3) = \left[ \pi\mu \gamma(b^2) b^{2-2b^2} \right]^{(Q - \sum_i \A_i)/b} 
\\
&~~~~~~~~~~~~~~~~~~~~\times { \Upsilon'_b(0) \Upsilon_b(2\A_1)\Upsilon_b(2\A_2)\Upsilon_b(2\A_3) \over \Upsilon_b({\textstyle\sum_i \A_i} - Q) \Upsilon_b(\A_1 + \A_2 - \A_3) \Upsilon_b(\A_2 + \A_3 - \A_1) \Upsilon_b(\A_3 + \A_1 - \A_2) } .
\fe
The special functions are given by the following
\ie
\label{Upsilon}
\gamma(x) &= {\Gamma(x)\over \Gamma(1-x)}\\
\log\Upsilon_b(x) &= \int_0^\infty dt~t^{-1}\left[\left({Q\over 2}-x\right)^2e^{-t} - {\sinh^2\left[\left({Q\over 2}-x\right){t\over 2}\right]\over \sinh{tb\over 2}\sinh{t\over 2b}}\right],~0<\text{Re}(x)<\text{Re}(Q).
\fe
Note in particular that the upsilon function satisfies $\Upsilon_b(Q-x) = \Upsilon_b(x)$, which implies that $\Upsilon_b({Q\over2}+iP)$ is a real function of $P$. To extend $\Upsilon_b(x)$ beyond the range of its definition, one notes the following identities
\ie
\Upsilon_b(x+b) =& \gamma(bx) b^{1-2bx}\Upsilon_b(x)\\
\Upsilon_b(x+b^{-1}) =& \gamma(b^{-1}x) b^{{2 x\over b}-1}\Upsilon_b(x),
\fe
which can be proven by considering an integral representation of $\log\Gamma(x)$.
The function $\Upsilon_b(x)$ has simple zeroes at $x=0$, $x=Q$ as well as $x = mb + {n \over b}$ when $m$ and $n$ are both non-positive integers, and when $m$ and $n$ are both positive integers. It is instructive to rewrite the Liouville three-point function coefficient as a manifestly real function of the Liouville momenta $P_i = -i (\A_i - {Q\over2})$, since $P$ takes non-negative values for operators in the physical Hilbert space:
\ie
 {\mathcal C}(P_1, P_2, P_3) \equiv&
\left[ \pi\mu \gamma(b^2) b^{2-2b^2} \right]^{{Q\over2b}} \left(\prod_{j=1}^3S(\alpha_j)^{-\half}\right) C(\A_1, \A_2, \A_3)
\\
 = & { \Upsilon'_b(0) \over \Upsilon_b({Q\over2} + i\sum_j P_j) } \times \left[{\left(\Upsilon_b(2iP_1)\Upsilon_b(-2iP_1)\right)^{\half}\over \Upsilon_b\left({Q\over 2}+i(P_2+P_3-P_1)\right)} \times (\text{2 permutations}) \right],
\fe
where we have used that the reflection amplitude can also be written as
\ie
S(\alpha) =& \left[\pi \mu \gamma(b^2)b^{2-2b^2}\right]^{(Q-2\alpha)/b}{\Upsilon_b(2\alpha)\over \Upsilon_b(2\alpha-Q)}.
\fe
The statement that the structure constants (\ref{dozz}) satisfies crossing symmetry was established in \cite{Ponsot:1999uf}.  The four-point function is constructed from the DOZZ structure constants as
\ie
\label{4pt}
& \langle \mathcal{V}_{\alpha_1}(z, \bar z) \mathcal{V}_{\alpha_2}(0) \mathcal{V}_{\alpha_3}(1) \mathcal{V}_{\alpha_4}(\infty) \rangle 
\\
&= \left(\prod_{j=1}^4S(\alpha_j)^{-\half}\right)\int_0^\infty {dP \over \pi} C(\A_1, \A_2, {Q\over2}+iP) C(\A_3, \A_4, {Q\over2}-iP) \left|F_c^{\rm Vir}\left({\Delta_i\over 2}; {\Delta_\A\over 2}; z\right)\right|^2
\\
&= \left[ \pi\mu \gamma(b^2) b^{2-2b^2} \right]^{-{Q\over b}} 
\int_0^\infty {dP \over \pi} \, {\cal C}(P_1, P_2, P) {\cal C}(P_3, P_4, P) \left|F_c^{\rm Vir}\left({\Delta_i\over 2}; {\Delta_\A\over 2}; z\right)\right|^2.
\fe
Note that the OPE coefficients ${\cal C}(P_1, P_2, P)$ are real for real Liouville momenta $P_1, P_2, P$ provided $c>1$, even if $b$ is complex (when $1<c<25$). The $\mu$-dependent prefactor can be absorbed by redefinining the normalization of sphere correlators as well as that of the primary operators themselves.

Although modular invariance demands that the Liouville momentum $P$ is real for all primary operators ${\cal V}_\alpha$ in the Hilbert space (this is also seen directly from canonical quantization of Liouville theory on the cylinder), we may analytically continue $\alpha_i$ to purely imaginary $P_i$. The analytically continued~\eqref{dozz} continues to obey the crossing equation and unitarity, provided that poles of $C(\A_1, \A_2, {Q\over 2}+iP)C(\A_3, \A_4, {Q\over 2}-iP)$ in $P$ do not cross the $P$-integration contour. If a pole crosses the integration contour, the crossing invariant 4-point function would pick up a residue contribution which may violate unitarity. This is indeed the case, as seen in section~\ref{ulbounds}.

\section{The BTZ spectral density}\label{app:BTZSpectral}
In this section we will evaluate the modular spectral function of perturbative pure gravity, including one-loop corrections. Restricting to $\tau=i\beta$, we can write the BTZ contribution to the reduced partition function as the modular $S$ transformation of the vacuum character \cite{Maloney:2007ud, Giombi:2008vd}
\ie
\hat Z_{\rm BTZ}(\beta) =&  \beta^{-1/2} e^{4\pi \xi\over \beta}(1-e^{-{2\pi\over \beta}})^2\\
=& \int_{2\xi}^\infty d\Delta~\beta^{1/2}e^{-2\pi\beta(\Delta-2\xi)}\rho_{\rm BTZ}(\Delta)
\fe
where we have used that $|\eta(i/\beta)|^2 = \beta |\eta(i\beta)|^2$. Applying an inverse Laplace transform, we can derive the BTZ spectral density
\ie
\rho_{\rm BTZ}(\Delta) = 2\pi \sum_{n=0}^2 C_n I_0\left(4\pi\sqrt{(2\xi-n)(\Delta-2\xi)}\right),
\fe
where $C_0 = 1, C_1 = -2, C_2 = 1$. Note that the other known saddle points of the gravitational path integral, related by $SL(2,\mathbb{Z})$ transformations, are always non-perturbatively suppressed for purely imaginary $\tau$.

Thus the perturbative pure gravity modular spectral function, obtained from the thermal $AdS_3$ and Euclidean BTZ saddle points in the gravitational path integral, can be written as
\ie\label{eq:BTZSpectral}
f_{\rm mod}^{\rm BTZ}(\Delta_*) =& \left.{1\over \hat{Z}_{\text{AdS$_3$}}(\beta)+\hat{Z}_{\rm BTZ}(\beta)}\left[\hat{Z}_{\text{AdS$_3$}}(\beta)+\int_{2\xi}^{\Delta_*}d\Delta\beta^{1/2}e^{-2\pi\beta(\Delta-2\xi)}\rho_{\rm BTZ}(\Delta)\right]\right|_{\beta=1}\\
=& \half +{1\over 2 e^{4\pi \xi}(1-e^{-2\pi})^2}\int_{2\xi}^{\Delta_*}d\Delta e^{-2\pi(\Delta-2\xi)}\rho_{\rm BTZ}(\Delta).
\fe

We are interested in the behaviour of this function for $\Delta_*$ in a window of size $\sim\sqrt{c}$ about ${c\over 6}$ in the semiclassical limit. From the asymptotic form of the Bessel function, it is easy to see that for $y\sim\mathcal{O}(1)$, we have
\ie
\rho_{\rm BTZ}(\Delta_*={c\over 6}+y\sqrt{c}) \approx 2(1-e^{-2\pi})^2\sqrt{3\over 2c}e^{2\pi({c\over 6}+y\sqrt{c}-3y^2)}+\mathcal{O}(c^{-1}).
\fe
Defining $\bar{f}_{\rm mod}^{\rm BTZ}(y_*) = f_{\rm mod}^{\rm BTZ}(\Delta_* = {c\over 6}+y_*\sqrt{c})$, we 
end up with the modular spectral function
\ie\label{eq:BTZSpectralLargeC}
\bar{f}_{\rm mod}^{\rm BTZ}(y_*) 
\approx 
 {3\over 4}+{1\over 4}\text{Erf}(\sqrt{6\pi}y_*),
\fe
where we have kept only the leading terms in the semiclassical approximation. This is the same as the spectral function one would obtain from applying the ``naive" Cardy formula (\ref{specbtz}).

\section{Details of the numerical computations}

\subsection{Details of the solution of the semidefinite problem}\label{app:semidefinite}
Here we provide some details of the numerical computations of the bounds on the spectral functions, implemented using the SDPB package \cite{Simmons-Duffin:2015qma}. In practice, there are several truncations that must be made. First, we must restrict to a finite basis of linear functionals acting on the crossing equation, of total derivative order $N$. We must also approximate the Virasoro conformal blocks; we can only compute the blocks to a finite order $d_q$ in the elliptic nome $q(z)$ from Zamolodchikov's recurrence relation (reviewed in Appendix~\ref{App:Zam}). Finally, recall that the upper and lower bounds on the spectral function are derived as the minimal coefficients such that a certain set of positivity conditions (for instance (\ref{eq:FourPtUpperInequality}) or (\ref{eq:ModularUpperInequality})) can be satisfied by a linear combination of derivatives of the conformal blocks or characters evaluated at the crossing symmetric point. In practice, we can only impose the positivity conditions on the blocks or characters of a finite set of spins in the spectrum; we denote the maximal spin considered by $s_{\rm max}$. 

The truncation on spin means that for a fixed derivative order $N$, we will not have taken into account all inequalities that the coefficients in (\ref{eq:FourPtUpperInequality}) or (\ref{eq:ModularUpperInequality}) must satisfy to constitute a bound on the spectral function, leading to bounds that are in principle too strong. Meanwhile, the truncation to finite $d_q$ introduces a controlled error into the computation of the (derivatives of the) conformal blocks evaluated at the crossing-symmetric point. Thus to derive bounds at a fixed $N$, we must ensure that both $s_{\rm max}$ and $d_q$ are sufficiently large so that a bound exists and is stable against further increasing these parameters to within our numerical precision. It is worth emphasizing that while the truncations to finite $s_{\rm max}$ and $d_q$ are controlled approximations, when these parameters are sufficiently large the bounds derived using a fixed derivative order $N$ are rigorous. Of course, the optimal bounds are obtained in the $N\to\infty$ limit.

Let us begin by discussing the bounds on the sphere four-point spectral function in the case that there are only scalar primaries in the spectrum. Of course in this case we need not worry about the spin truncation. We should note that in practice, the inequalities we feed into semidefinite programming are not quite of the form (\ref{eq:FourPtUpperInequality},\ref{eq:FourPtLowerInequality}), for the simple reason that the Virasoro blocks are not polynomials in the dimension of the internal primary. To illustrate the procedure, we write
\ie\label{eq:BlockApproximation}
\mathcal{F}_{12;0,\Delta}(z,\bar z) =& (256 q\bar q)^{{\Delta\over 2}-\xi}\mathcal{P}_{12}(\Delta;q,\bar q)+\mathcal{O}(q^{d_q},\bar q^{d_q})\\
\fe
where $\mathcal{P}_{12}(\Delta;q,\bar q)$ is a binomial in $q,\bar q$ with $\Delta$-dependent coefficients. Derivatives of the blocks can then be cast in terms of
\ie
\left.\partial_z^n\partial_{\bar z}^m(q\bar q)^{{\Delta\over 2}-\xi}\mathcal{P}_{12}(\Delta;q,\bar q)\right|_{z=\bar z=\half} = {(16e^{-\pi})^\Delta\over Q(\Delta)}P_{n,m}(\Delta)
\fe
where $P_{n,m}(\Delta)$ is a polynomial in $\Delta$ and
\ie
Q(\Delta) = \prod_{i}(\Delta-\Delta_i)^2,
\fe
where $\Delta_i$ are the locations of the poles kept at the given order of approximation in the computation of the Virasoro block. Importantly, the prefactor $(16e^{-\pi})^\Delta Q^{-1}(\Delta)$ is non-negative for unitary values of the internal dimension. The positivity conditions (\ref{eq:FourPtUpperInequality}), (\ref{eq:FourPtLowerInequality}) then amount to the following 
\ie
(y_{0,0}-\Theta(\Delta_*-\Delta))P_{0,0}(\Delta) + \sum_{1\le m+n\le N,\text{ odd}}y_{m,n}P_{m,n}(\Delta)\ge & 0,~\Delta\ge 0\\
(w_{0,0}+\Theta(\Delta_*-\Delta))P_{0,0}(\Delta) + \sum_{1\le m+n\le N,\text{ odd}}w_{m,n}P_{m,n}(\Delta)\ge & 0,~\Delta\ge 0.
\fe

\begin{figure}[h!]
\centering
\includegraphics[width=.5\textwidth]{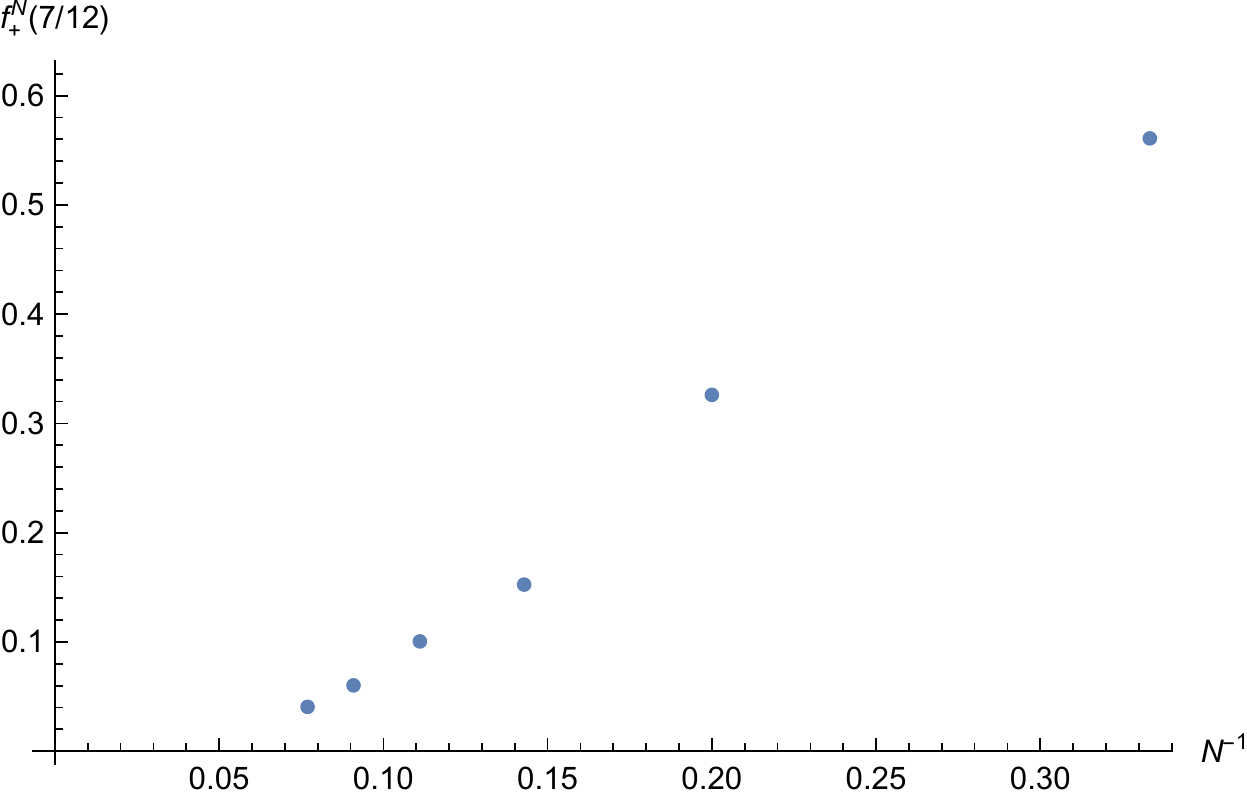}
\caption{The upper bounds  $f^+_N(\Delta_*={7\over 12})$ for $c=8$, $\Delta_\phi = {7\over 12}$ as a function of the inverse derivative order.}\label{fig:LiouvilleBoundsConvergence}
\end{figure}

In the case that there are only scalar primaries in the spectrum, we must take particular care to ensure that $d_q$ is sufficiently large, for the reason that the four-point function when decomposed into Virasoro blocks truncated at a finite order in $q$ would appear to have contributions from primaries of nonzero spin. Upper and lower bounds on the spectral function in this case can only be found numerically at a fixed derivative order $N$ when $d_q$ is sufficiently large. Empirically, for central charges and derivative orders in the ranges considered in section~\ref{ulbounds}, we find that $d_q = 4N$ is sufficient to compute stable bounds on the spectral function. To illustrate the convergence of the bounds as the derivative order is increased, Figure~\ref{fig:LiouvilleBoundsConvergence} shows the upper bound on the spectral function $f_+^N(\Delta_* = {7\over 12})$ as a function of $N^{-1}$ for $c=8$ with the external operator dimensions at the Liouville threshold. It is clear that we have not been able to access $N$ sufficiently large so that extrapolation to the $N\to\infty$ limit can be reliably performed.

Since the $q$-truncation order is the bottleneck for the speed of the numerical computations, this limits the range of derivative orders we are able to consider in computing bounds on the spectral function. For this reason, it is convenient to consider bounds obtained by further truncating the basis of linear functionals to $\left.\partial_z^n\partial_{\bar z}^m\right|_{z=\bar z=\half}$ with $m+n\le N$ and either $m\le 1$ or $n\le 1$. This basis leads to weaker bounds at a fixed $N$, but renders bounds at larger $N$ accessible. For instance, as shown in section~\ref{reducedbasis}, we are able to compute bounds on the spectral function up to $N=25$ with $d_q = N+9$ using linear functionals in this reduced basis. However, we caution that it is not clear that the $N\to\infty$ limit of the bounds obtained using this reduced basis of linear functions converges to that of the full-basis bounds.

We now turn to the bounds on the modular spectral function. The implementation of the positivity conditions (\ref{eq:ModularUpperInequality}) and (\ref{eq:ModularLowerInequality}) proceeds similarly as in deriving bounds on the four-point spectral function; here, for each spin one simply factors out $\left.q^{{\Delta+s\over 2}-\xi}\bar q^{{\Delta-s\over 2}-\xi}\right|_{z=\bar z=\half}$ to reduce derivatives of the reduced Virasoro characters to polynomials in the primary operator dimension. Now, although the Virasoro characters are known exactly, one must contend with the fact that the positivity conditions can only be imposed on a finite set of spins. Empirically, for values of the central charge up to those considered in section~\ref{sec:LargeC} ($c\sim \mathcal{O}(10^2)$), the truncation $s_{\rm max} = N+10$ is sufficient to ensure stable bounds on the modular spectral function.

\subsection{Details of the solution of the linear problem}

\label{app:linear}
In this subsection we discuss some details of the numerical evaluation of the inner products relevant for solving the linear problem in seciton~\ref{linearprob}. Note that the norm of $v$ is given by
\beq
	\crr{v,v}=\int_0^\infty d\Delta C^4_{12;0,\Delta}\frac{f_p(\Delta)}{f_v(\Delta)}.
\eeq
Finiteness of this norm requires the spectrum to be continuous and $ C^4_{12;0,\Delta}\frac{f_p(\Delta)}{f_v(\Delta)}$ to be locally integrable and decaying sufficiently quickly. If $C^4_{12;0,\Delta}$ gives a convergent OPE expansion for $|z|<1$, it should decay at infinity at least as $16^{-2\Delta}$. The decay condition is therefore automatically satisfied if $\frac{f_p(\Delta)}{f_v(\Delta)}$ grows slower than $16^{2\Delta}$.

We need also to ensure that $p_{n,m}$ have finite norm. The norm is
\beq
	\crr{p_{n,m},p_{n,m}}=\int_0^\infty (\partial^n_z\Fcal_{12;0,\Delta})^2(\partial^m_{\bar z}\Fcal_{12;0,\Delta})^2\frac{f_v(\Delta)}{f_p(\Delta)}d\Delta,
\eeq
the integrand behaves as 
\beq
	\text{polynomial}\times (16q)^{2\Delta}\frac{f_v(\Delta)}{f_p(\Delta)},
\eeq
where $q=e^{-\pi}$, and we get that the ratio $\frac{f_v(\Delta)}{f_p(\Delta)}$ should grow slower than $(16q)^{-2\Delta}$. We then choose this ratio to be
\beq
	\frac{f_v(h)}{f_p(h)}=(16q)^{-2\Delta}e^{-\lambda \Delta},\label{eq:measure}
\eeq
where $\lambda\in(0,2\pi)$. We will set $\lambda=\pi$.

Let us describe the details of the calculation. We want to compute the Gram matrix $\crr{p_{n,m},p_{n',m'}}$. For this, we need to be able to compute integrals with the Virasoro conformal blocks. 
Recalling~\eqref{eq:BlockApproximation}, the inner product $\crr{p_{n,m},p_{n',m'}}$ is given by
\beq	
	\crr{p_{n,m},p_{n',m'}}=\int_0^\infty P_{n,m}(\Delta)P_{n',m'}(\Delta)\frac{e^{-\lambda \Delta}}{Q^2(\Delta)}d\Delta.
\eeq
It is a standard fact that such integrals can be evaluated in terms of incomplete gamma functions. However, we want to do this efficiently, since $P$ and $Q$ are high-degree polynomials.\footnote{For example, in order to do calculations for $c=8,\,\Delta_\phi=\frac{c-1}{12},\,N=25$, it is necessary to compute the $q$-expansion of the conformal blocks to the order $q^{100}$ (see below). At this order $\deg P_{25,0}=679$ and $Q$ has $327$ zeroes.} The computation of products $P_{n,m}(\Delta)P_{n',m'}(\Delta)$ can be optimized by means of fast Fourier transform. After the product is computed, it suffices to compute the integrals 
\beq
\int_0^\infty \frac{\Delta^k e^{-\lambda \Delta}}{Q^2(\Delta)}d\Delta.
\eeq 
To do that, we first write
\beq
	Q^{-2}(\Delta)=\sum_i \sum_{k=1}^4 \frac{\alpha_{i,k}}{(\Delta-\Delta_i)^k},
\eeq
thus reducing the problem to the integrals of the form 
\beq
	\int_{0}^\infty \frac{\Delta^ne^{-\lambda \Delta}}{(\Delta-\Delta_i)^k}d\Delta.
\eeq
Reduction to incomplete gamma function is immediate if we shift $\Delta$ by $\Delta_i$. However, in this case we need to expand $(\Delta+\Delta_i)^n$ which produces $n$ terms, and $n$ can be large. Instead we write
\al{
	\int_{0}^\infty \frac{\Delta^ne^{-\lambda \Delta}}{(\Delta-\Delta_i)^k}d\Delta
	&=\frac{e^{-\lambda \Delta_i}\Gamma(n+1)}{\Gamma(k)}\sum_{l=0}^{k-1}\binom{k-1}{l}(-\lambda)^{k-1-l}\Delta_i^{l-n}\Gamma(l-n;-\Delta_i\lambda).
}
Here $k$ is bounded by $4$, so we get a compact sum. The parameter $n$ does enter into the incomplete gamma function, but it satisfies a recursive relation which allows one to compute it as $n$ increases, effectively making the complexity of computation of this integral $O(1)$ for every value of $n$.

Having found the Gram matrix, it is immediate to find the coefficients of the expansion of $v_N$ in the basis $p_{n,m}$; they are given by the first row of the inverse of the Gram matrix. Computation of $P_N\theta_{\Delta_*}$ proceeds similarly, except that now we need to know all the inner products $\crr{p_{n,m},\theta_{\Delta_*}}$. These can be computed just as above, shifting everything by $\Delta_*$. 
In practice, we find that the basis of $p_{n,m}$ is ill-conditioned and thus we need to know the Gram matrix to a high precision. This typically demands a large $q$-truncation order $d_q$. For example, in Figure~\ref{fig:linear1}(a) and Figure~\ref{fig:linear2} in section~\ref{linearprob}, values of $d_q$ between $60$ and $100$ were used. In Figure~\ref{fig:linear1}(b), however, we needed to go to $d_q=200$. In general, the required value of $d_q$ grows with $N$, similarly to what we observed in semidefinite problems. On the other hand, the linear method computes faster than the semidefinite one, which allows us to study much higher values of $N$. 

There is a small subtlety in the computation of $v_N$ for the mixed correlator, due to the fact that $Q(\Delta)$ has a double zero at $\Delta=0$, which in the case $\Delta_1\neq \Delta_2$ is not cancelled by zeroes of $P_{n,m}$. In this case (e.g. in the Figure~\ref{fig:linear2}) we have  tried two approaches. The first approach is introducing lower bound on the intermediate scaling dimension $\Delta_{gap}$ below the Liouville threshold. The second approach is to modify equation~\eqref{eq:measure} as 
\beq
	\frac{f_v(h)}{f_p(h)}=(16q)^{-2\Delta}e^{-\lambda \Delta}\Delta^4.\label{eq:modified_measure}
\eeq
While the obtained results differ slightly at small $N$, already at $N=13$ both provide equally good approximations for the Liouville spectral functions in Figure~\ref{fig:linear2}.

\subsubsection{Numerical checks of completeness}
\label{app:linear_completeness}

\begin{figure}[h!]
\centering
\subfloat{\includegraphics[width=.49\textwidth]{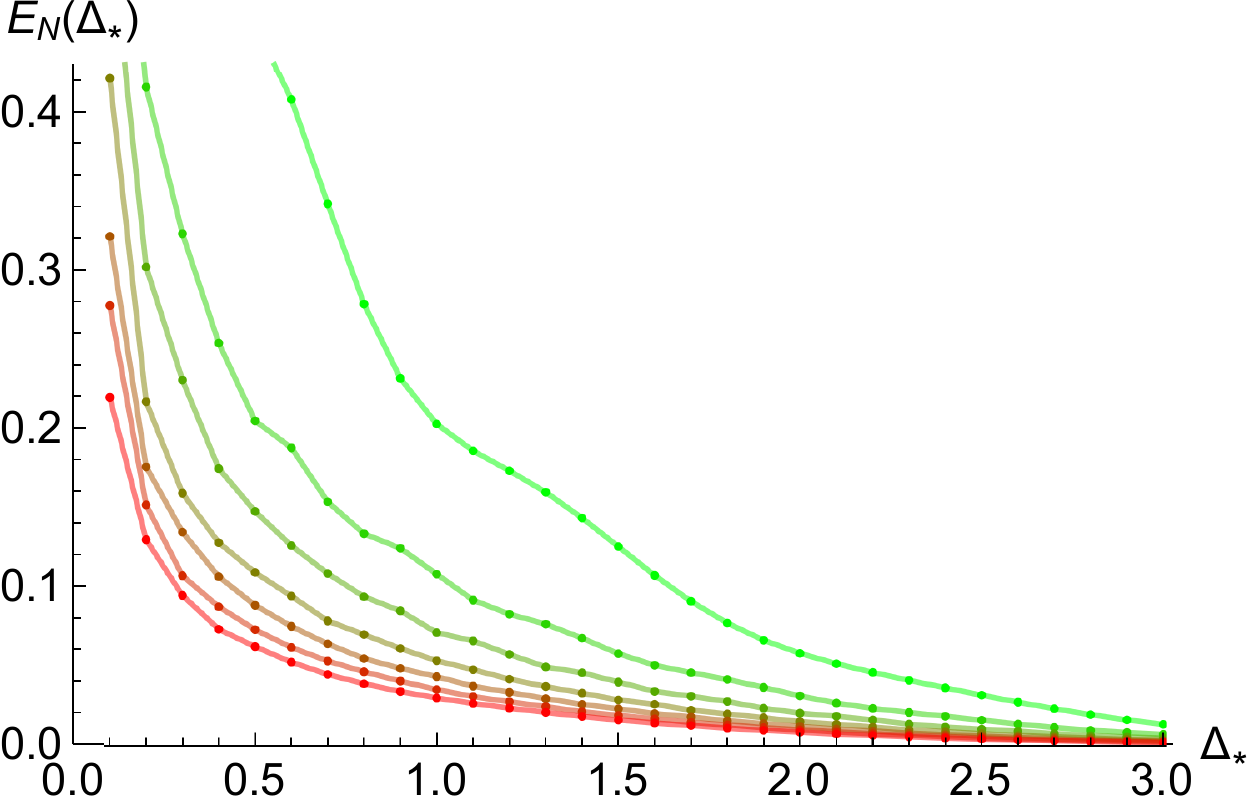}}~
\subfloat{
\includegraphics[width=.49\textwidth]{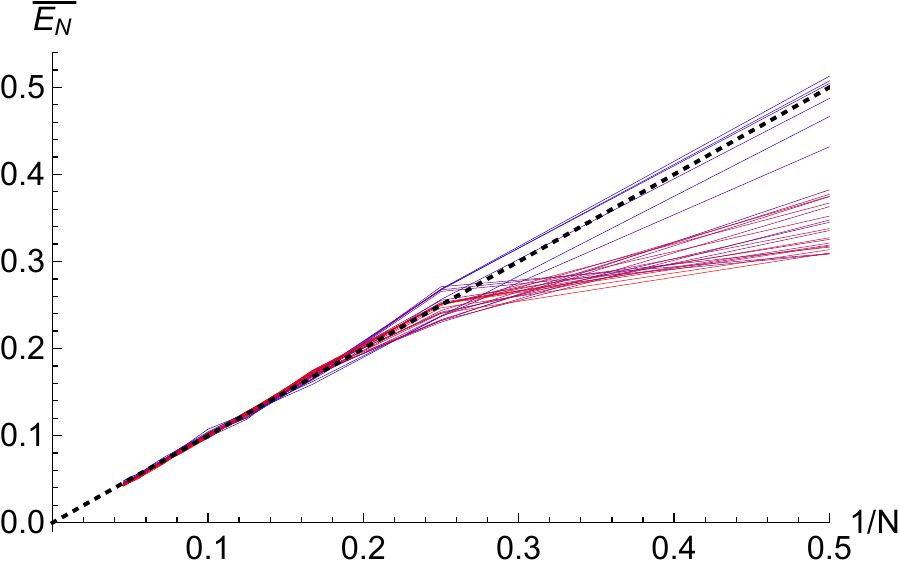}}
\caption{Plots of approximation errors for $n+m$ even. \textbf{Left}: $E_N$ as a function of $\Delta_*$ for $N$ from $4$ (green) to $28$ (red) in steps of $4$. \textbf{Right}: Normalized $E_N$ as a function of $N^{-1}$ for $\Delta_*$ from $0.4$ (blue) to $1.5$ (red). The dashed black line is shown for comparison and has slope $1$.}\label{fig:evenEN}
\end{figure}

\begin{figure}[h!]
\centering
\subfloat{\includegraphics[width=.49\textwidth]{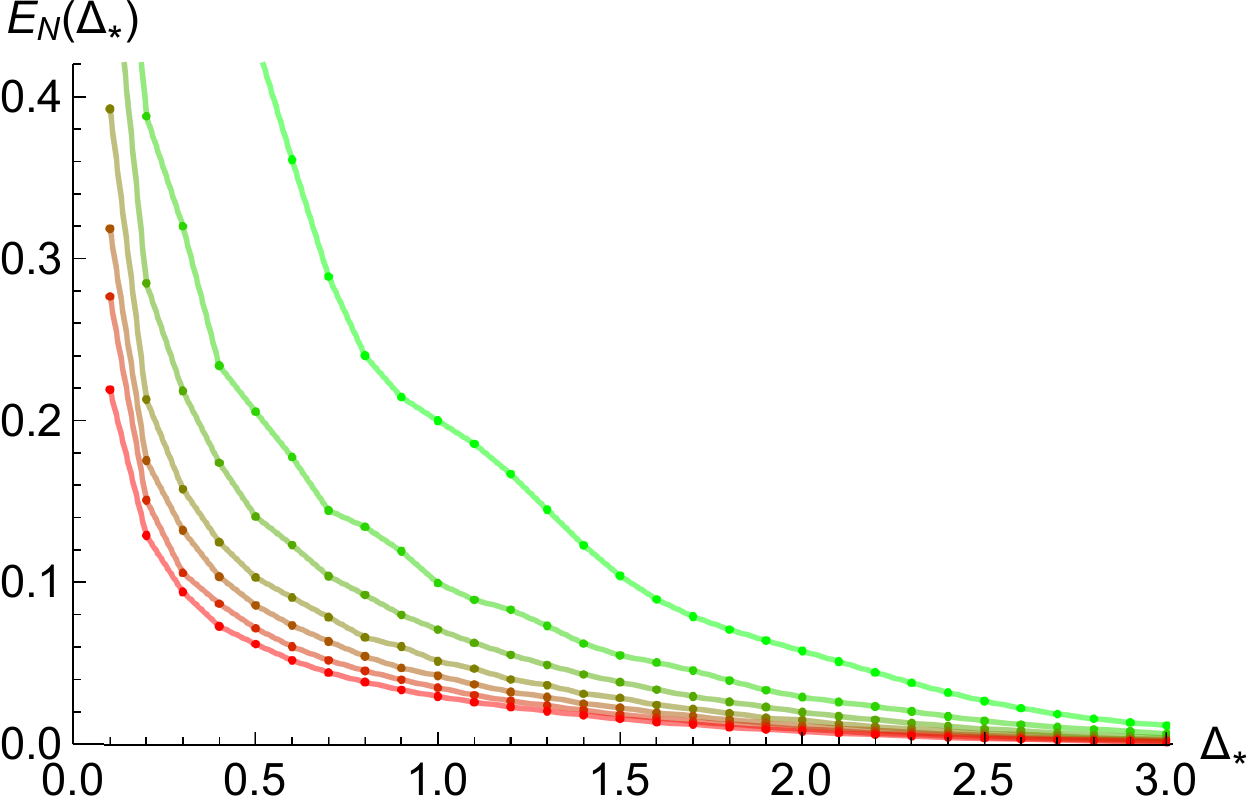}}~
\subfloat{
\includegraphics[width=.49\textwidth]{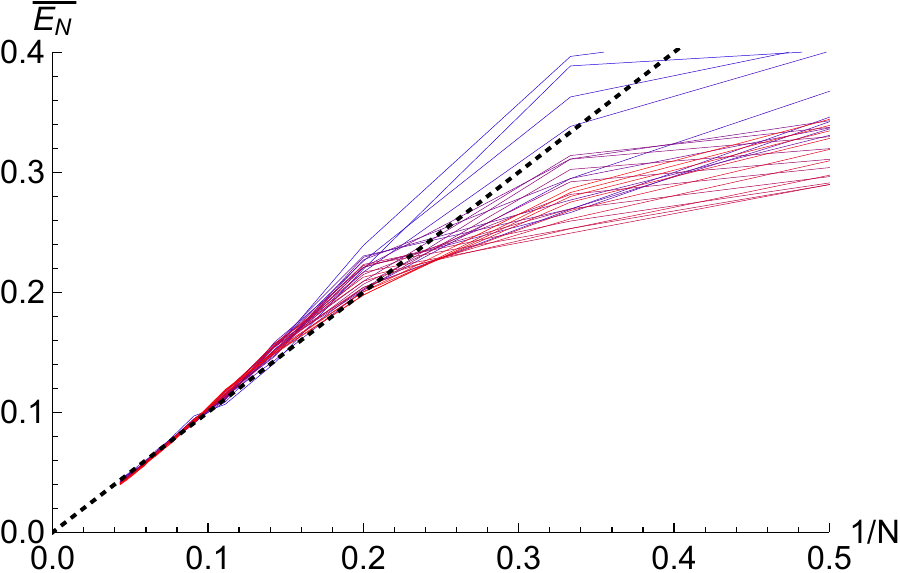}}
\caption{Plots of approximation errors for $n+m$ odd. \textbf{Left}: $E_N$ as a function of $\Delta_*$ for $N$ from $5$ (green) to $29$ (red) in steps of $4$. \textbf{Right}: Normalized $E_N$ as a function of $N^{-1}$ for $\Delta_*$ from $0.4$ (blue) to $1.5$ (red). The dashed black line is shown for comparison and has slope $1$.}\label{fig:oddEN}
\end{figure}

Here we consider the question of completeness of the systems
\al{
	\mathcal{B}_{even}&=\{\left.\partial_z^n \partial_{\bar z}^m F_{12;0,\Delta}\right|_{z=\bar z= {1\over 2}}, n,m\in \mathbb{Z}_{\geq 0}, n+m~{\rm even}\},\\
	\mathcal{B}_{odd}&=\{\left.\partial_z^n \partial_{\bar z}^m F_{12;0,\Delta}\right|_{z=\bar z= {1\over 2}}, n,m\in \mathbb{Z}_{\geq 0}, n+m~{\rm odd}\}\cup \{\left.F_{12;0,\Delta}\right|_{z=\bar z= {1\over 2}}\}
}
with respect to the measure described above. In both cases we attempt to approximate the step functions $\theta_{\Delta_*}\simeq P_N \theta_{\Delta_*}$, where $P_N$ is the projection onto the subspace spanned by elements of either system with $n+m\leq N$, and compute the residual errors
\beq
	E_N=\frac{|(1-P_N)\theta_{\Delta_*}|^2}{|\theta_{\Delta_*}|^2}.
\eeq
We do this for a range of $\Delta_*$ in the case of the mixed correlator with $c=8$, $\Delta_1=\Delta_0,\,\Delta_2=\frac{12}{7}\Delta_0$. The results are shown in Figure~\ref{fig:evenEN} for $\mathcal{B}_{even}$ and in Figure~\ref{fig:oddEN} for $\mathcal{B}_{odd}$, consistent with the completeness of both bases. In the plots of $E_N$ as a function of $N^{-1}$, we have rescaled $E_N(\Delta_*)$ by an $N$-independent factor for each sample value of $\Delta_*$ (denoted by $\overline{E_N}$) so that for all $\Delta_*$ the slope of the linear fit with $N^{-1}$ (which appears to be valid asymptotically for large $N$) is approximately $1$.

\subsection{Bounds from a reduced basis of linear functionals}
\label{reducedbasis}

\begin{figure}[h!]
\centering
\subfloat{\includegraphics[width=.49\textwidth]{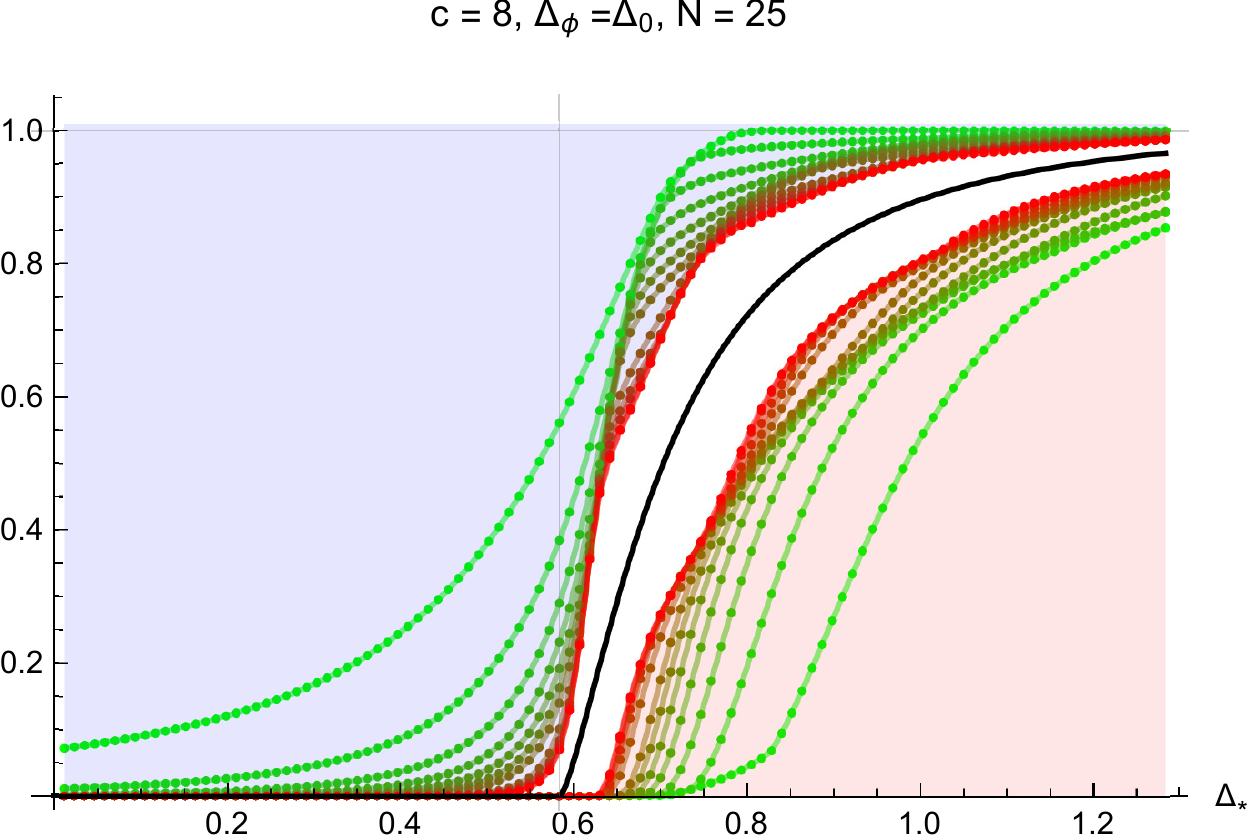}}~
\subfloat{\includegraphics[width=.49\textwidth]{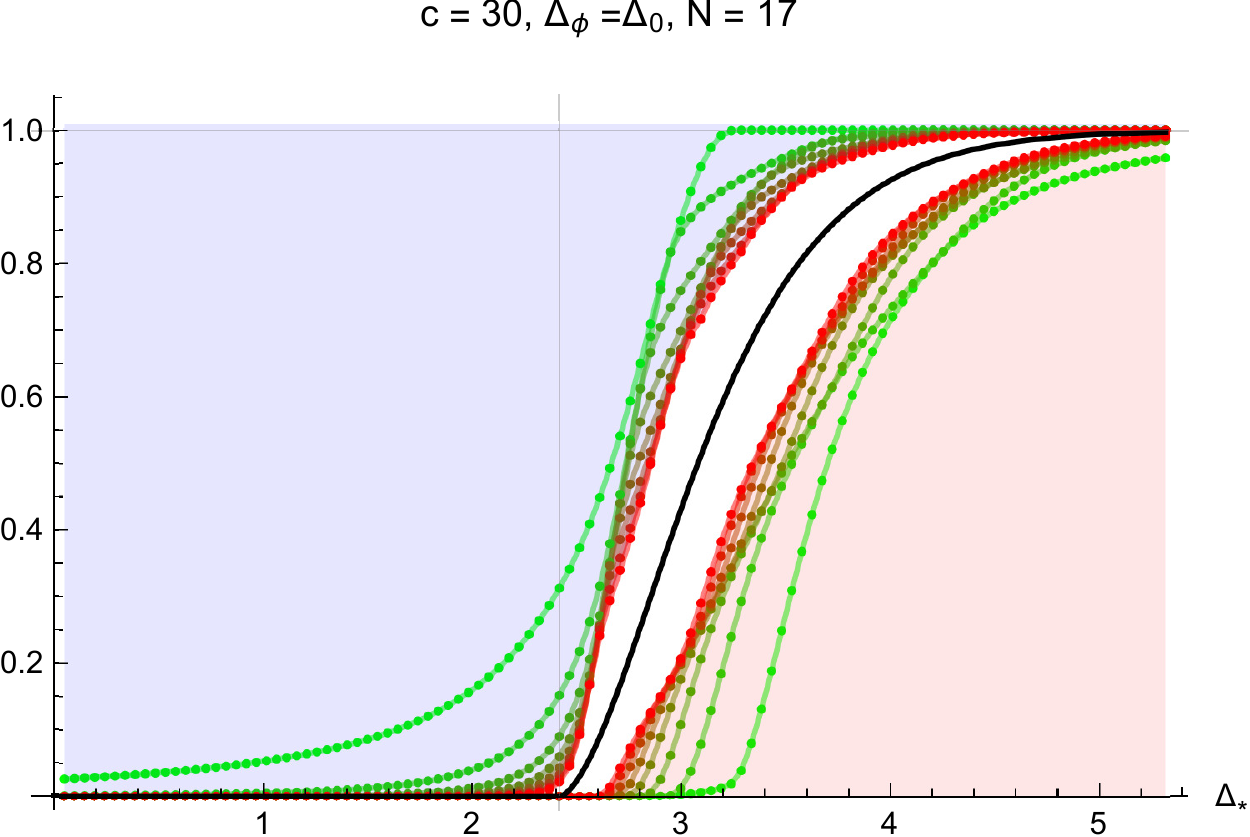}}
\caption{Upper and lower bounds on the spectral function from from linear functionals in the reduced basis, assuming only scalar primaries and $\Delta_\phi = {c-1\over 12}$ for $c=8$ (left) and $c=30$ (right). The black curve denotes the corresponding spectral function of Liouville theory.}\label{fig:LiouvilleBoundsReduced}
\end{figure}

Here we consider the bounds on the scalar-only spectral function using the following reduced basis of linear functionals
\ie
\label{mnres}
\left.\partial_z^n\partial_{\bar z}^m\right|_{z=\bar z=\half},~~~m\leq 1 {\rm ~or~} n\leq 1,~~m+n\le N.
\fe
Figure~\ref{fig:LiouvilleBoundsReduced} shows the reduced-basis bounds for $c=8$ and $c=30$ with $\Delta_{\phi} = \Delta_0$.
Clearly at fixed $N$ the bounds obtained using the reduced basis would be weaker, but due to the simplicity of the reduced basis it is now possible to access bounds at higher $N$ within the same computing time. This provides a useful arena to study the convergence of the bounds at large derivative orders, with the caveat that the $N\to\infty$ limit of the bounds obtained from the reduced basis are likely weaker than the optimal bounds from the most general linear functionals. If the latter is the case, one may eventually need to relax the restriction on ${\min}(m,n)$ in~\eqref{mnres}.

\bibliographystyle{JHEP}
\bibliography{SpectralSubmission.bib}

\end{document}